\documentclass[11pt,a4paper]{article}
\pdfoutput=1
\usepackage[pdftex]{graphics}
\usepackage{jheppub}
\usepackage{amsmath,amssymb,amsfonts}
\usepackage[force]{feynmp-auto}
\usepackage{cancel}
\usepackage{caption}
\usepackage{subcaption}

\newcommand{\be}{\begin{eqnarray}}
\newcommand{\ee}{\end{eqnarray}}
\newcommand{\nn}{\nonumber}
\newcommand{\bn}{\begin{enumerate}}
\newcommand{\en}{\end{enumerate}}
\newcommand{\bl}{\begin{align}}
\newcommand{\el}{\end{align}}

\def\IR{\mathbb{R}}
\def\IZ{\mathbb{Z}}

\def\CI{\mathcal{I}}

\def\CN{\mathcal{N}}
\def\CO{\mathcal{O}}

\def\a{\alpha}
\def\b{\beta}
\def\g{\gamma}
\def\e{\epsilon}
\def\ve{\varepsilon}
\def\z{\zeta}
\def\th{\theta}

\def\k{\kappa}
\def\l{\lambda}
\def\m{\mu}
\def\n{\nu}
\def\r{\rho}

\def\s{\sigma}

\def\t{\tau}

\def\w{\omega}
\def\G{\Gamma}
\def\D{\Delta}
\def\Th{\Theta}
\def\L{\Lambda}

\def\det{{\rm det}}

\def\bz{\bar{z}}

\def\da{{\dot{\a}}}
\def\db{{\dot{\b}}}

\def\jmath{{j}}

\def\d{{\delta}}
\def\be{{\bar{\epsilon}}}

\def\mem{\hspace{0.1em}}

\def\Re{{\operatorname{Re}}}
\def\Im{{\operatorname{Im}}}

\newcommand{\wrap}[1]{{\smash{#1}\vphantom{\b}}}

\def\mem{\hspace{0.1em}}

\def\hnem{\hspace{-0.05em}}
\def\hhem{\hspace{0.025em}}

\def\mdot{{\mem\cdot\mem}}

\usepackage{graphicx}
\usepackage[export]{adjustbox}

\usepackage{bm}

\def\rmx{{
    \scalebox{1.2}[1]{$\mathrm{x}$}\kern-0.625em\scalebox{1.2}[1]{$\mathrm{x}$}
}}
\def\rmy{{
    \scalebox{1.2}[1]{$\mathrm{y}$}\kern-0.645em\scalebox{1.2}[1]{$\mathrm{y}$}\kern-0.02em
}}
\def\rmz{{
    \kern0.055em\scalebox{1.2}[1]{$\mathrm{z}$}\kern-0.528em\scalebox{1.2}[1]{$\mathrm{z}$}\kern0.04em
}}

\newcommand{\dbar}{
    d\kern-.20em\makebox[0pt][l]{$\bar{}$}\kern.20em
}
\newcommand{\deltabar}{
    \delta\kern-.20em\makebox[0pt][l]{$\bar{}$}\kern.20em
}

\usepackage{mathtools}

\def\minie{{\textstyle\frac{1}{2}}}

\newcommand{\lambdabar}{
    \lambda\kern-.20em\makebox[0pt][l]{$\bar{}$}\kern.20em
}

\def\siF{{\smash{{*}^{-1}\hnem F}\vphantom{F}}}

\def\si{{{*}^{-1}\hnem}}
\def\J{{\smash{J^\star}\vphantom{J}}}

\def\mdot{{\mem\cdot\mem}}

\usepackage[usenames,dvipsnames]{xcolor}
\definecolor{jh}{RGB}{0,156,25}

\newcommand{\rred}[1]{\textcolor{red}{\; #1\; }}
\usepackage{soul}

\title{Massive twistor worldline in electromagnetic fields} 

\author[a]{Joon-Hwi Kim}
\author[b]{Jung-Wook Kim}
\author[c,d,e]{Sangmin Lee} 

\affiliation[a]{Department of Physics, California Institute of Technology, 
\\
1200 E California Blvd, Pasadena, CA 91125, U.S.A.}
\affiliation[b]{Max Planck Institute for Gravitational Physics (Albert Einstein Institute),\\
Am M\"uhlenberg 1, D-14476 Potsdam, Germany}  
\affiliation[c]{Department of Physics and Astronomy, Seoul National University, \\
1 Gwanak-ro, Gwanak-gu, Seoul 08826, Korea}
\affiliation[d]{Center for Theoretical Physics, Seoul National University, \\
1 Gwanak-ro, Gwanak-gu, Seoul 08826, Korea}
\affiliation[e]{College of Liberal Studies, Seoul National University, \\
1 Gwanak-ro, Gwanak-gu, Seoul 08826, Korea}

\abstract{
We study the (ambi-)twistor model for spinning particles interacting via electromagnetic field, as a toy model for studying classical dynamics of gravitating bodies including effects of both spins to all orders. 
We compute the momentum kick and spin kick up to one-loop order and show precisely how they are encoded in the classical eikonal. 
The all-orders-in-spin effects are encoded as a dynamical implementation of the Newman-Janis shift, and we find that the expansion in both spins can be resummed to simple expressions in special kinematic configurations, at least up to one-loop order. 
We confirm that the classical eikonal can be understood as the generator of canonical transformations that map the in-states of a scattering process to the out-states. 
We also remark that cut contributions for converting worldline propagators from time-symmetric to retarded 
amount to the iterated action of the leading eikonal at one-loop order.
}

\emailAdd{joonhwi@caltech.edu, jung-wook.kim@aei.mpg.de, sangmin@snu.ac.kr}

\begin{document}
\maketitle

\section{Introduction}

Chandrasekhar has remarked that ``(t)he black holes of nature are the most perfect macroscopic objects there are in the universe: [...] they are the simplest objects as well.''~\cite{Chandrasekhar:1985kt} Can we idealise these simplest objects of the universe and make them even simpler? Since electromagnetic interactions are simpler than gravitational interactions, let us phrase this question more concretely in the context of electromagnetism. What would be the description of the simplest charged, massive, spinning (macroscopic) objects moving on a background electromagnetic field?

One class of charged spinning (macroscopic) objects that can be called ``simplest'' is known in the literature as root-Kerr particles~\cite{Arkani-Hamed:2019ymq}, which possess spin-induced multipole moments of Kerr-Newman black holes~\cite{Chung:2019yfs, Scheopner:2023rzp}. They can be called simplest in the sense that they correspond to the classical spin limit of ``minimal coupling'' defined by the high-energy limit~\cite{Arkani-Hamed:2017jhn}, and that all multipole moments are generated by the Newman-Janis shift~\cite{Newman:1965tw}, where the position of the particle sourcing the gravitational/electromagnetic field is complexified and shifted in the imaginary spin direction. In its original formulation, the Newman-Janis shift only applies to stationary solutions of the Einstein(-Maxwell) equations, therefore the answer to the question posed in the previous paragraph would only be complete when the Newman-Janis shift is generalised to dynamical worldlines of spinning bodies. 

In this work, improving upon the ideas of ref.~\cite{Guevara:2020xjx}, 
we argue that the twistorial description of relativistic spherical tops~\cite{Kim:2021rda} qualifies as a complete answer.
The authors showed in ref.~\cite{Kim:2021rda} that the spherical top model \cite{Hanson:1974qy} for a relativistic spinning particle is equivalent to a massive twistor model (similar but not identical to ref.~\cite{Fedoruk:2014vqa}) in the absence of interactions. The attempt to couple the twistor model to a background field was initiated in ref.~\cite{Kim:2023aff}. 
A complete description of the twistor model minimally coupled to electromagnetic field is given here.

We use the model to compute scattering observables, such as the momentum kick and the spin kick, at low orders in perturbation theory while maintaining exact spin-dependence. Following the nomenclature of ref.~\cite{Bern:2023ccb}, we call the perturbation theory ``post-Lorentzian" (PL) expansion, 
where $n$-PL order terms are suppressed by $(q_1 q_2)^n$ where $q_1$, $q_2$ are the electric charges of two interacting particles. 
When organised diagrammatically through Feynman-like diagrams, $n$-PL order terms involve $(n-1)$-loop momentum integrals, although the diagrams themselves have no loops.
This is a toy model for studying the gravitational case, where the dynamics is organised in the post-Minkowskian (PM) expansion while keeping the exact spin-dependence.

We stress that exact spin-dependence is not only of theoretical interest, but is also of phenomenological interest. 
The previous sentence may sound odd to a person familiar with post-Newtonian (PN) calculations: 
In the PN expansion spin effects are formally counted as 1PN, since the corrections take the form of $a/r$, and for compact objects such as black holes the spin length scales as the horizon scale $a \sim Gm$. For computing the gravitational waveforms\textemdash which are the quantities directly relevant for observations\textemdash the 4.5PN corrections seem to be good enough, at least for non-spinning quasi-circular equal mass binaries; the 4.5PN contributions add less-than-a-radian correction to the $\sim 10^{3 - 5}$ cumulative gravitational wave cycles in the detector frequency bands~\cite{Blanchet:2023bwj}. The spin effect corrections to the conservative dynamics has already been computed to 5PN order~\cite{Kim:2021rfj, Kim:2022pou, Kim:2022bwv, Levi:2022dqm, Levi:2022rrq, Mandal:2022nty, Mandal:2022ufb}. Why would we need all-orders-in-spin effects if they are going to be smaller than what is already known, which already seems to be sufficient for observations?\footnote{Caution: The PN expansion is known to converge best for equal-mass quasi-circular orbits, and convergence of the best-case scenario does not guarantee convergence in other regions of the parameter space. For example, even 5PN may not be enough to reduce systematic errors below the level of statistical errors~\cite{Owen:2023mid}.}

One reason all-orders-in-spin results can be of interest is because in practical applications the perturbative results need to be resummed for a better accuracy~\cite{Buonanno:1998gg, Antonelli:2019ytb, Khalil:2022ylj, Damour:2022ybd, Rettegno:2023ghr, Buonanno:2024vkx}.\footnote{One may also recall that the revival of interest in the PM expansion was partly kindled by the search for alternative resummation schemes of the gravitational two-body dynamics~\cite{Damour:2016gwp}.} The resummations reorganise the perturbative expansion by leveraging the knowledge of singularity structures that the non-perturbative answer is expected to possess. 
Therefore, all-orders-in-spin calculations may reveal singularity structures we can take advantage of in the resummations, which were not visible at the lowest spin orders. This can be used, for example, in improving effective-one-body based waveform models which are known to perform worse for extremal black hole spins~\cite{Ramos-Buades:2023ehm,Pompili:2023tna}, where spin effects are resummed as geodesic motion on a deformed Kerr geometry~\cite{Khalil:2023kep}. Understanding the mechanism behind the effectiveness of the resummation will be useful in motivating alternative resummation schemes for spin effects, which may yield better accuracy.

From the viewpoint of resummations, an all-orders-in-spin result that is as rigid as possible and as simple as possible while keeping essential features of the dynamics will be the most useful, since we are interested in the singularity structures of all-orders-in-spin dynamics; any additional structures or free parameters may obscure the singularity structures that we wish to dig up from all-orders-in-spin results. This motivation brings us back to the question raised in the beginning of this manuscript;  what is the simplest spinning object that interacts with the background Maxwell field? 
The motivation also limits the inputs of the theory to all-orders-in-spin multipole moment information encoded by the Newman-Janis shift.
The expectation is that while the dynamics of the spinning particle may deviate from that of physical black holes from $\CO(q^2)$, the singularity structures of the all-orders-in-spin dynamics are still captured by the twistor worldline model.

The twistor worldline model predicts surprisingly simple singularity structures in special kinematic configurations, which we expect to be shared by scattering dynamics of physical black holes. For example, the 2PL \emph{aligned-spin eikonal}  \eqref{eq:one-loop_eik_as_full} resums to the simple expression
\begin{align}
    \chi_{(2,\text{aligned})} &= \frac{(q_1 q_2)^2 \left( b^2 + \frac{(\z - 2)\g}{(\g^2 - 1)} \, \e[b, v_1, v_2, a] + \frac{\g^2 (1 - \z) + \z}{\g^2 - 1} \, a^2 \right)}{32 \pi m_1 \sqrt{\g^2 - 1} \, (b^2 - a^2)^{3/2}} + \left( 1 \leftrightarrow 2 \right) \,, \label{eq:one-loop_eik_as_full_copy}
\end{align}
where the impact parameter $b^\m \sim (x_1^\m - x_2^\m)_\perp$ is defined by the covariant spin supplementary conditions (SSC), $a^\m = a_1^\mu + a_2^\m$ is the sum of the spin-length vectors, and $\z$ is the ratio parameter defined by $a_1^\m = \z a^\m$. To the best of authors' knowledge, this is the first observation of spin effect resummation in binary dynamics at the next-to-leading order (NLO) in the coupling constant expansion, where the model is free of unphysical behaviour and the spins of both constituents are included to all orders.\footnote{The known spin-resummed NLO scattering angles reported in the literature~\cite{Guevara:2018wpp,Menezes:2022tcs} are based on the Compton amplitudes that develop unphysical behaviour from cubic (electromagnetism) or quintic (gravity) order in spin. The spin-resummed results reported by ref.~\cite{Brandhuber:2024bnz} should be considered as leading order effects in the $R^3$ coupling expansion.}

In addition to the construction of the interacting twistor model, and application of the model to compute observables to the 2PL order, another key result of this work is the clarification of the \emph{classical eikonal}'s role. 
In the Hamiltonian formulation of binary dynamics, the classical eikonal is defined as a suitable classical limit of the quantum eikonal phase that acts as the generator of canonical transformations, mapping the incoming scattering states to the outgoing states. The scattering states are defined in the phase space of free particles, and the eikonal encodes the interactions such that it produces all scattering observables through canonical transformations. 

The manuscript is organised as follows. In section \ref{sec:twistor-EM} we review the massive twistor model and couple it to background electromagnetic fields. In section \ref{sec: scattering observables} we compute scattering observables using equations of motion. We set up WQFT formulation of the model in section \ref{sec:TWQFT}, and use it to compute Compton amplitudes in section \ref{sec:Compton} and the classical eikonal in section \ref{sec:Eikonal}. We conclude our studies and propose future directions in section \ref{sec:discussion}.

\paragraph{Note added}
While this work was being completed, ref.~\cite{Gonzo:2024zxo} appeared. Their proposal for how to extract classical observables from the radial action overlaps with our discussion on the classical eikonal as the scattering generator in section~\ref{sec:eikonal-generator}.

 
\section{Massive twistor in electromagnetic field} \label{sec:twistor-EM}

\subsection{Free theory}

In ref.~\cite{Kim:2021rda}, we proposed a massive twistor model 
and showed its equivalence to the Hanson-Regge spherical top model at the free theory level. 
A general discussion of how to couple the twistor model to background fields was given in ref.~\cite{Kim:2023aff}. 
Here, we give a brief review of these two main references and clarify some aspects of the twistor model 
before specialising to the minimal coupling to a background electromagnetic field. 

\begin{figure}[htbp]
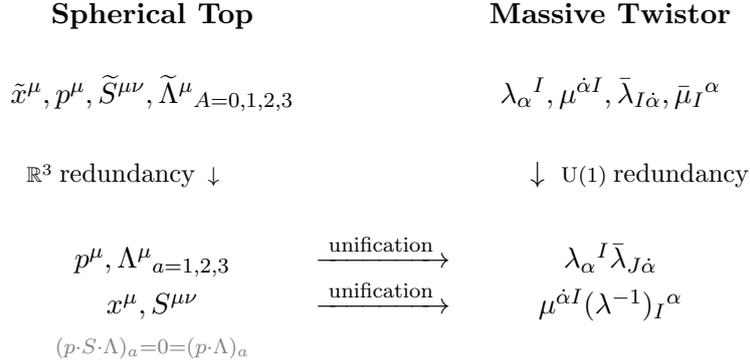

    \centering
    \begin{equation}
\begin{array}{ccccc}
     \mbox{\bf Spherical Top}  &  & & &  \mbox{\bf Massive Twistor} 
     \\
     \\
     \tilde{x}^\mu, p^\mu, \widetilde{S}^{\mu\nu},\widetilde{\Lambda}^\mu{}_{A=0,1,2,3} & & & & \lambda_\alpha{}^I, \mu^{\dot{\alpha} I}, \bar{\lambda}_{I\dot{\alpha}},\bar{\mu}_I{}^\alpha 
     \\
     \\
      \scriptstyle{\mathbb{R}^3} \; \mbox{\small redundancy} \;\; \downarrow \qquad & & & & \qquad \downarrow \;\;  \scriptstyle{\mathrm{U(1)}}\; \mbox{\small redundancy}
     \\
     \\
     p^\mu,\Lambda^\mu{}_{a=1,2,3} & & \xrightarrow[]{\mathrm{\; unification\; }} & & \lambda_\alpha{}^I \bar{\lambda}_{J\dot{\alpha}}
     \\
     x^\mu,S^{\mu\nu} & & \xrightarrow[]{\mathrm{\; unification\; }} & & \mu^{\dot{\alpha} I} (\lambda^{-1})_I{}^\alpha 
     \\
     \textcolor{gray}{\scriptstyle{(p\cdot S\cdot \Lambda)_a = 0 = (p\cdot \Lambda)_a}} & & & &
\end{array}
\nonumber 
\end{equation}
    \caption{Equivalence between the spherical top model and the massive twistor model.}
    \label{fig:free-equivalence}
\end{figure}

A widely used model for describing relativistic spinning particles is the Hanson-Regge spherical top \cite{Hanson:1974qy}, which uses the variables $( p_\m , \tilde{x}^\m , \tilde{\L}^\m{}_A , \tilde{S}_{\m\n} )$ to describe a spinning body; the momentum $p_\m$, the position $\tilde{x}^\m$, the body-fixed tetrad $\tilde{\L}^{\m}{}_A$ ($A=0,1,2,3$) describing the orientation of the body, and the spin tensor $\tilde{S}_{\m\n}$. 
The tilde notation emphasises the fact that under the ``spin-gauge" redundancy \cite{Steinhoff:2015ksa}, the variables $( \tilde{x}, \tilde{\L}, \tilde{S} )$ have some gauge dependency while $p$ is gauge-invariant. It was shown in ref.~\cite{Kim:2021rda} that the gauge orbit is $\mathbb{R}^3$ 
and the ``origin" of the orbit corresponds to the covariant gauge conditions: $p^\m S_{\m\n} \L^\n{}_a = 0 = p_\m \L^\m{}_a$ ($a=1,2,3$).

The twistor variables describe the same dynamics with less gauge redundancy. Their global symmetry groups are the superconformal SU$(2,2)$ (to be broken by the mass-shell condition) and the massive little-group SU$(2)$ . 
The gauge redundancy on the twistor side is U(1): $( \l, \m ) \rightarrow e^{i\theta} ( \l, \m )$, $( \bar{\l}, \bar{\m} ) \rightarrow e^{-i\theta} ( \bar{\l}, \bar{\m} )$, $\theta \in \mathbb{R}$. 

The equivalence of the two models is clearly articulated in terms of the gauge-invariant coordinates on both sides. 
The twistor model unifies the variables $( p_\m , \L^\m{}_a )$ using the hermitian bi-linear products of $( \l_\a{}^I , \bar{\l}_{I \dot\a} )$:
\begin{align}
    p^\m = \frac{1}{2} \bar{\s}^{\m \dot\a \a} \l_\a{}^I \bar{\l}_{I \dot\a} \,, \quad 
    \Lambda_{(IJ)}^\m = \frac{\bar{\s}^{\m \dot\a \a} \l_{\a (I} \bar{\l}_{J) \dot\a}}{\sqrt{2} m} \,,
\end{align}
in a similar vein to massive spinor-helicity variables~\cite{Conde:2016izb,Arkani-Hamed:2017jhn}.
An advantage of the Hamiltonian formulation is that nothing stops us from unifying momenta $p_\mu$ and ``generalised positions" $\Lambda^\m{}_a$.
The remaining variables $( x^\m , S_{\m\n} )$ are unified and mapped to 
a set of gauge-invariant complex variables $ \m^{\dot\a I} (\l^{-1})_I{}^\a $ and their complex conjugates.
This is in contrast to the supersymmetric worldline description~\cite{Gibbons:1993ap,Jakobsen:2021lvp,Jakobsen:2021zvh} 
where the spin tensor $S_{\mu\nu}$ is realised as a bi-linear in Grassmann variables $( \Psi_A, \bar{\Psi}_A )$ and 
the tetrad $\L^\m{}_a$ is not visible.

A prominent feature of the twistor model is that there is no spin-gauge redundancy 
in the first place, so that no discussion of the spin supplementary condition (SSC) is ever needed. 
The only constraint yet to be imposed is the mass-shell condition $p^2 + m^2 = 0$, 
which is common to the spherical top model and the twistor model. 
Figure~\ref{fig:free-equivalence} summarises the mapping between the two models before imposing the mass-shell constraint.

\subsubsection{Constraints revisited} 

Our conventions for spinors and twistors are slightly different from those of ref.~\cite{Kim:2021rda}; see appendix~\ref{sec:conventions} for details. 
Let us mention a few key relations. The fundamental Poisson brackets of the spherical top model include 
\begin{align}
    \{ x^\m, p_\n \} = \delta^\m_\n \;\; \Rightarrow \;\; \{ x^{\dot\a \a} , p_{\b \dot\b} \} = - 2 \delta^{\dot\a}_{\dot\b} \delta^\a_\b  \,.
\label{eq:def_PB_top}
\end{align}
The fundamental Poisson brackets of the twistor model are
\begin{align}
    \{ \bar{\m}_{I}{}^{\a} , \l_\b{}^{J} \} = \delta^\a_\b \delta_I^J \,,\quad \{ \m^{\dot\a I} , \bar{\l}_{J \dot\b} \} = \delta_{\dot\b}^{\dot\a} \delta^I_J \,.
\label{eq:def_PB_twi}
\end{align}
The incidence relations relating the two models read
\begin{align}
    \m^{\dot\a I} = \frac{1}{2} z^{\dot\a \b} \l_{\b}{}^{I} \,,
    \quad \bar{\m}_{I}{}^{\a} = \frac{1}{2} \bar{\l}_{I \dot\b} \bar{z}^{\dot\b\a} \,, 
    \label{eq:incidence_new_copy}
\end{align}
where the complexified position variable $z$ defined by the incidence relation 
is mapped to the spherical top variables through 
\begin{align}
    z^\mu = x^\mu + i y^\mu \,,
    \quad  y^\mu = \frac{1}{m^2} \ve^{\m\n\r\s} p_\n S_{\r\s} \,.
    \label{eq:z-x-y}
\end{align}
The imaginary part $y^\m$ of the complexified position variable $z^\m$ is related to the spin-length vector $a^\mu = s^\m/m$ 
widely used in the literature by $y^\m = - a^\m$. The extra sign is to respect the standard relation of non-relativistic spins: $\{ s^i, s^j \} = \epsilon^{ijk} s^k$. 

In our current conventions, the free action of ref.~\cite{Kim:2021rda} reads 
\begin{align} 
\begin{split}
    &S_{\text{free}} = \int \left[ \l_\a{}^I d \bar{\m}_I{}^\a + \bar{\l}_{I\dot \a} d\m^{\dot\a I} + \frac{1}{2} \left( \bar{\k} (\Delta - m) + \k ( \bar{\Delta} - m)\right) d\s \right] \,, 
    \\
     &\Delta = \det (\l) = - \frac{1}{2} \e^{\a\b} \e_{IJ} \l_\a{}^I \l_\b{}^J \,,\quad  
    \bar{\Delta} = \det (\bar{\l}) = \frac{1}{2} \e^{IJ} \e^{\dot\a \dot\b} \bar{\l}_{I \dot\a} \bar{\l}_{J \dot\b} \,.
\end{split}
    \label{eq:tw_free_action}
\end{align}
The Lagrange multipliers $( \kappa, \bar{\kappa} )$ enforce the conditions,
\begin{align}
    \Delta = m = \bar{\Delta} \,, 
\end{align}
which in turn imply the mass-shell condition $-p^2 = \Delta \bar{\Delta} = m^2$. 
After a suitable gauge-fixing of $(\kappa, \bar{\kappa})$, the ``Hamiltonian" $\mathrm{Re}(\kappa) (\Delta + \bar{\Delta})/2$ 
generates the worldline time-evolution for all dynamical variables 
which matches the expectation from the spherical top model. 

Despite its success, it turns out that the free action in \eqref{eq:tw_free_action} is not suitable 
for the transition to the interacting theory, and we 
propose an alternative action:
\begin{align}
        S_\mathrm{free} = \int \left[ \l_\a{}^I \dot{\bar{\m}}_I{}^\a + \bar{\l}_{I\dot \a} \dot{\m}^{\dot\a I} - \kappa^0\phi_0 - \kappa^1 \phi_1 \right] d\sigma \,, 
        \label{eq:tw_free_action1}
\end{align}
where 
\begin{align}
    \phi_0 = \frac{1}{2}(m^2 - \Delta \bar{\Delta}) = \frac{1}{2} (p^2 +m^2) \,,
    \quad 
    \phi_1 = \frac{1}{2i} ( \bar{\l}_{I\da} \m^{\da I} - \bar{\m}_I{}^\a \l_\a{}^I ) = p \cdot y\,.
    \label{phi0-phi1-new}
\end{align}
For later purposes, we also propose the gauge-fixing conditions, 
\begin{align}
    \chi^0 = - \frac{1}{2} (\bar{\l}_{I\da} \m^{\da I} + \bar{\m}_I{}^\a \l_\a{}^I ) \,,
    \quad 
    \chi^1 = \frac{i}{2} \log(\Delta/\bar{\Delta}) \,.
    \label{chi0-chi1-new}
\end{align}
We recognise $\phi_0$ as the mass-shell constraint in twistor variables, 
and $\phi_1$ as the generator of the U(1) gauge orbit. Aside from how to implement the mass-shell constraint, the main difference between 
the two proposals is that \eqref{eq:tw_free_action1} contains 
the U(1) gauge generator while \eqref{eq:tw_free_action} contains a U(1) gauge-fixing condition in the form 
\begin{align}
    \frac{\mathrm{Im}(\kappa)}{2} (\Delta - \bar{\Delta}) \,.
    \label{gauge-fixing-old}
\end{align}

To understand why the old proposal \eqref{eq:tw_free_action} is problematic and why it still yielded the free equations of motion correctly, let us revisit the general theory of constrained Hamiltonian dynamics, 
building upon appendix A of ref.~\cite{Kim:2021rda}. 
The system consists of Hamiltonian $H$, symplectic form $\omega$, abelian gauge generators $\phi_A$ and gauge fixing functions $\chi^A$. 
The minimal requirements are  
\begin{align}
    \left\{\phi_A, H\right\}=0, \quad
    \left\{\phi_A, \phi_B\right\}=0 \quad 
    C^A{}_B :=\left\{\chi^A, \phi_B\right\}, \quad 
    \operatorname{det}(C) \neq 0 \,.
    \label{minimal-req}
\end{align}
In general, the gauge-fixed action is written as 
\begin{align}
    S = \int d t\left(p_i \dot{q}^i-H-\kappa^A \phi_A-\bar{\kappa}_A \chi^A\right) \,.
    \label{free-action-typical}
\end{align}
The Lagrange multipliers $( \kappa^A, \bar{\kappa}_A )$ enforce the constraints $\phi_A = 0 = \chi^A$. 
The variation of $S$ with respect to the dynamical variables $( p_i, q^i )$ gives the equations of motion. 
The time evolution of a generic function $f(p_i,q^i)$ is computed from the Poisson bracket 
\begin{align}
    \frac{df}{dt} =   \frac{\partial f}{\partial q^i} \dot{q}^i + \frac{\partial f}{\partial p_i} \dot{p}_i  = \{ f , H + \kappa^A \phi_A + \bar{\kappa}_A \chi^A \} \,.
\end{align}
For consistency, the equations of motion should not induce change of $\phi_A$ and $\chi^A$ in time: 
\begin{align}
    \frac{d\phi_A}{dt} = 0 = \frac{d\chi^A}{dt} \,.
    \label{consistency-phi-chi}
\end{align}
The vanishing of $d\phi_A/dt$ and the minimal requirements \eqref{minimal-req} imply $\bar{\kappa}_A = 0$. Hence, we do not see directly $\bar{\kappa}_A$ in the final form of the equations of motion. The vanishing of $d\chi^A/dt$ implies $\kappa^A = -(C^{-1})^A{}_B \{ \chi^B, H\}$. 
We can either choose some $\chi^A$ to fix $\kappa^A$, or prescribe some $\kappa^A$ to fix $\chi^A$ implicitly.

If some of the $\kappa^A$ multipliers can be set to zero without violating the requirements \eqref{minimal-req}, 
the corresponding $\phi_A$ will not directly contribute to the equations of motion and 
it may look permissible to exchange the roles of $\phi_A$ and $\chi^A$. 
That is precisely what happened to the free twistor model. 
But, as soon as we add interaction terms in the action, the distinction between the gauge generators $\phi_A$ and 
the gauge fixing conditions $\chi^A$ becomes evident. 
By the very definition of gauge redundancy, the interaction terms are required to Poisson-commute with all $\phi_A$ in a sense to be specified below. On the contrary, there is no reason for the interaction terms to commute with $\chi^A$. 

To conclude, in view of the general theory of constrained Hamiltonian system where the gauge generators and gauge-fixing conditions 
play different roles, we need a new proposal for the free action \eqref{eq:tw_free_action1} to incorporate interaction terms.

\subsubsection{Regge trajectory} \label{sec:regge_traj}

Due to rotational kinetic energy, the mass of a spinning top is in general not a constant,
but rather a function 
of the spin-magnitude $W^2 = (y\cdot p)^2 - y^2 p^2$. 
The derivative $m' = dm/d(W^2)$ is colloquially called the ``Regge trajectory". 
For a free spinning particle, 
the angular velocity is $(2m')$ times the spin, 
so $(2m')^{-1}$ is the relativistic rotational inertia \cite{Kim:2021rda}. 

In most applications, where we do not keep track of the angular velocity and focus on the $( x,y,p )$ variables, the Regge trajectory does not affect the dynamics. 
Specifically, 
the equations of motion for $( x,y,p )$ are independent of $m'$. 
In the rest of this section, we will see, from a few different but related angles, how the Regge trajectory $m'$ decouples from the dynamics of $( x,y,p )$. 
From the next subsection on, we will set $m'=0$ to simplify computations.

\subsubsection{Dirac bracket and effective phase space} 
\label{sec:eff-Dirac-bracket}

To describe the physical phase space of the free twistor theory, 
it is convenient to construct the Dirac bracket. 
The ingredients are the unconstrained Poisson bracket \eqref{eq:def_PB_twi}, the gauge generators \eqref{phi0-phi1-new}, and the gauge-fixing functions \eqref{chi0-chi1-new}. 

Let us first consider the case of vanishing Regge trajectory ($m'=0$). Using 
\begin{align}
    C^A{}_B = \{ \chi^A, \phi_B \} = \begin{pmatrix}
        \D \bar{\D} & 0 \\ 
        0 & 1 
    \end{pmatrix} \,, 
    \label{C-twistor}
\end{align}
we construct the Dirac bracket in the standard way, 
\begin{align}
 \{ f, g \}_* = \{ f, g \} -(C^{-1})^A{}_B \left( \{ f, \phi_A \} \{\chi^B, g \} -  \{ g, \phi_A \} \{\chi^B, f \}\right) \,.
 \label{poisson-vs-dirac}
\end{align}
The non-vanishing brackets among the twistor variables are  
\begin{align}
\begin{split}
        \{ \bar{\m}_{I}{}^{\a} , \l_\b{}^{J} \}_* &= \delta^\a_\b \delta_I^J - \frac{1}{2} \l_\b{}^J (\l^{-1})_I{}^\a\,,
        \\
        \{ \m^{\dot\a I} , \bar{\l}_{J \dot\b} \}_* &= \delta_{\dot\b}^{\dot\a} \delta^I_J - \frac{1}{2} \bar{\l}_{J\db}  (\bar{\l}^{-1})^{\da I} \,, 
        \\
        \{ \bar{\m}_{I}{}^{\a} , \bar{\m}_{J}{}^{\b}\}_* &= -\frac{1}{2} \left[ \bar{\m}_{I}{}^{\a} (\l^{-1})_{J}{}^{\b} - \bar{\m}_{J}{}^{\b} (\l^{-1})_{I}{}^{\a}   \right] \,,
        \\
        \{ \m^{\dot\a I} , \m^{\dot\b J}  \}_* &= -\frac{1}{2} \left[ \m^{\dot\a I} (\bar{\l}^{-1})^{\dot\b J} - \m^{\dot\b J} (\bar{\l}^{-1})^{\dot\a I}  \right] \,.
\end{split}
\label{DB_twi}
\end{align}
We can certainly use these to compute the Dirac brackets among the U(1)-invariant composite variables $( x^\m, y^\m, p^\m )$. 
But, since we are interested only in the dynamics of $( x^\m, y^\m, p^\m )$ variables in the bulk of this paper, we can take a shortcut. 

We can regard the 16-dimensional unconstrained phase space 
as a fibre bundle with the base coordinates $( x,y,p )$  
and the fibre SU(2)$\times$U(1). 
The original Poisson bracket \eqref{eq:def_PB_twi} projected onto the $( x,y,p )$ base space can be written as 
\begin{align}
\begin{split}
   \{ \;\; , \;\; \}_\circ &= \eta^{\m\n} \frac{\partial}{\partial x^\m} \wedge \frac{\partial}{\partial p^\n} - \frac{y^\m p^\n +p^\m y^\n - \eta^{\m\n} (y\cdot p) }{p^2}  \frac{\partial}{\partial x^\m} \wedge \frac{\partial}{\partial y^\n}   
   \\
   &\quad - \frac{\epsilon^{\m\n}[y,p]}{2p^2} \left( \frac{\partial}{\partial y^\m} \wedge \frac{\partial}{\partial y^\n} + \frac{\partial}{\partial x^\m} \wedge \frac{\partial}{\partial x^\n}\right) \,.
\end{split}
\label{Poisson-bracket-effective}
\end{align}

The transition from the Poisson bracket to the Dirac bracket 
can be done within this effective description. 
Since $( x,y,p )$ are all U(1) gauge invariant, 
we can simply disregard the pair $( \phi_1, \chi^1 )$ and take into consideration  
\begin{align}
    \phi_0 = \frac{1}{2}(p^2 + m^2) \,,
    \quad 
    \chi^0 = - x\cdot p \,.
\end{align}
The resulting Dirac bracket on the $( x,y,p )$ base space is 
\begin{align}
\begin{split}
   \{ \;\; , \;\; \}_* &= \left( \eta^{\m\n} - \frac{p^\m p^\n}{p^2} \right) \frac{\partial}{\partial x^\m} \wedge \frac{\partial}{\partial p^\n} - \frac{y^\m p^\n - \eta^{\m\n} (y\cdot p) }{p^2}  \frac{\partial}{\partial x^\m} \wedge \frac{\partial}{\partial y^\n}   
   \\
   &\quad - \frac{\epsilon^{\m\n}[y,p]}{2p^2} \left( \frac{\partial}{\partial y^\m} \wedge \frac{\partial}{\partial y^\n} + \frac{\partial}{\partial x^\m} \wedge \frac{\partial}{\partial x^\n}\right) - \frac{x^{[\m} p^{\n]}}{p^2} \frac{\partial}{\partial x^\m} \wedge \frac{\partial}{\partial x^\n} \,.
\end{split}
\label{Dirac-bracket-effective}
\end{align}
It agrees perfectly with the full Dirac bracket \eqref{DB_twi} evaluated over $(x,y,p)$. 

Let us summarise the content of the effective description. The phase space is effectively nine dimensional: twelve coordinates $( x^\m, y^\m, p^\m )$ with three constraints,
\begin{align}
    p^2 + m^2 = 0 \,,
    \quad 
    p\cdot y = 0 \,,
    \quad 
    p\cdot x = 0 \,.
    \label{free-constraints}
\end{align}
By construction, the Dirac brackets of the constraints automatically vanish:
\begin{align}
    \{ f , p^2 +m^2 \}_* = 0 \,,
    \quad 
    \{ f , p\cdot y \}_* = 0 \,,
    \quad 
    \{ f , p\cdot x \}_* = 0 \,. 
    \label{Dirac-bracket-constraint}
\end{align}
Incidentally, the Dirac bracket is consistent with fixing the magnitude of spin:
\begin{align}
      \{ f , y^2  \}_* = 0 \,. 
       \label{bonus-constraint}
\end{align}
So, the phase space is eight dimensional in a sense, 
although the conservation of $y^2$ is of a dynamical origin
and shouldn't be put on an equal footing as the true constraints. 

In later sections, we will study scattering processes of the twistor particles, where the interaction is (asymptotically) turned off at the infinite past and future. It makes sense to use the (effective) free phase space for each particle to describe the asymptotic scattering states.


For completeness, we also compute the Dirac bracket 
when the Regge trajectory is non-trivial ($m'\neq 0$). 
Remarkably, the $C$-matrix \eqref{C-twistor} is not altered by $m'$ at all. 
It remains to compute the $(m')$-dependent terms from $\{f, \phi_0\}\{\chi^0, g \}$ in \eqref{poisson-vs-dirac}. 

In terms of the twistor variables, the  function $\chi^0$ is given by 
\begin{align}
    \chi^0 = -x\cdot p = \frac{1}{2} \left( \m^{\da I} \bar{\l}_{I \da} + 
    \bar{\m}_I{}^\a \l_\a{}^I \right) \,.
\end{align}
It follows that 
\begin{align}
\begin{aligned}
    \{\chi^0 , \l_\a{}^I \} &= + \frac{1}{2} \l_\a{}^I \,,
    &\quad 
    \{\chi^0 , \m^{\da I} \} &= - \frac{1}{2} \m^{\da I} \,,
    \\
    \{\chi^0 , \bar{\l}_{I\da} \} &= + \frac{1}{2} \bar{\l}_{I\da}
    &\quad 
    \{\chi^0 , \bar{\m}_I{}^\a \} &= - \frac{1}{2}\bar{\m}_I{}^\a \,.
\end{aligned}
\label{chi0-twi}
\end{align}
The $W$ tensor is defined as \cite{Kim:2021rda} 
\begin{align}
 W_{I J} = - i\left(\mu^{\dot{\alpha}}{ }_{(I} \bar{\lambda}_{J) \dot{\alpha}}-\lambda_{\alpha(I} \bar{\mu}_{J)}{ }^\alpha\right) \,.
\end{align}
It is normalized such that 
\begin{align}
    W^2 = \frac{1}{2} W_K{}^L W_L{}^K = (y\cdot p)^2 - y^2 p^2 \,.
\end{align}
The $(m')$-dependent terms from $\{f, \phi_0\}$ come from 
\begin{align}
    \{ f , m(W^2)^2/2 \} = mm' \{ f, W^2 \}  \,. 
\end{align}
It is useful to note that 
\begin{align}
\begin{aligned}
    \{\l_\a{}^I , W^2 \} &= -i  \l_\a{}^K W_K{}^I \,,
    &\quad 
    \{\m^{\da I} , W^2 \} &= -i  \m^{\da K} W_K{}^I  \,,
    \\
    \{\bar{\l}_{I\da} , W^2 \} &= +i W_I{}^K \bar{\l}_{K\da} \,,
    &\quad 
    \{ \bar{\m}_I{}^\a , W^2 \} &= +i W_I{}^K  \bar{\m}_K{}^\a\,.
\end{aligned}
\label{twi-Wsq}
\end{align}

The full Dirac bracket consists of two parts, 
\begin{align}
    \{f, g \}_* = \{f, g \}^0_* + \{f, g \}^\prime_* \,, 
\end{align}
where $\{X, Y \}^0_*$ is the result for $m'=0$ stated in \eqref{DB_twi} 
and $\{X, Y \}^\prime_*$ denotes the $(m')$ terms. 
Using \eqref{chi0-twi} and \eqref{twi-Wsq}, 
it is straightforward to show that the $(m')$ terms are 
\begin{align}
\begin{aligned}
\{ \lambda_\alpha{ }^I, \lambda_\beta{ }^J \}'_* 
&= \frac{i m'}{2 m}\left[ + (\lambda_\alpha{ }^K W_K{ }^I ) \lambda_\beta{ }^J-\lambda_\alpha{ }^I\left(\lambda_\beta{ }^K W_K{ }^J\right)\right] \,,
\\
\{ \lambda_\alpha{ }^I, \bar{\lambda}_{J \dot{\beta}} \}'_* 
&= \frac{i m'}{2 m}\left[ + (\lambda_\alpha{ }^K W_K{ }^I ) \bar{\lambda}_{J \dot{\beta}}+\lambda_\alpha{ }^I (W_J{ }^K \bar{\lambda}_{K \dot{\beta}} )\right]   \,,
\\
\{ \lambda_\alpha{ }^I, \mu^{\dot{\beta} J} \}'_* 
&= \frac{im'}{2  m}\left[- (\lambda_\alpha{ }^K W_K{ }^I ) \mu^{\dot{\beta} J} - \lambda_\alpha{ }^I (\mu^{\dot{\beta} K} W_K{ }^J )\right] \,,
\\
\{ \lambda_\alpha{ }^I, \bar{\mu}_J{ }^\beta \}'_* 
&=  \frac{i m'}{2  m}\left[- (\lambda_\alpha{ }^K W_K{ }^I ) \bar{\mu}_J{ }^\beta + \lambda_\alpha{ }^I (W_J{ }^K \bar{\mu}_K{ }^\beta )\right] \,,
\\
\{ \mu^{\dot{\alpha} I}, \mu^{\dot{\beta} J} \}'_* 
&=  \frac{i m'}{2 m}\left[- (\mu^{\dot{\alpha} K} W_K{ }^I ) \mu^{\dot{\beta} J} + \mu^{\dot{\alpha} I} (\mu^{\dot{\beta} K} W_K{ }^I )\right] \,,
\\
\{\mu^{\dot{\alpha} I}, \bar{\mu}_J{ }^\beta \}'_* 
&=  \frac{i m'}{2 m}\left[ - (\mu^{\dot{\alpha} K} W_K{ }^I) \bar{\mu}_J{ }^\beta - \mu^{\dot{\alpha} I} (W_J{ }^K \bar{\mu}_K{ }^\beta )\right] \,,
\end{aligned} 
\label{DB_twi_Regge}
\end{align}
and their complex conjugates. 
They agree perfectly with Dirac brackets computed in ref.~\cite{Kim:2021rda} despite the differences in the choice of constraints. 

Comparing \eqref{DB_twi_Regge} with \eqref{DB_twi}, it may appear that the non-vanishing Regge trajectory complicates
the Dirac bracket significantly. 
Fortunately, when we focus on the effective $( x,y,p )$ phase space, the complication disappears completely and the Dirac bracket \eqref{Dirac-bracket-effective} remains valid. 
In essence, the reason is that the $(m')$-terms in \eqref{DB_twi_Regge} 
all involve infinitesimal shifts along the SU(2) little-group, 
but the $( x,y,p )$ variables are little-group scalars.

\subsection{Interacting theory} 

In this subsection, we explain how to couple the twistor particle to a background electromagnetic field minimally using the Newman-Janis shift. 

\subsubsection{Symplectic perturbation theory} 

Given a Hamiltonian system defined by \eqref{minimal-req} and \eqref{free-action-typical}, 
we may consider two types of deformations. 
A familiar way is to deform the Hamiltonian, 
\begin{align}
    H = H^\circ + q\, H' \,, 
\end{align}
where $q$ is a continuous perturbation parameter. 
We demand that the gauge generators are independent of the deformation. The requirement \eqref{minimal-req} implies that 
\begin{align}
    \{ \phi_A , H^\circ \} = 0 = \{ \phi_A , H' \} \,.
\end{align}
Then all the statements around \eqref{consistency-phi-chi} remain valid for any value of $q$. 

An alternative way to deform the theory, which we adopt for our twistor model,  
is to keep the Hamiltonian fixed and deform the symplectic form \cite{Kim:2023aff}:
\begin{align}
    \omega = \omega^\circ + q\, \omega' \,.
\end{align}
This is equivalent to keeping the Hamiltonian fixed and deforming the Poisson brackets used in equations of motion (EOM). 
For a non-spinning particle, a key feature of the symplectic perturbation viewpoint is that $\omega' = F = dA$ is manifestly invariant under the gauge transformation of the Maxwell field and that we do not need to distinguish the two notions of momenta (canonical vs kinetic). 
This feature will generalise straightforwardly to our twistor model. 

Again, we demand that the gauge generators are independent of $q$. 
We should make sure that the requirements $\{ \phi_A , H \} = 0 = \{\phi_A , \phi_B \}$ hold 
with the deformed symplectic form. Perturbatively, 
\begin{align}
    \{ f , g \} = \{ f , g\}_\circ - q \{ f , \zeta^m \}_\circ \omega'_{mn} \{ \zeta^n , g \}_\circ + \mathcal{O}(q^2) \,, 
\end{align}
where $\zeta^m$ is an arbitrary coordinate system on the phase space and $\{ \bullet, \bullet \}_\circ$ is the Poisson bracket defined by $\w^\circ$. 

For a random choice of $\omega'$, the perturbation term has no reason to vanish:
\begin{align}
    - q \{ \phi_A , \zeta^m \}_\circ \omega'_{mn} \{ \zeta^n , \phi_B \}_\circ \neq 0 \,.
\end{align}
But, if we change coordinates (at least locally), 
$\zeta^m \rightarrow (z^i , w^a )$   
where $z^i$ are gauge-invariant while $w^a$ are gauge-dependent, and take $\omega' = \frac{1}{2} \omega'_{ij} dz^i \wedge dz^j$ with no $dw^a$ components, it is immediately clear that $\{ \phi_A , H \} = 0 = \{\phi_A , \phi_B \}$ hold \emph{exactly}. 

Coming back to our twistor model, recall that the new free action \eqref{eq:tw_free_action1} carries two gauge generators 
$\phi_0$, $\phi_1$ in \eqref{phi0-phi1-new}.
Since the Poisson bracket is antisymmetric, the only issue is whether $\{ \phi_0 , \phi_1 \} = 0$ continues to hold 
in the interacting theory. 
But, as long as $\omega'$ consists only of U$(1)$-gauge invariant coordinates such as $p_\m$ and $z^\m = x^\m + i y^\m$, 
$\{ \phi_0, \phi_1 \} = 0$ would follow trivially.

A common feature of a relativistic particle theory, shared by the spherical top model, is that the mass-shell constraint $\phi_0 = (p^2+m^2)/2$ also generates the worldline time-evolution. 
Ignoring other gauge generators temporarily, the equation of motion of the free theory is 
\begin{align}
    \frac{df}{d\s} = \kappa^0 \{ f , \phi_0 \}_\circ \,.
\end{align}
There are two equivalent ways to express the equation of motion of the interacting theory. 
One is based on the infinite series \cite{Kim:2023aff}, 
\begin{align}
\begin{split}
        \omega^{-1} &= (\omega^\circ)^{-1} - q (\omega^\circ)^{-1} \omega' (\omega^\circ)^{-1} + q^2 (\omega^\circ)^{-1} \omega' (\omega^\circ)^{-1} \omega' (\omega^\circ)^{-1} + \cdots  
\\
\Longrightarrow \quad 
    \frac{1}{\kappa^0} \frac{df}{d\s} &= \{ f , \phi_0 \}_\circ - q \{ f, z^i \}_\circ \omega'_{ij} \{ z^j , \phi_0 \}_\circ 
    \\
    &\hskip 2cm + q^2 \{ f, z^i \}_\circ \omega'_{ij} \{ z^j , z^k \}_\circ \omega'_{kl} \{ z^l , \phi_0 \}_\circ + \cdots \,.
\end{split}
         \label{symp-pert-long}
\end{align}
The other is more compact, 
\begin{align}
\begin{split}
    \omega^{-1} &= (\omega^\circ)^{-1} - q (\omega^\circ)^{-1} \omega' \omega^{-1} 
    \\
    \Longrightarrow
    \quad 
     \frac{1}{\kappa^0} \frac{df}{d\s} &= \{ f , \phi_0 \}_\circ - q \{ f, z^i \}_\circ \omega'_{ij} \{ z^j , \phi_0 \} \,, 
\end{split}
    \label{symp-pert-short}
\end{align}
but less explicit in that the last term involves the Poisson bracket of the interacting theory. 
Of course, \eqref{symp-pert-short} can be iterated to reproduce \eqref{symp-pert-long}.

\subsubsection{Minimal coupling via Newman-Janis shift} 
\label{sec:NJ-eom}

The unification of $( x^\m , S_{\m\n} )$ into the complex coordinates $( z ,\bar{z} )$ defined by \eqref{eq:incidence_new_copy} and \eqref{eq:z-x-y} opens up an avenue for implementing the Newman-Janis shift~\cite{Newman:1965tw} into the dynamics. 
It led one of the authors to the concept of ``spinspacetime"~\cite{Kim:2023vgb} which unifies spacetime and spin at the fundamental level, which has refined Newman's early idea~\cite{Newman:1974fr}. 
Our twistor model is a new member of existing attempts to incorporate the Newman-Janis shift into the dynamics~\cite{Guevara:2020xjx,Alessio:2023kgf}.  

In the twistor model, the root-Kerr particle is defined by the minimal extension via Newman-Janis (NJ) shift from Minkowski spacetime to spinspacetime~\cite{Kim:2023aff,Kim:2023vgb}.
To introduce the minimal extension, 
we begin with the usual coupling of a charged scalar particle to the Maxwell field 
and split it into two pieces,\footnote{For related (but not equivalent) massive twistor models and their electromagnetic couplings, see refs.~\cite{Bette:2004ip,deAzcarraga:2014hda} and references therein.} 
\begin{align}
    S_{\text{int}} = 
    q \int A_\m (x) dx^\m = q \int A^+_\m (x) dx^\m + q \int A^-_\m (x) dx^\m \,,
\end{align}
where $q$ is the charge and the convention fixed by 
$A_\mu = (-\phi, \vec{A})$.
The superscripts $(\pm)$ denote the self-dual and anti-self-dual parts in the sense that, under the hodge dual $*$ on the field-strengths,
\begin{align}
    F^{\pm} = dA^{\pm} \,,
    \quad 
    *F^\pm = \pm i F^\pm \,.
\end{align}
The minimal NJ shift correlates (anti-)self-duality and holomorphy of the complexified spacetime as~\cite{Guevara:2020xjx}
\begin{align}
    S_{\text{int}} = q \int A^+_\m (z) dz^\m + q \int A^-_\m (\bar{z}) d\bar{z}^\m = - \frac{q}{2} \int A^+_{\a \dot\a} (z) dz^{\dot\a \a} - \frac{q}{2} \int A^-_{\dot\a \a} (\bar{z}) d \bar{z}^{\dot\a \a} \,. 
    \label{eq:int_NJ_shift}
\end{align}
In the parlance of the symplectic perturbation theory, 
\begin{align}
\begin{split}
    \omega' &= (\omega')^+ + (\omega')^- = \frac{q}{2} F^+_{\mu\nu}(z)  dz^\mu \wedge dz^\nu 
    + \frac{q}{2} F^-_{\mu\nu}(\bar{z})  d\bar{z}^\mu \wedge d\bar{z}^\nu \,. 
\end{split}
\end{align}
Compared to the non-spinning case where $\omega' = \textstyle{\frac{1}{2}} F_{\m\n}(x) dx^\m \wedge dx^\n$, 
the Newman-Janis shift introduces a complicated non-linear deformation depending on the $y^\m$ coordinate. 
But, the invariance under the Maxwell gauge transformation is still manifest.

As usual, the coupling \eqref{eq:int_NJ_shift} plays a dual role; it enters the equation of motion of the charged particle, and it acts as a source term in the Maxwell's equations. 
We focus on the former aspect in this and the next section, 
leaving the latter aspect to later sections.

\subsubsection{Equations of motion} 

Adding up the free part \eqref{eq:tw_free_action1} and the interaction part \eqref{eq:int_NJ_shift}, the full action of the root-Kerr particle becomes 
\begin{align}
    S = \int \left[ \l_\a{}^I \dot{\bar{\m}}_I{}^\a + \bar{\l}_{I\dot \a} \dot{\m}^{\dot\a I} - \frac{\kappa^0}{2} (m^2 - \Delta \bar{\Delta}) - \kappa^1 W_0 + q A^+_\m(z)  \dot{z}^\m + q A^-_\m(\bar{z}) \dot{\bar{z}}^\m \right] d\sigma \,. \label{eq:action_full}
\end{align}
The variation of the action with the vanishing Regge trajectory ($m'=0$) gives 
\begin{align}
\begin{aligned}
    \delta S &= \delta \lambda_{\a} {}^{I} \left[ \dot{\bar{\m}}_I{}^\a + \frac{\k^0 \D \bar{\D}}{2} (\l^{-1})_I{}^\a + \frac{\k^1}{2i} \bar{\m}_I{}^\a \right] 
    \\ &\phantom{=}
    + \delta \bar{\l}_{I\dot\a} \left[ \dot{\m}^{\dot\a I} + \frac{\k^0 \D \bar{\D}}{2} (\bar{\l}^{-1})^{\dot\a I} - \frac{\k^1}{2i} {\m}^{\dot\a I} \right]
    \\ &\phantom{=} + \delta \m^{\dot\a I} \left[ - \dot{\bar{\l}}_{I \dot\a} - \frac{\k^1}{2i} {\bar{\l}}_{I \dot\a} \right] 
    + \delta \bar{\m}_I{}^\a \left[ - \dot{\l}_\a{}^I + \frac{\k^1}{2i} {\l}_\a{}^I \right]
    \\ &\phantom{=} 
    + \delta z^\m \left[ q F_{\m\n}^+ (z) \dot{z}^\n \right] + \delta \bar{z}^\m \left[ q F_{\m\n}^- (\bar{z}) \dot{\bar{z}}^\n \right]
    \\ 
    &= \delta \lambda_{\a} {}^{I} \left[ \dot{\bar{\m}}_I{}^\a + \frac{\k^0 \D \bar{\D}}{2} (\l^{-1})_I{}^\a + \frac{\k^1}{2i} \bar{\m}_I{}^\a + q F_{\m\n}^+ (z) \dot{z}^\n \s^\m_{\b \dot\b} \m^{\dot\b J} (\l^{-1})_J{}^\a (\l^{-1})_I{}^\b \right]
    \\ &\phantom{=} + \delta \bar{\l}_{I\dot\a} \left[ \dot{\m}^{\dot\a I} + \frac{\k^0 \D \bar{\D}}{2} (\bar{\l}^{-1})^{\dot\a I} - \frac{\k^1}{2i} {\m}^{\dot\a I} + q F_{\m\n}^- (\bar{z}) \dot{\bar{z}}^\n \s^\m_{\b \dot\b} \bar{\m}_J{}^{\b} (\bar{\l}^{-1})^{\dot\b I} (\bar{\l}^{-1})^{\dot\a J} \right]
    \\ &\phantom{=} - \delta \m^{\dot\a I} \left[  \dot{\bar{\l}}_{I \dot\a} + \frac{\k^1}{2i} {\bar{\l}}_{I \dot\a} + q F_{\m\n}^+(z) \dot{z}^\n \s^\m_{\a \dot\a} (\l^{-1})_I{}^\a \right]
    \\ &\phantom{=} - \delta \bar{\m}_I{}^\a \left[ \dot{\l}_\a{}^I - \frac{\k^1}{2i} \l_\a{}^I + q F_{\m\n}^-(\bar{z}) \dot{\bar{z}}^\n \s^\m_{\a \dot\a} (\bar{\l}^{-1})^{\da I} \right] \,.
\end{aligned}
\end{align}
In the last step, we used the incidence relations to express $\delta (z,\bar{z})$ in terms of $\delta (\m,\l,\bar{\m},\bar{\l})$. 
To specify the equations of motion completely, we have to fix the Lagrange multipliers $\k^{0,1}$. 
A convenient choice to be used throughout this paper is 
\begin{align}
    \k^0 = \frac{1}{m} \,, \quad \k^1 = 0 \,.
\end{align}
With this choice, the EOM for the twistor variables are 
\begin{align}
\begin{aligned}
   \dot{\bar{\m}}_I{}^\a &= - \frac{m}{2} (\l^{-1})_I{}^\a - q F_{\m\n}^+ (z) \dot{z}^\n \s^\m_{\b \dot\b} \m^{\dot\b J} (\l^{-1})_J{}^\a (\l^{-1})_I{}^\b  \,,
   \\
   \dot{\m}^{\dot\a I} &= - \frac{m}{2} (\bar{\l}^{-1})^{\dot\a I} - q F_{\m\n}^- (\bar{z}) \dot{\bar{z}}^\n \s^\m_{\b \dot\b} \bar{\m}_J{}^{\b} (\bar{\l}^{-1})^{\dot\b I} (\bar{\l}^{-1})^{\dot\a J} \,,
   \\
   \dot{\bar{\l}}_{I \dot\a} &= - q F_{\m\n}^+(z) \dot{z}^\n \s^\m_{\a \dot\a} (\l^{-1})_I{}^\a \,, 
   \\
   \dot{\l}_\a{}^I  &= - q F_{\m\n}^-(\bar{z}) \dot{\bar{z}}^\n \s^\m_{\a \dot\a} (\bar{\l}^{-1})^{\da I} \,.
\end{aligned}
\label{eom-twistor} 
\end{align}
The equations for the U$(1)$ gauge invariant variables $(p, z, \bar{z})$ are, 
in the vector notation, 
\begin{align}
\begin{aligned}
    \dot{p}_\mu &= q F^+_{\m\n}(z) \dot{z}^\nu + q F^-_{\m\n}(\bar{z})\dot{\bar{z}}^\n \,,
    \\
    \dot{z}^\m &= \frac{ p^\m}{m} + \frac{2iq}{m^2} \left[ y^\m p^\n + p^\m y^\n  
    + i \e^{\m\n\r\s}y_\r p_\s \right] F^-_{\n\l} (\bar{z}) \dot{\bar{z}}^\l  \,,
    \\ 
    \dot{\bar{z}}^\m &= \frac{ p^\m}{m} - \frac{2iq}{m^2} \left[ y^\m p^\n + p^\m y^\n  
    - i \e^{\m\n\r\s}y_\r p_\s \right] F^+_{\n\l} ({z}) \dot{{z}}^\l \,,
\end{aligned}
\label{eom-zzbar-new}
\end{align}
where we applied the constraints $y\cdot p = 0$, $\Delta \bar{\Delta} = m^2$ after deriving the equations. 
We may turn on the Regge trajectory $(m' \neq 0)$ 
and repeat deriving the equations. 
Not surprisingly, 
the twistor equations receive new terms proportional to $(m')$, 
but \eqref{eom-zzbar-new} remains unchanged.

The appearance of time-derivatives on the RHS of \eqref{eom-zzbar-new} may look peculiar. 
But, it is a generic feature of the symplectic perturbation theory explained earlier. It is easy to see how \eqref{eom-zzbar-new} fits with the general structure of \eqref{symp-pert-short}:
\begin{align}
    \{ f, \Delta \bar{\Delta} \} = \{ f, \Delta \bar{\Delta} \}_\circ - q\{ f, z^\n \}_\circ F^+_{\n\l}\{ z^\l, \Delta \bar{\Delta} \} - q\{ f, \bar{z}^\n \}_\circ F^-_{\n\l}\{ \bar{z}^\l, \Delta \bar{\Delta} \} \,.
    \label{zig-zag-eom}
\end{align}
Finally, turning on the Regge trajectory ($m'\neq 0$) modifies \eqref{eom-twistor} slightly, but leaves \eqref{eom-zzbar-new} unchanged. In what follows, we take \eqref{eom-zzbar-new} as the fundamental EOM for the twistor particle and use it to compute physical observables. 

Unless the background takes very special values, 
it would be impossible to solve the EOM exactly. We approach the problem as a perturbative expansion in $q$. 
The EOM truncated up to the 2PL order is given by 
\begin{align}
    \begin{split}
        \dot{x}^\m &= \frac{p^\m}{m} - i \frac{q}{m} (F^+ - F^-)^\m{}_\n y^\n  
        \\
        &\quad 
        + \frac{4 q^2}{m^3} p^\m  (y F^+  F^- y)
        + \frac{2 q^2}{m^3}  y^\m \left[ (p  F^+  F^-  y) + (p F^-  F^+ y) \right] + \CO(q^3) \,,
        \\
        \dot{y}^\m &= \frac{q}{m} (F^+ + F^-)^\mu{}_\nu y^\nu - i \frac{4q^2}{m^3}  \left[  (p F^- y) (F^+)^\m{}_\n  -  (p F^+  y)(F^-)^\m{}_\n  \right] y^\n + \CO(q^3) \,,
        \\
        \dot{p}^\m &= \frac{q}{m} (F^+ + F^-)^\mu{}_\nu p^\nu 
        + \CO(q^3) \,.
    \end{split}
    \label{EOM-xyp-2PL}
\end{align} 
Here the scalar product of vectors and tensors are defined as, for example,  
\begin{align}
    (pF^+F^-y) = p_\m (F^+)^\m{}_\n (F^-)^\n{}_\r y^\r \,,
    \quad 
     (p F^- y) = p_\m  (F^-)^\m{}_\n y^\n \,.
\end{align}

\section{Scattering observables} 
\label{sec: scattering observables}

We compute the scattering observables (velocity kick and spin kick) of the massive spinning bodies in the conservative sector up to the 2PL order;  
2PL is the lowest order where our model may disagree with other models implementing the (dynamical) Newman-Janis shift. 
Before we compute the observables of our twistor model, we present a general description of the scattering observables that is valid in any (Hamiltonian) worldline model.

\subsection{Classical eikonal as scattering generator} \label{sec:eikonal-generator}

The eikonal phase has proved useful in quantum and classical scattering in field theory, gravity and string theory; see ref.~\cite{DiVecchia:2023frv} for a comprehensive review.\footnote{See ref.~\cite{Luna:2023uwd} for a comparison of different approaches to spinning observables beyond the leading order.} 
In the context of amplitudes-based methods for classical gravitational scattering, several variants of the eikonal phase are available in the literature such as the HEFT phase~\cite{Brandhuber:2021eyq}, the radial action \cite{Bern:2021dqo}, and the exponential representation \cite{Damgaard:2021ipf,Damgaard:2023ttc} just to mention a few. 

Here, we introduce the notion of ``classical eikonal" which is a particular classical avatar of the eikonal phase.
Simply put, the classical eikonal is the generator of the canonical transformation 
between the initial states and the final states of the scattering problem.
It is motivated by the classical limit of an $S$-matrix, 
but in can be defined purely within classical mechanics. 
The definition is valid in any worldline model in Hamiltonian formulation. 
The relevant phase space is that of free particles, just as the scattering states of a quantum scattering process are defined on the free Hilbert space. 

\paragraph{Scattering generator} 

We first recall the KMOC \cite{Kosower:2018adc} method of extracting a classical observable from a quantum theory: 
\begin{align}
    \Delta O = \lim_{\hbar \rightarrow 0} \left[ \langle \psi | \hat{S}^\dagger \hat{O} \hat{S} |\psi \rangle - \langle \psi | \hat{O} |\psi \rangle \right] \,.
\end{align}
We can trade the unitary operator $\hat{S}$ for a hermitian operator $\hat{\chi}$ as \cite{Damgaard:2021ipf,Damgaard:2023ttc}
\begin{align}
    \hat{S} = e^{i\hat{\chi}/\hbar} \,.
\end{align}
Forgetting about the state $|\psi\rangle$ and working in the Heisenberg picture, we have 
\begin{align}
\begin{split}
     \hat{O}' =  \hat{S}^\dagger \hat{O} \hat{S}  &=  e^{-i\hat{\chi}/\hbar}  \hat{O} e^{i\hat{\chi}/\hbar}  
     \\
     &=  \hat{O} + \frac{1}{i\hbar} [ \hat{\chi} , \hat{O} ] + \frac{1}{2(i\hbar)^2} [  \hat{\chi} , [ \hat{\chi} , \hat{O} ] ] + \cdots \,.
\end{split}
\label{evolution-quantum}
\end{align}
Following Dirac's correspondence, 
\begin{align}
   \lim_{\hbar \rightarrow 0}  \frac{1}{i\hbar} [ \hat{X} , \hat{Y} ] =  \{ X, Y \} \,, 
\end{align}
we deduce that the classical limit of \eqref{evolution-quantum} should give 
\begin{align}
      O' =  O + \{ \chi, O \} + \frac{1}{2}  \{ \chi , \{ \chi, O \} \} + \cdots \,. 
      \label{evolution-classical}
\end{align}
The classical quantity $\chi$, which we call ``scattering generator",  is to be understood as a function on the phase space of a 
Hamiltonian system. 
The scattering generator defines a canonical transformation that maps a free phase space at past infinity to another free phase space at future infinity. 
For a constrained Hamiltonian system with gauge generators and gauge-fixing conditions, 
$\chi$ must be gauge-invariant. If we also demand that $O$ is gauge-invariant, 
the difference between the Dirac bracket and the Poisson bracket shown in \eqref{poisson-vs-dirac} vanishes, 
so we may work with the Poisson bracket \eqref{Poisson-bracket-effective}.

We used the classical limit of quantum mechanics to motivate the existence of $\chi$. But, it is certainly possible to argue for its existence purely within classical mechanics. The Hamiltonian time evolution generates infinitesimal canonical transformation at each moment in time. Integrating the time evolution from past infinity to future infinity would result in a finite canonical transformation. Provided that the general relation between a Lie group and its Lie algebra holds for the (infinite dimensional) group of all canonical transformations on the free phase space, any finite canonical transformation could be written in a ``conjugation" form as in \eqref{evolution-classical}.

The master formula \eqref{evolution-classical} can compute any scattering observable. 
Perturbatively, with 
\begin{align}
    \chi = \chi_{(1)} + \chi_{(2)} + \chi_{(3)} + \cdots \,, 
\end{align}
it produces systematically all $n$-PL impulse formulas 
\begin{align}
\begin{split}
     \Delta_{(1)} O &=  \{ \chi_{(1)} , O \} \,,
      \\
      \Delta_{(2)} O &=  \{ \chi_{(2)} , O \} + \frac{1}{2} \{ \chi_{(1)} , \{ \chi_{(1)} , O \} \} \,,  
           \\
      \Delta_{(3)} O &=  \{ \chi_{(3)} , O \} + \frac{1}{2} \{ \chi_{(2)} ,  \{ \chi_{(1)} , O \} \} + \frac{1}{2} \{ \chi_{(1)} ,   \{\chi_{(2)} , O \} \} 
      \\
      &\hskip 3cm + \frac{1}{6} \{ \chi_{(1)} ,   \{\chi_{(1)} , \{\chi_{(1)} , O \} \} \}  \,. 
\end{split}
\label{impulse-nPL}
\end{align}

An important feature of the master formula \eqref{evolution-classical} is that, once all the constraints of the system are taken into account by a suitable Dirac bracket, the preservation of the constraints is guaranteed to all order in perturbation theory. 
For the twistor model, recall from section~\ref{sec:eff-Dirac-bracket} that the Dirac bracket satisfies 
\begin{align}
    \{ f , p^2 \}_* = \{ f , y^2 \}_* = \{ f , y\cdot p \}_* = 0 \,.
\end{align} 
When $f$ is gauge invariant, the Dirac bracket can be replaced by the Poisson bracket. 
Setting $f=\chi = \chi_{(1)} + \chi_{(2)} + \cdots$, since 
each $\chi_{(n)}$ scale differently with the coupling constant, we deduce that at each $n$, 
\begin{align}
    p \cdot \{ \chi_{(n)} , p \} = 0 \,,
    \quad 
    y \cdot \{ \chi_{(n)} , y \} = 0 \,,
    \quad 
     y \cdot \{ \chi_{(n)} , p \} +   p \cdot \{ \chi_{(n)} , y \} = 0 \,.
     \label{eikonal-ortho}
\end{align}
From a perturbative point of view, 
the iteration terms in \eqref{impulse-nPL} are essential for consistency. 
For example, at 2PL, 
the iteration term in \eqref{impulse-nPL} ensures the mass-shell condition:
\begin{align}
\begin{split}
     2 p\cdot  \Delta_{(2)} p &= 2 p_\m \{ \chi_{(2)} , p^\m \} + p_\m\{ \chi_{(1)} , \{ \chi_{(1)} , p^\m  \} \}
     \\
     &= \{ \chi_{(1)} , p_\m \{ \chi_{(1)} , p^\m  \} \} -  \{ \chi_{(1)} , p_\m  \}\{ \chi_{(1)} , p^\m  \} 
     = -  (\Delta_{(1)} p)^2 \,,
\end{split}
\label{2PL-mass-shell}
\end{align}
where we used \eqref{eikonal-ortho} and the Leibniz rule for the Poisson brackets.
In view of \eqref{eikonal-ortho} and \eqref{2PL-mass-shell}, when we compute the 2PL observables later in this section, 
we will call $\{ \chi_{(2)}, O\}$ and $\frac{1}{2} \{ \chi_{(1)} , \{ \chi_{(1)} , O \} \}$ 
``transverse" and ``longitudinal" (or ``iteration"), respectively. 

\subsubsection{Non-spinning example}

Arguing for the existence of $\chi$ is one thing, 
giving an algorithm to compute $\chi$ is another.
It seems plausible that the ``WQFT eikonal" $\chi_\mathrm{WQFT}$ \cite{Mogull:2020sak} coincides with our classical eikonal $\chi$. 
We will verify this expectation up to 2PL by explicit computations.
Proving the equivalence to all orders in perturbation theory is an open question, which requires incorporation of bremsstrahlung effects.

Before we compute the scattering generator and observables of the twistor model, to illustrate the ideas in a simpler setting, we review the binary dynamics of non-spinning particles in electromagnetism following ref.~\cite{Kosower:2018adc}, 
which shows how to separate the transverse term $\{ \chi_{(2)} , p \}$ from 
the longitudinal term $\frac{1}{2} \{ \chi_{(1)} , \{ \chi_{(1)} , p \} \}$. 
The solution of the EOM involves retarded Green functions on the worldline. 
When the solution is split into the time-symmetric part and the time-anti-symmetric part, 
the former gives the transverse term while the latter gives the iteration term.

The 1PL momentum kick is well known: 
\begin{align}
    \Delta_{(1)} p_{1}^\m = - q_1 q_2 \gamma \int_k  (ik^\m) \frac{ e^{ik\cdot b}}{k^2} \deltabar(v_1\cdot k) \deltabar(v_2\cdot k)  \,.
     \label{spinless-1PL-impulse}
\end{align}
where the velocity vectors are defined as $v_a^\m := p_a^\m / m_a$ 
and the relative boost is defined as $\gamma := - (v_1\cdot v_2)$.
The impact parameter $b^\m$ is the projection of the relative position $x_{12}^\m = x_1^\m - x_2^\m$ onto the plane transverse to $v_1$ and $v_2$.

The 1PL momentum kick \eqref{spinless-1PL-impulse} satisfies \eqref{impulse-nPL} rather trivially as 
\begin{align}
    \Delta_{(1)} p_{1}^\m  = \eta^{\m\n} \frac{\partial \chi_{(1)}}{\partial x_1^\n} = \{ \chi_{(1)} , p_{1}^\m \} \,, 
\quad 
    \chi_{(1)} = - q_1 q_2 \gamma \int_{k_\perp}  \frac{ e^{ik\cdot b}}{k^2}   \,, 
    \label{spinless-1PL-eikonal}
\end{align}
where the integral over the transverse plane is defined as 
\begin{align}
    \int_{k_\perp} = \int \dbar^4 k\,  \deltabar(v_1\cdot k) \deltabar(v_2\cdot k) \,.
\end{align}

As we proceed to the 2PL, it is useful to recall that the computation 
can be organized according to the mass ratio:
\begin{align}
        \Delta_{(2)} p_1^\mu = \frac{1}{m_1} K^\mu_{1,1} +  \frac{1}{m_2} K^\mu_{1,2} \,,
        \quad
        \Delta_{(2)} p_2^\mu = \frac{1}{m_1} K^\mu_{2,1} +  \frac{1}{m_2} K^\mu_{2,2} \,.
\end{align}
The vectors $K^\m_{a,b}$ are integrals independent of the masses. The conservation of $(p_1^\m + p_2^\m)$ requires 
that
\begin{align}
    K^\m_{1,1} + K^\m_{2,1} = 0 \,,
    \quad 
    K^\m_{1,2} + K^\m_{2,2} = 0 \,. 
\end{align}
The exchange symmetry between the two particles relate the integrals by 
\begin{align}
    K^\m_{1,1} \leftrightarrow K^\m_{2,2} \,,
    \quad 
    K^\m_{1,2} \leftrightarrow K^\m_{2,1} \,. 
\end{align}
So, it suffices to compute $K^\m_{1,1}$. 
To do so, we may work in the probe limit $(m_2 \rightarrow \infty)$ \cite{Vines:2018gqi}. 
In the rest of this section, we will work in the probe limit until further notice.

The 2PL EOM of a spinless particle in the probe limit is given by  
\begin{align}
     \dot{v}_{(2)}^\mu &= \frac{q}{m} F^\m{}_\n \dot{x}_{(1)}^\n + \frac{q}{m} x_{(1)}^\l \partial_\l F^\m{}_\n v_{(0)}^\n  \,,
\end{align}
where $F$ means $F_{(0)}$. 
Using the Bianchi identity, we may rewrite it as 
\begin{align}
         \dot{v}_{(2)}^\mu &= \frac{q}{m} (\partial^\m F_{\l\n}) x_{(1)}^\l v_{(0)}^\n  + \frac{d}{dt} \left( \frac{q}{m} F^\m{}_\n x_{(1)}^\n \right)   \,.
\end{align}
The total derivative on the RHS does not contribute to the momentum kick, so that 
\begin{align}
\begin{split}
     \Delta_{(2)} p_1^\m &= q \int d\s (\partial^\m F_{\l\n}) x_{(1)}^\l v_{(0)}^\n 
     \\
    &= \frac{q^2}{m} \int_{-\infty}^\infty d\s (\partial^\m F_{\l\n}) v_{(0)}^\n  \left[ \int_{-\infty}^\s d\s' \int_{-\infty}^{\s'} d\s''  F^\l{}_\r v_{(0)}^\r \right]  \,. 
\end{split}
\end{align}
Performing Fourier transform of ($F_{\l\n}$, $F^\l{}_\r$) by ($e^{i\ell\cdot x}$, $e^{ik\cdot x}$) respectively, 
and switching to $q = k + \ell$, we obtain 
\begin{align}
     \Delta_{(2)} p_1^\m = \frac{(q_1 q_2)^2}{m_1} \int_{q_\perp} e^{i q \cdot b} 
    \int_{\ell} \frac{\deltabar(v_2 \cdot \ell)}{\ell^2(q-\ell)^2} 
     (i\ell^\mu) \left[ 1+ \gamma^2 \frac{\ell\cdot (q-\ell)}{(v_1\cdot \ell + i0^+)^2} \right]  \,.
    \label{review:spinless-probe}
\end{align}
The $i0^+$ prescription comes from the retarded Green function on the worldline. 
To separate the time-symmetric parts from the time-anti-symmetric parts, 
as explained in ref.~\cite{Kosower:2018adc}, we apply the exchange $\ell \leftrightarrow q - \ell$ to get 
\begin{align}
     \Delta_{(2)} p_1^\m = \frac{(q_1 q_2)^2}{m_1} \int_{q_\perp} e^{i q \cdot b} 
    \int_{\ell} \frac{\deltabar(v_2 \cdot \ell)}{\ell^2(q-\ell)^2} 
     (iq^\mu - i\ell^\mu ) \left[ 1+ \gamma^2 \frac{\ell\cdot (q-\ell)}{(v_1\cdot \ell - i0^+)^2} \right]  \,.
    \label{review:spinless-time-reversed}
\end{align}
The next step is to take the average of \eqref{review:spinless-probe} and \eqref{review:spinless-time-reversed} and 
write the result as 
\begin{align}
    \Delta_{(2)} p_1^\m = \left. \Delta_{(2)} p_1^\m \right|_\mathrm{tr} + \left. \Delta_{(2)} p_1^\m \right|_\mathrm{iter} \,.
    \label{time-symm-anti}
\end{align}
The time-symmetric part carries an integrand proportional to $q^\m$ so that 
\begin{align}
\begin{split}
      \left. \Delta_{(2)} p_1^\m \right|_\mathrm{tr} &=  \frac{(q_1 q_2)^2}{2m_1} \int_{q_\perp} e^{i q \cdot b} 
    \int_{\ell} \frac{\deltabar(v_2 \cdot \ell)}{\ell^2(q-\ell)^2} 
     (iq^\mu) \left[ 1+ \gamma^2 \frac{\ell\cdot (q-\ell)}{(v_1\cdot \ell)^2} \right] \,. 
\end{split}
\end{align}
It yields the 2PL eikonal via $\left. \Delta_{(2)} p_1^\m \right|_\mathrm{tr} = \{ \chi_{(2)}, p_1^\m \}$. 
Omitting the $i0^+$ prescription that is no longer needed, and restoring the symmetry between the two particles, 
we obtain
 \begin{align}
\begin{split}    
      \chi_{(2)} &= \frac{(q_1 q_2)^2}{2m_1} \int_{q_\perp} e^{i q \cdot b}
    \int_\ell \frac{\deltabar(v_2 \cdot \ell)}{\ell^2(q-\ell)^2} 
    \left[ 1+ \gamma^2 \frac{\ell\cdot (q-\ell)}{(v_1\cdot \ell)^2} \right] + (1\leftrightarrow 2) \,.
\end{split}
\label{spinless-chi2}
\end{align}
It agrees perfectly with the WQFT eikonal~\cite{Wang:2022ntx}. 
The time-anti-symmetric part of \eqref{time-symm-anti} is  
\begin{align}
\begin{split}
     \left. \Delta_{(2)} p_1^\m \right|_\mathrm{iter} &=   \frac{ (q_1 q_2)^2\gamma^2 }{2m_1} \int_{q_\perp} e^{i q \cdot b} 
    \int_{\ell} \frac{\deltabar(v_2 \cdot \ell)  }{k^2 \ell^2 } (k \cdot \ell)
      \left[ \frac{i\ell^\mu }{(v_1\cdot \ell + i0^+)^2} - \frac{i\ell^\mu }{(v_1\cdot \ell - i0^+)^2} \right]  
    \\
    &=  - \frac{  (q_1 q_2)^2\gamma^2}{2m_1} \int_{q_\perp} e^{i q \cdot b} 
    \int_{\ell} \frac{\deltabar(v_2 \cdot \ell)  }{k^2 \ell^2} (k \cdot \ell) \ell^\m \, \deltabar'(v_1 \cdot \ell) \,, 
\end{split}
\label{spinless-2PL-antisymm}
\end{align}
where $k=q-\ell$ and we used the identity, 
\begin{align}
    \frac{i}{x + i0^+} -  \frac{i}{x - i0^+} = \deltabar(x) 
    \quad 
    \Longrightarrow 
    \quad 
    \frac{i}{(x + i0^+)^2} -  \frac{i}{(x - i0^+)^2} = - \deltabar'(x) \,.
\end{align}
We may compare it with the iteration term. From \eqref{spinless-1PL-eikonal} and \eqref{spinless-1PL-impulse}, we find
\begin{align}
    \frac{1}{2} \{ \chi_{(1)} , \{ \chi_{(1)} , p_1^\m \} \} = 
    \frac{i}{2} (q_1q_2)^2 \left\{ \gamma \int_{k_\perp} \frac{e^{ik\cdot b}}{k^2} , \gamma \int_{\ell_\perp} \ell^\mu \frac{e^{i\ell\cdot b}}{\ell^2} \right\} \,, 
    \label{spinless-2PL-iteration}
\end{align}
In computing the bracket in \eqref{spinless-2PL-iteration}, 
we note that 
\begin{align}
 \{ \deltabar(v_1\cdot k) \deltabar(v_2\cdot k)  e^{ik\cdot (x_1-x_2)} , \gamma \} = \deltabar(v_1\cdot k) \deltabar(v_2\cdot k) e^{ik\cdot (x_1-x_2)} \frac{i k \cdot \left( - p_2 + p_1 \right) }{m_1m_2} = 0 \,.
\end{align}
So, the only non-vanishing contributions come from 
\begin{align}
    \{  e^{ik\cdot (x_1-x_2)} , \deltabar(v_1\cdot \ell) \deltabar(v_2\cdot \ell) \}  \quad \mbox{and} \quad 
    \{  \deltabar(v_1\cdot k) \deltabar(v_2\cdot k) ,  e^{i\ell\cdot (x_1-x_2)}  \} \,.
\end{align}
Collecting the contributions, we obtain (in the probe limit) 
\begin{align}
\begin{split}
     &\frac{1}{2} \{ \chi_{(1)} , \{ \chi_{(1)} , p_1^\m \} \}  
    = - \frac{1}{2} (q_1q_2)^2\gamma^2 \int_{k,\ell}  e^{i(k+\ell)\cdot b}  \frac{k\cdot \ell}{k^2 \ell^2 } \mathcal{F}(k,\ell) \ell^\mu  \,,
    \\
    &\mathcal{F}(k,\ell) = \deltabar(v_1\cdot k) \deltabar(v_2\cdot k) 
    \frac{\deltabar^{\prime}(v_1 \cdot \ell) \deltabar(v_2 \cdot \ell)}{m_1} 
    - (k \leftrightarrow \ell) \,.
\end{split} 
\label{1PL-to-2PL-almost} 
\end{align}
Using the identities, 
\begin{align}
    \delta(x+y) \delta(y) = \delta(x) \delta(y) \,,
    \quad 
    \delta(x+y)\delta'(y) = \delta(x) \delta'(y) - \delta'(x) \delta(y) \,,
\end{align}
we can simplify \eqref{1PL-to-2PL-almost} a bit further and obtain 
\begin{align}
\begin{split}
     \frac{1}{2} \{ \chi_{(1)} , \{ \chi_{(1)} , p_{1\m} \} \} 
     = - \frac{  (q_1 q_2)^2\gamma^2}{2m_1} \int_{q_\perp} e^{i q \cdot b}  \int_\ell \frac{k\cdot \ell}{k^2 \ell^2  } \ell_\mu 
     \deltabar'(v_1\cdot\ell) \deltabar(v_2\cdot \ell) 
     \,,
\end{split} 
\label{1PL-to-2PL} 
\end{align}
in perfect agreement with \eqref{spinless-2PL-antisymm}.

In the paragraphs above, we manipulated the integrands of the Fourier integrals to separate the iteration term. 
In the non-spinning case, it is easy to perform the Fourier integral and compute the brackets in position space. 
We write the impact parameter as
\begin{align}
     b^\m &= x_{12}^\m + \frac{\g (x_{12} \cdot v_2) - (x_{12} \cdot v_1)}{\g^2 - 1} v_1^\m + \frac{\g (x_{12} \cdot v_1) - (x_{12} \cdot v_2)}{\g^2 - 1} v_2^\m \,,\, 
\end{align}
where $x_{12}^\m = x_1^\m - x_2^\m$. 
The 1PL non-spinning eikonal is, after the Fourier integral, 
\begin{align}
    \chi_{(1)} &= \frac{q_1 q_2 \g}{4 \pi \sqrt{\g^2 - 1}} \left[ \frac{1}{\e} + \log \left( \frac{b^2}{b_0^2} \right) \right] \,, 
\end{align}
where $\epsilon$ is the dimensional regularisation parameter $D = 4 - 2\e$ and $b_0^2$ is a scheme-dependent constant. The precise values of $\epsilon$ and $b_0$ are irrelevant. 
The relevant brackets are 
\begin{align}
    &\{ b^2 , \g \} = 0 \,,
    \quad 
    \{ b^2 , p_1^\m \} \doteq 2 b^\m \,,
    \quad
    \{ b^2 , b^\m \} \doteq 2 b^2 \left( \frac{\g v_2^\m - v_1^\m}{(\g^2 - 1) m_1} - \frac{\g v_1^\m - v_2^\m}{(\g^2 - 1) m_2} \right) \,,
\end{align}
where $\doteq$ denotes equivalence up to mass-shell constraint $p_i^2 + m_i^2 = 0$. 
Thus,
\begin{align}
\begin{split}
   \{ \chi_{(1)} , \{ \chi_{(1)} , p_1^\m \} \} &= \left[ \frac{q_1 q_2 \g}{4 \pi \sqrt{\g^2 - 1}} \right]^2 \{ \log b^2 , \{ \log b^2 , p_1^\m \} \} 
    \\ 
    &= \left[ \frac{q_1 q_2 \g}{4 \pi \sqrt{\g^2 - 1}} \frac{1}{b^2} \right]^2 \{ b^2 , \{ b^2 , p_1^\m \} \} 
    \doteq 2 \left[ \frac{q_1 q_2 \g}{4 \pi \sqrt{\g^2 - 1}} \frac{1}{b^2} \right]^2 \{ b^2 , b^\m \} 
    \\ 
    &\doteq \left[ \frac{q_1 q_2 \g}{2 \pi \sqrt{\g^2 - 1}} \frac{1}{\sqrt{b^2}} \right]^2 \times \left( \frac{\g v_2^\m - v_1^\m}{(\g^2 - 1) m_1} - \frac{\g v_1^\m - v_2^\m}{(\g^2 - 1) m_2} \right) 
    \\ &= \left| \{ \chi_{(1)} , p_1^\m \} \right|^2 \times \left( \frac{\g v_2^\m - v_1^\m}{(\g^2 - 1) m_1} - \frac{\g v_1^\m - v_2^\m}{(\g^2 - 1) m_2} \right) \,.  
\end{split}
\end{align}
This term generates the longitudinal impulse needed to preserve the mass-shell condition, 
\begin{align}
    p_{1\m} \, \{ \chi_{(1)} , \{ \chi_{(1)} , p_1^\m \} \} &= - \{ \chi_{(1)} , p_1^\m \}^2 \,.
\end{align}

\subsection{Momentum kick and spin kick from EOM}

In the scattering problem of two spinning particles, 
the momentum kick and the spin kick are the main scattering observables. 
In this subsection, we compute them up to the 2PL order by solving the EOM perturbatively. 
The perturbation computes the deviation from the free (straight line, constant spin) trajectory for each particle, 
\begin{align}
\begin{split}
 z^\mu(\sigma) &= (x_{(0)}^\mu + i y_{(0)}^\mu + v_{(0)}^\mu \sigma) + \delta z^\mu(\sigma) \,, 
\\
p^\mu(\sigma) &= m (v_{(0)}^\mu + \delta v^\mu(\sigma) ) \,.   
\end{split}
\label{background-worldline}
\end{align}
Without loss of generality, and taking constraints into consideration, we demand that the constant parameters satisfy 
\begin{align}
v_{(0)} \cdot v_{(0)} = -1 \,,\quad (x_{(0)} + i y_{(0)})\cdot v_{(0)} = 0 \,.
\end{align}
The perturbative solution is arranged in the PL order as 
\begin{align}
\begin{split}
   \delta z^\mu &= z_{(1)}^\mu + z_{(2)}^\mu + \cdots \,,  
   \\
   \delta v^\mu &= v_{(1)}^\mu + v_{(2)}^\mu + \cdots  \,, 
\end{split}
\end{align}
where $X_{(n)}$ is proportional to $q^n$. 
To avoid clutter, we will often omit the subscript $X_{(0)}$ from background values. 

\subsubsection{1PL}

At 1PL, the equations of motion are reduced to 
\begin{align}
\label{pert-eom-1PL-v}
       \dot{v}_{(1)}^\mu &= \frac{q}{m} (F_{(0)}^+ + F_{(0)}^-)^\m{}_\n v_{(0)}^\n \,,
       \\
       \dot{z}_{(1)}^\mu  &= v_{(1)}^\mu + \frac{2iq}{m}  (F_{(0)}^-)^\m{}_\n y_{(0)}^\n \,.
\label{pert-eom-1PL-z}
\end{align}
It is understood that $F^\pm_{(0)}$ here are evaluated along the background worldline \eqref{background-worldline}.

In computing the impulse of particle 1 to the 1PL order, we may treat particle 2 as a stationary source. 
The field-strength measured at the position of particle 1 is given by 
\begin{align}
\begin{split}
        F^+_{\m\n}(z_1(\s_1) ) = i \frac{q_2}{2} \int_k \deltabar(v_2\cdot k) \frac{(k\wedge v_2)_{\m\n} - i \epsilon_{\m\n}[ k,  v_2]} {k^2} e^{ik\cdot (z_1(\s_1) - \bar{z}_2(0))}  \,,
        \\
        F^-_{\m\n}(\bar{z}_1(\sigma_1)) = i \frac{q_2}{2} \int_k \deltabar(v_2\cdot k) \frac{(k\wedge v_2)_{\m\n} + i \epsilon_{\m\n} [k, v_2] }{k^2} e^{ik\cdot (\bar{z}_1(\s_1) - z_2(0))}   \,, 
\end{split}
\label{field-strength-fixed}
\end{align}
where the worldline time-dependence from $z_2^\m(\s_2) = (x_2^\m + i y_2^\m + v_2^\m \s_2)$ has been integrated out to leave  $\deltabar(v_2\cdot k)$ behind. 
The wedge and the epsilon notations mean
\begin{align}
\begin{split}
   &(a\wedge b)_{\m\n} = a_\m b_\n - b_\m a_\n \,, 
    \\
    &\e_{\m\n}[a,b] = \e_{\m\n\r\s} a^\r b^\s \,, 
    \quad 
    \e_{\m}[a,b,c] = \e_{\m\n\r\s} a^\n b^\r c^\s \,, 
    \quad 
    \e[a,b,c,d] = \e_{\m\n\r\s} a^\m b^\n c^\r d^\s \,. 
\end{split}
\end{align}
Contracting $F^{\pm}_{\m\n}$ with $v_1^\n$ and integrating \eqref{pert-eom-1PL-v} over $\s_1$, we obtain the velocity kick, 
\begin{align}
\begin{aligned}
    \D_{(1)} v_1^\m &= 
    - \frac{q_1 q_2}{2m_1} \int_{k_\perp} \frac{i \gamma k^\m + \e^{\m}[k,v_1,v_2]}{k^2} e^{i k \cdot (b+iy)} 
    \\ 
    &\qquad    
    - \frac{q_1 q_2}{2m_1} \int_{k_\perp} \frac{i \gamma k^\m - \e^{\m}[k,v_1,v_2]}{k^2}  e^{i k \cdot (b-iy) } 
    \\
    &= 
    - \frac{q_1 q_2}{m_1} \int_{k_\perp}  \left[ (ik^\m) \cosh(k\cdot y) \gamma  - \e^{\m}[k,v_1,v_2] \sinh(k\cdot y) \right] \frac{ e^{ik\cdot b}}{k^2}  \,.
\end{aligned} 
\label{1PL-impulse}
\end{align}
The argument of the exponential is decomposed as 
\begin{align}
    z_{1}^\m - \bar{z}_{2}^\m = (x_{1}^\m - x_2^\m) + i (y_1^\m + y_2^\m) =: x_{12}^\m + i y^\m \,, 
\end{align}
and $x_{12}^\m$ is projected onto the impact parameter vector $b^\mu$ by $\deltabar(v_1\cdot k) \deltabar(v_2\cdot k)$. Note how the spin sum $(y_1^\mu + y_2^\mu)$ arises from the \emph{difference} between complex spinspacetime coordinates.

Similarly, we can compute the spin kick and find 
\begin{align}
\begin{aligned}
    \D_{(1)} y_1^\m 
    &=  \frac{q_1 q_2}{m_1} \int_{k_\perp} \left[ i (k\wedge v_2)^\m{}_\n y_1^\n \cosh(k\cdot y) -  \e^{\m}[ k, v_2, y_1]  \sinh(k\cdot y) \right] \frac{e^{i k \cdot b}}{k^2}  \,.
\end{aligned}
\label{1PL-spin-kick}
\end{align}
It is easy to verify orthogonality and conservation of magnitudes of velocity/spin at 1PL:
\begin{align}
\begin{aligned}
    \D (y_1 \cdot v_1) &= (\D y_1 \cdot v_1) + (y_1 \cdot \D v_1) = 0 \,,
    \\ \D (v_1^2) &= 2 (\D v_1 \cdot v_1) = 0 \,,
    \\ \D (y_1^2) &= 2 (\D y_1 \cdot y_1) = 0 \,.
\end{aligned}
\end{align}

The 1PL observables can be compared with predictions of QED in the classical limit. We use the results of ref.~\cite{Bern:2023ity} as the reference. When truncated to linear order in $y$, \eqref{1PL-impulse} and \eqref{1PL-spin-kick} are found to be consistent with (4.45) and (4.46) of ref.~\cite{Bern:2023ity}, under the conditions $C_i = 1$, $D_i = 0$, and the covariant spin supplementary condition. 
Note that the spin tensor kick $\D S^{\m\n}$ reported by ref.~\cite{Bern:2023ity} receives contributions from both \eqref{1PL-impulse} and \eqref{1PL-spin-kick}.

The 1PL observables can be reproduced by the classical eikonal as \cite{Guevara:2019fsj}
\begin{align}
\begin{split}
     \Delta_{(1)} p_1^\m &= \{ \chi_{(1)} , p_1^\m \} = \eta^{\m\n} \frac{\partial}{\partial x_1^\n}   \chi_{(1)} \,,
    \\ 
     \Delta_{(1)} y_1^\m &= \{ \chi_{(1)} , y_1^\m \}
     = \frac{1}{m_1} \left[ v_1^\mu y_1^\nu  \frac{\partial}{\partial x_1^\nu} 
    +  \e^{\m\n}[ v_1, y_1] \frac{\partial}{\partial y_1^\n } \right] \chi_{(1)} \,.
\end{split}
\label{1PL-eikonal-to-impulse}
\end{align}
An explicit form of the 1PL eikonal is 
\begin{align}
    \chi_{(1)} &= - q_1 q_2 \int_{k_\perp} \left[ \cosh(k\cdot y) \gamma - i \frac{\sinh(k\cdot y)}{k\cdot y} \epsilon[k,v_1,v_2,y]  \right] \frac{ e^{ik\cdot b}}{k^2} \,.
    \label{1PL-eikonal}
\end{align}
To reproduce \eqref{1PL-impulse} and \eqref{1PL-spin-kick} from \eqref{1PL-eikonal} via \eqref{1PL-eikonal-to-impulse}, 
we need the identity, 
\begin{align}
    \frac{k^\m \e[k,v_1,v_2,y]}{k\cdot y} \approx - \e^\m[k,v_1,v_2] \,. 
    \label{kdoty-identity}
\end{align}
The approximate equality ($\approx$) means that we may impose the transversality condition $k\cdot v_1 = 0 = k\cdot v_2$ and ignore ultra-local ($\propto k^2/k^2$) terms inside the integral. 
With this understanding, the identity above can be derived from the 4d Schouten identity,
\begin{align}
\begin{split}
    &a^\m \e[b,c,d,e] +  b^\m \e[c, d, e, a] + c^\m \e[d,e,a,b]
    \\
    &\qquad  + d^\m \e[e,a,b,c] + e^\m \e[a,b,c,d] = 0 \,.
\end{split}
\label{4d-schouten}
\end{align}
In what follows, we will not distinguish the approximate equality from the strict equality. 
Inside the integrands of the Fourier integrals, we will take the liberty to set 
\begin{align}
    k^2 = 0\,,
    \quad 
    \ell^2 = 0 \,,
    \quad 
    k\cdot v_2 = 0 = \ell \cdot v_2 \,,
    \quad
    (k+\ell) \cdot v_1 = 0\,. 
\end{align}

\subsubsection{2PL}  \label{sec:2PL spin kick}

The 2PL equations directly relevant for the impulse computation are 
\begin{align}
\begin{split}
       \dot{v}_{(2)}^\mu &= \frac{q}{m} (F^+_{(0)} + F^-_{(0)} )^\m{}_\n v_{(1)}^\n + \frac{q}{m} ( F^+_{(1)} + F^-_{(1)} )^\m{}_\n v_{(0)}^\n  \,,
       \\
       \dot{y}_{(2)}^\mu  &= \frac{q}{m} (F^+_{(0)} + F^-_{(0)})^\mu{}_\nu y_{(1)}^\nu + \frac{q}{m} (F^+_{(1)} + F^-_{(1)})^\mu{}_\nu y_{(0)}^\nu 
       \\
       & \quad - i \frac{4q^2}{m^2}  \left[  (v F^- y) (F^+)^\m{}_\n  -  (v F^+  y)(F^-)^\m{}_\n  \right]_{(0)} y_{(0)}^\n  \,.
\end{split}
\label{pert-eom-2PL}
\end{align}

\paragraph{2PL velocity kick} 
As we saw in the non-spinning case, 
it is sufficient to work in the probe limit $(m_1/m_2 \rightarrow 0)$, 
and we can use the Bianchi identity and discard a total derivative to get 
\begin{align}
     \Delta_{(2)} p^\m &= q \int d\s (\partial^\m F^+_{\l\n}) z_{(1)}^\l v_{(0)}^\n + q \int d\s (\partial^\m F^-_{\l\n}) \bar{z}_{(1)}^\l v_{(0)}^\n \,.
\end{align}
Next, using \eqref{pert-eom-1PL-z}, we can turn $(z_{(1)}, \bar{z}_{(1)})$ into integrals,  
\begin{align}
\begin{split}
     \Delta_{(2)} p^\m &= \Delta_{(2v)} p^\m + \Delta_{(2y)} p^\m \,,
     \\
   \Delta_{(2v)} p^\m  &= \frac{q^2}{m} \int_{-\infty}^\infty d\s \partial^\m (F^+ + F^-)_{\l\n} v_{(0)}^\n  \left[ \int_{-\infty}^\s d\s' \int_{-\infty}^{\s'} d\s''  (F^+ + F^-)^\l{}_\r v_{(0)}^\r \right] \,,
    \\
   \Delta_{(2y)} p^\m  &= \frac{2iq^2}{m}  \int_{-\infty}^\infty d\s  (\partial^\m F^+_{\l\n}) v^\n_{(0)} \int_{-\infty}^\s d\s' (F^-)^\l{}_\r y_{(0)}^\r 
    \\
    &\quad - \frac{2iq^2}{m}  \int_{-\infty}^\infty d\s  (\partial^\m F^-_{\l\n}) v^\n_{(0)} \int_{-\infty}^\s d\s' (F^+)^\l{}_\r y_{(0)}^\r\,. 
\end{split}
\end{align}
We divided the computation into two parts. The ``y-part" ($\Delta_{(2y)} p^\m $) is linear in $y_1$ aside from the $y$-dependence in the exponential factors. The ``v-part" ($\Delta_{(2v)} p^\m $) is independent of $y_1$ aside from the exponential factors. While we replace $v^\m_{(1)}$ by $\dot{z}_{(1)}$ or $\dot{\bar{z}}_{(1)}$, we also encounter terms proportional to $( F^+)^\m{}_\n  (F^-)^\n{}_\l - ( F^-)^\m{}_\n  (F^+)^\n{}_\l$, but they vanish identically.
Let us analyse the two parts one by one.

For the y-part, after using the field-strengths \eqref{field-strength-fixed} and integrating over the worldline, we obtain 
\begin{align}
   \Delta_{(2y)} p_1^\m  &= \frac{(q_1q_2)^2}{m_1}\int_{q_\perp} e^{i q \cdot b} \int_{\ell} \frac{\deltabar(v_2 \cdot \ell) (i\ell^\mu) }{\ell^2 k^2(ik\cdot v_1+0^+) } 
      \left( \mathrm{ch}\!\boxminus C_y + \mathrm{sh}\!\boxminus  S_y  \right) \,,
\label{2PL-impulse-y}
\end{align}
where we set $q= k + \ell$ as before and introduce shorthand notations, 
\begin{align}
\begin{split}
\mathrm{ch}\boxminus = \cosh[(k-\ell)\cdot y]\,, &\quad 
\mathrm{sh}\boxminus = \sinh[(k-\ell)\cdot y] \,, 
\\
\mathrm{ch}\boxplus = \cosh[(k+\ell)\cdot y]\,, &\quad 
\mathrm{sh}\boxplus = \sinh[(k+\ell)\cdot y] \,.
\end{split}
\end{align}
The functions $C_y$, $S_y$ are 
\begin{align}
\begin{split}
     C_y 
     &= - \e[k,\ell,y_1,v_1] + 2 (v_2\cdot y_1) \e[k,\ell,v_1,v_2] \,, 
     \\
     S_y &= i (\ell\cdot v_1) [(k-\ell)\cdot y_1] + 2 i \g (v_2\cdot y_1) (k\cdot \ell)  
     \,.
\end{split}
\label{2PL-impulse-y-N}
\end{align}

We divide the v-part further into the same helicity contribution ($\D_{(2vs)} p^\m$) 
and the opposite helicity contribution ($\D_{(2vo)} p^\m$).
The opposite helicity part is 
\begin{align}
\begin{split}
   \Delta_{(2vo)} p_1^\m  &= \frac{(q_1q_2)^2}{m_1} \int_{q_\perp} e^{i q \cdot b} \int_{\ell} \frac{\deltabar(v_2 \cdot \ell) (i\ell^\mu) }{\ell^2 k^2(ik\cdot v_1+0^+)^2 } 
      \left( \mathrm{ch}\!\boxminus \, C_{vo} +  \mathrm{sh}\!\boxminus \, S_{vo}  \right) \,,
\\
     &\quad C_{vo} = - (\g^2-\textstyle{\frac{1}{2}}) (k\cdot \ell) +  (k\cdot v_1)(\ell \cdot v_1) \,, 
     \quad
     S_{vo} = i \g \,\e[k,l,v_1, v_2] \,.
\end{split}
\label{2PL-impulse-vo}
\end{align}
The same helicity part is quite simple: 
\begin{align}
   \Delta_{(2vs)} p_1^\m  &= \frac{(q_1q_2)^2}{m_1} \int_{q_\perp} e^{i q \cdot b} \int_{\ell} \frac{\deltabar(v_2 \cdot \ell) (i\ell^\mu)  }{\ell^2 k^2(ik\cdot v_1+0^+)^2 } (-\textstyle{\frac{1}{2}})(k\cdot \ell) \mathrm{ch}\!\boxplus  \,.
\label{2PL-impulse-vs}
\end{align}
In the non-spinning limit $(y\rightarrow 0)$, it cancels against the $+\frac{1}{2} (k\cdot \ell)$ term of $C_{vo}$ in \eqref{2PL-impulse-vo}, 
in agreement with \eqref{review:spinless-probe}.

\paragraph{2PL eikonal}

As in the non-spinning example, we can extract the 2PL eikonal from the 2PL momentum kick. 
The key idea \cite{Kosower:2018adc} is to apply the exchange $\ell \leftrightarrow k = q-\ell$ to the integrand, 
\begin{align}
\begin{split}
       \Delta_{(2)} p_1^\m  = \int_{q_\perp} e^{iq\cdot b} \int_{\ell} (i\ell^\m) \, \mathcal{J}(k,\ell) 
= \int_{q_\perp} e^{iq\cdot b} \int_{\ell} (iq^\m - i\ell^\m) \, \mathcal{J}(\ell,k) \,.
\end{split}
\end{align}
Since $q\cdot v_1 = k \cdot v_1 + \ell\cdot v_1 = 0$, 
the exchange flips the $i0^+$ prescription for the worldline Green function. 
Taking the average of the two expressions and taking the term proportional to $q^\m$ in the integrand, 
we separate the transverse part of the momentum kick, from which 
we read off the 2PL eikonal equipped with the time-symmetric $i0^+$, 
\begin{align}
\begin{split}
    \left. \Delta_{(2)} p_1^\m \right|_\mathrm{tr} &= \frac{1}{2} \int_{q_\perp} e^{iq\cdot b} \int_{\ell} (iq^\m) \, \mathcal{J}(\ell,k)  = \{ \chi_{(2)} , p_1^\m \} = \eta^{\m\n} \frac{\partial}{\partial x^\n_1} \chi_{(2)} 
    \\
  &\quad \Longrightarrow \quad 
  \chi_{(2)} = \frac{1}{2} \int_{q_\perp} e^{iq\cdot b} \int_{\ell} \mathcal{J}(\ell,k)  \,.   
\end{split}
\label{2PL-kick-chi2}
\end{align}
The remaining $\ell^\m$ terms in the integrand should be matched against the iteration term, 
\begin{align}
  \left. \Delta_{(2)} p_1^\m \right|_\mathrm{iter} = \frac{1}{2} \int_{q_\perp} e^{iq\cdot b} \int_{\ell} (i\ell^\m) \, \left[ \mathcal{J}(k,\ell) - \mathcal{J}(\ell,k) \right]
  = 
  \frac{1}{2} \{ \chi_{(1)} , \{ \chi_{(1)} , p_1^\m \} \} \,.
  \label{2PL-kick-iteration}
\end{align}

We have computed $\Delta_{(2)} p_1^\m$ from the EOM. It is straightforward to split it into $\left. \Delta_{(2)} p_1^\m \right|_\mathrm{tr}$ and $\left. \Delta_{(2)} p_1^\m \right|_\mathrm{iter}$, 
and then read off $\chi_{(2)}$ from $\left. \Delta_{(2)} p_1^\m \right|_\mathrm{tr}$. 
The final result for $\chi_{(2)}$ is 
\begin{align}
\begin{split}
        \chi_{(2)} &= \chi_{(2)1} + \chi_{(2)2} \,,
        \\
        \chi_{(2)1} &= - \frac{(q_1q_2)^2}{2m_1} \int_{q_\perp} e^{i q \cdot b} \int_{\ell} \frac{\deltabar(v_2\cdot \ell )}{k^2\ell^2(\ell\cdot v_1)^2} 
        \left[ \mathcal{V}_1 + \mathcal{Y}_1  \right] \,,
        \\
       \mathcal{V}_1 &=  \left[ - (\g^2-\textstyle{\frac{1}{2}} ) (k\cdot \ell) +  (k\cdot v_1)(\ell \cdot v_1)  \right] \cosh[(k-\ell)\cdot y]
       \\
       &\quad 
       - \textstyle{\frac{1}{2}} (k\cdot \ell) \cosh[(k+\ell)\cdot y]
       + i \g \,\e[k,\ell,v_1, v_2] \sinh[(k-\ell)\cdot y] \,,
       \\
       \mathcal{Y}_1 &= -i (\ell\cdot v_1) \left[ - \e[k,\ell,y_1,v_1] + 2 (v_2\cdot y_1) \e[k,\ell,v_1,v_2]  \right] \cosh[(k-\ell)\cdot y]
       \\
       &\quad + (\ell\cdot v_1)  \left[  (\ell\cdot v_1) [(k-\ell)\cdot y_1] + 2  \g (v_2\cdot y_1) (k\cdot \ell)  \right] \sinh[(k-\ell)\cdot y] \,.
\end{split}
\label{2PL-eikonal-final}
\end{align}
The other half of the answer, $\chi_{(2)2}$, can be obtained from $\chi_{(2)1}$ by the exchange of particle labels $(1\leftrightarrow 2)$. 
Verifying the iteration relation \eqref{2PL-kick-iteration} is also straightforward, but involves a lengthy computation; see appendix~\ref{app:iteration} for details.

\paragraph{2PL spin kick}

To complete the 2PL story, let us compute the 2PL spin kick from the EOM and confirm that it can be reproduced by the 2PL eikonal we obtained earlier. 
We begin with a copy of the 2PL EOM for spin from \eqref{pert-eom-2PL},
\begin{align}
\begin{split}
       \dot{y}_{(2)}^\mu  &= \frac{q}{m} (F^+_{(0)} + F^-_{(0)})^\mu{}_\nu y_{(1)}^\nu + \frac{q}{m} (F^+_{(1)} + F^-_{(1)})^\mu{}_\nu y_{(0)}^\nu 
       \\
       & \quad - i \frac{4q^2}{m^2}  \left[  (v F^- y) (F^+)^\m{}_\n  -  (v F^+  y)(F^-)^\m{}_\n  \right]_{(0)} y_{(0)}^\n  \,.
\end{split}
\label{pert-eom-2PL-copy}
\end{align}
Using the 1PL EOM, we turn the terms on the RHS to integrals, 
\begin{align}
\begin{split}
     \Delta_{(2)} y^\m &= \Delta_{(2a)} y^\m + \Delta_{(2b)} y^\m + \Delta_{(2c)} y^\m + \Delta_{(2d)} y^\m\,,
     \\
   \Delta_{(2a)} y^\m  &= \frac{q^2}{m^2} \int_{-\infty}^\infty d\sigma (F^+ + F^-)^\m{}_\n \int_{-\infty}^\sigma d\s' (F^+ + F^-)^\n{}_\l y_{(0)}^\l \,,
     \\
   \Delta_{(2b)} y^\m  &= \frac{q^2}{m^2} \int_{-\infty}^\infty d\sigma \partial_\l (F^+ + F^-)^\m{}_\n y_{(0)}^\n
   \left[ \int_{-\infty}^\s d\s' \int_{-\infty}^{\s'} d\s''  (F^+ + F^-)^\l{}_\r v_{(0)}^\r \right] \,,
     \\
   \Delta_{(2c)} y^\m  &= \frac{2iq^2}{m^2} \int_{-\infty}^\infty d\sigma \partial_\l (F^+)^\m{}_\n y_{(0)}^\n \int_{-\infty}^\s d\s' (F^-)^\l{}_\r y_{(0)}^\r  
   \\
   &\quad - \frac{2iq^2}{m^2} \int_{-\infty}^\infty d\sigma \partial_\l (F^-)^\m{}_\n y_{(0)}^\n \int_{-\infty}^\s d\s' (F^+)^\l{}_\r y_{(0)}^\r \,, 
      \\
   \Delta_{(2d)} y^\m  &= - i \frac{4q^2}{m^2}   \int_{-\infty}^\infty d\sigma  \left[  (v F^- y) (F^+)^\m{}_\n  -  (v F^+  y)(F^-)^\m{}_\n  \right]_{(0)} y_{(0)}^\n \,.
\end{split}
\label{many-terms-for-y}
\end{align}
Both (b) and (c) terms come from the second term on the RHS of \eqref{pert-eom-2PL-copy}. 

Again, it suffices to work in the probe limit; we keep using the Fourier integral \eqref{field-strength-fixed} of the field-strength produced by a fixed source. 
After worldline time integrals, we reach an expression of the form 
\begin{align}
    \Delta_{(2)} y_1^\m = \frac{(q_1q_2)^2}{m_1^2} \int_{q_\perp} e^{iq\cdot b} \int_\ell \frac{\deltabar(v_2\cdot \ell)}{k^2\ell^2(ik\cdot v_1+0^+)^2} \CN^\m \,.
    \label{2PL spin kick overview}
\end{align}
The numerator $\CN^\m$ can be computed separately for each term in \eqref{many-terms-for-y}. 
For (a) and (b) terms, 
we also distinguish the same/opposite helicity contributions. 
\begin{align}
\begin{split}
      \CN^\m_{(2ao)} =  \left( \mathrm{ch}\!\boxminus C_{ao} +  \mathrm{sh}\!\boxminus S_{ao}  \right)^\m \,,
      &\quad 
       \CN^\m_{(2as)}  =  \left( \mathrm{ch}\!\boxplus C_{as} +  \mathrm{sh}\!\boxplus  S_{as}  \right)^\m \,,
    \\  
     \CN^\m_{(2bo)} = \left( \mathrm{ch}\!\boxminus C_{bo} +  \mathrm{sh}\!\boxminus S_{bo} \right)^\m \,,
     &\quad 
         \CN^\m_{(2bs)} = \left( \mathrm{ch}\!\boxplus  C_{bs} +  \mathrm{sh}\!\boxplus  S_{bs}  \right)^\m \,,
          \\
      \CN^\m_{(2c)} =  \left( \mathrm{ch}\!\boxminus C_c +  \mathrm{sh}\!\boxminus S_c  \right)^\m \,,
    &\quad 
     \CN^\m_{(2d)} = \left( \mathrm{ch}\!\boxminus C_d +  \mathrm{sh}\!\boxminus S_d  \right)^\m \,,
\end{split}
\label{C-S-12}
\end{align}
It is straightforward to evaluate all $C$, $S$ functions in \eqref{C-S-12}. 
The intermediate steps involve many terms, but after some cancellations, the final results are often quite simple. 
For instance, the same helicity sector gives 
\begin{align}
    \begin{split}
        C^\m_{as} + C^\m_{bs} &= - \frac{i}{2} (k\cdot \ell) \left[ v^\m_1 (\ell\cdot y_1)- y^\m_1 (\ell\cdot v_1)   \right] \,,
        \\
        S^\m_{as} + S^\m_{bs} &= - \frac{1}{2} (k\cdot \ell) \e^\m[v_1,y_1,\ell] \,.
    \end{split}
\end{align}
After collecting and simplifying the terms, using the symmetrisation of the $i0^+$ prescription, 
we extract the transverse part of the spin kick and check whether the result agrees with 
\begin{align}
\begin{split}
    \{ \chi_{(n)} , y_1^\m \} &= \frac{1}{m_1} \left[ v_1^\mu y_1^\nu  \frac{\partial}{\partial x_1^\nu} 
    +  \e^{\m\n}[ v_1, y_1] \frac{\partial}{\partial y_1^\n } \right] \chi_{(n)} \,.
\end{split}
\label{eikonal-to-impulse-copy1}
\end{align}
Again, the procedure is straightforward, and we confirm the agreement, but the calculations tend to be lengthy. We give some details in appendix~\ref{app:spin kick more}.

 
\section{Twistor WQFT} \label{sec:TWQFT}

Worldline Quantum Field Theory (WQFT)~\cite{Mogull:2020sak} is a means of organising classical equations of motion in a diagrammatic way that resembles Feynman diagrams of particle physics. 
The simplest WQFT model action consists of two parts; the bulk part which describes field degrees of freedom (DOFs) on the background spacetime, and one-dimensional sigma model that takes the background spacetime as the target space, where the latter is interpreted as the worldline action of a point particle moving on the background spacetime. The DOFs are decomposed into the background value (which satisfies the classical equations of motion) and fluctuations from the background value, and diagrammatic techniques developed for perturbative QFT calculations are applied to the field fluctuations. The background value for the background spacetime is usually taken to be the flat Minkowski spacetime, and the background value for the worldline is usually taken to be the straight trajectory of a free particle; $x^\m(\s) = b^\m + v^\m \s + \delta x^\m (\s)$. The fluctuation DOFs are evaluated as a perturbative series of the coupling constant, which in the gravitational case is taken to be the Newton's constant $G$.

\subsection{Worldline Feynman rules} \label{sec:Feynman}

When we apply the WQFT method to our twistor model, one novelty is that the twistor variables $( \l,\bar{\l}, \m, \bar{\m} )$ are the fundamental variables, and the target space of the worldline is the twistor space indirectly related to the background spacetime by the incidence relations.  
It is natural to express the propagators and vertex factors in terms of the twistor variables. 
But, it is often convenient to use the vector variables $( z, \bar{z} )$ in intermediate steps. 
We will use the incidence relations to switch between the twistor variables (``twistor picture") 
and the vector variables (``spacetime picture") whenever necessary. 

\paragraph{Classical limit} 

When $\hbar$ is restored, the fundamental variables have the dimensions
\begin{align}
    [ \l_\a{}^I ] = [\bar{\l}_{I \dot\a}] = [M^{1/2}] \,,\, [\m^{\dot\a I}] = [\bar{\m}_I{}^\a] = [M^{1/2} L] \,,\, [A_\m] = [M^{1/2} L^{-1/2}] \,,
\end{align}
and the coupling becomes dimensionful: $[q] = [M^{1/2} L^{1/2}]$. The action also becomes $i S \to i S/\hbar$, therefore the interaction vertices are weighted by $\hbar^{-1}$ while the propagators (both $\langle \l \bar{\m} \rangle$ type and $\langle \m \m \rangle$ type) are weighted by $\hbar$. We have no other $\hbar$ scaling if we only use frequency $\w$ and wavenumber four-vector $k^\m$ in momentum space, which is allowed because there is no ``mass'' in any of the fluctuation fields. Therefore, at a given $q$ order (which determines the number of vertices coupling to the photon field), the classical contribution is determined from the diagrams with the least number of propagators that makes the diagram connected, which is equivalent to the statement that tree diagrams determine the classical physics. The tree diagrams, however, generally have momentum integrals similar to loop integrals of quantum field theory.

\subsubsection{Background-fluctuation expansion} 

In the non-spinning WQFT, the expansion around a straight line trajectory is done by 
\begin{align}
    x^\mu(\sigma) = b_0^\mu + v^\mu \sigma + \delta x^\mu(\sigma) \,, 
    \quad 
    v^2 = -1 \,, 
    \quad
    b_0\cdot v = 0 \,.
\end{align}
After the NJ-shift, the expansion is generalised as 
\begin{align}
     z^\mu(\sigma) = b_0^\mu + i y_0^\mu + v^\mu \sigma + \delta z^\mu(\sigma) \,, 
    \quad 
     y_0 \cdot v = 0 \,.
    \label{z-fluctuation}
\end{align}

We should rephrase the background-fluctuation expansion in terms of twistor variables. 
As we explained in section~\ref{sec:regge_traj}, we may assume a flat Regge trajectory $m' = 0$. 
The resulting equations of motion for the free action \eqref{eq:tw_free_action1} are 
\begin{align}
\begin{aligned}
    \frac{\delta S_{\text{free}}}{\delta \bar{\m}_I{}^\a} &= 0 = - \frac{d \l_\a{}^I}{d \s} + \frac{\k^1}{2i} {\l}_\a{}^I
    \\ \frac{\delta S_{\text{free}}}{\delta {\m}^{\dot\a I}} &= 0 = -\frac{d \bar{\l}_{I \dot\a}}{d \s} - \frac{\k^1}{2i} {\bar{\l}}_{I \dot\a}
    \\ \frac{\delta S_{\text{free}}}{\delta {\l}_{\a}{}^I} &= 0 = \frac{d \bar{\m}_I{}^\a }{d \s} + \frac{\k^0 \D \bar{\D}}{2} (\l^{-1})_I{}^\a + \frac{\k^1}{2i} \bar{\m}_I{}^\a
    \\ \frac{\delta S_{\text{free}}}{\delta \bar{\l}_{I \dot\a}} &= 0 = \frac{d {\m}^{\dot\a I}}{d \s} + \frac{\k^0 \D \bar{\D}}{2} (\bar{\l}^{-1})^{\dot\a I} - \frac{\k^1}{2i} {\m}^{\dot\a I} \,.
\end{aligned}
\end{align}
Just like in the previous section, we fix the Lagrange multipliers as $\k^0 = 1/m$ and $\k^1 = 0$. We introduce the background values $\ell_\a{}^I$ and $\bar{\ell}_{I\dot\a}$ satisfying the conditions
\begin{align}
    \det(\ell) = \det(\bar{\ell}) = m \,,\quad  
    \ell_\a{}^I \bar{\ell}_{I\dot\a} = - m v_{\a \dot\a} \,,
\end{align}
where $v^\m$ is the normalised velocity vector introduced in
\eqref{z-fluctuation}. 
The following relations satisfied by inverse matrices are useful in calculations.
\begin{align}
    (\ell^{-1})_I{}^\a = - \frac{\e_{IJ} \e^{\a\b} \ell_\b {}^J}{\det(\ell)} \,,\quad 
    (\bar{\ell}^{-1})^{\dot\a I} = \frac{\e^{\dot\a \dot\b} \e^{IJ} \bar{\ell}_{J \dot\b}}{\det(\bar{\ell})} \,,\quad  
    (\bar{\ell}^{-1})^{\dot\a I} (\ell^{-1})_I{}^\a = - \frac{v^{\dot\a \a}}{m} \,.
    \label{eq:inv_lambda}
\end{align}
The twistor variables are expanded as
\begin{align}
\begin{aligned}
    \l_\a{}^I &\to \ell_\a{}^I + \l_\a{}^I (\s) \,,
    \\ \bar{\l}_{I \dot\a} &\to \bar{\ell}_{I \dot\a} + \bar{\l}_{I \dot\a} (\s) \,,
    \\ \m^{\dot\a I} &\to c^{\dot\a I} - \frac{m}{2} (\bar{\ell}^{-1})^{\dot\a I} \s + \m^{\dot\a I} (\s) \,,
    \\ \bar{\m}_I{}^{\a} &\to \bar{c}_{I}{}^{\a} - \frac{m}{2} (\ell^{-1})_I{}^\a \s + \bar{\m}_I{}^{\a} (\s) \,.
\end{aligned} \label{eq:twistor_expansion_def}
\end{align}
The background-fluctuation expansion for $z^\m$ and $\bar{z}^\m$ can be determined from the incidence relations \eqref{eq:incidence_new}, 
\begin{align}
\begin{aligned}
    z^{\dot\a \a} &\to + 2 c^{\dot\a I} (\ell^{-1})_I{}^\a - m (\bar{\ell}^{-1})^{\dot\a I} (\ell^{-1})_I{}^\a \s + \delta z^{\dot\a \a} (\s) = z_0^{\dot\a \a} + v^{\dot\a \a} \s + \delta z^{\dot\a \a} (\s) \,,
    \\ 
    \bar{z}^{\dot\a \a} &\to + 2 (\bar{\ell}^{-1})^{\dot\a I} \bar{c}_{I}{}^\a - m (\bar{\ell}^{-1})^{\dot\a I} (\ell^{-1})_I{}^\a \s + \delta \bar{z}^{\dot\a \a} (\s) = \bar{z}_0^{\dot\a \a} + v^{\dot\a \a} \s + \delta \bar{z}^{\dot\a \a} (\s) \,.
\end{aligned}
\end{align}
The relation between the fluctuation fields are determined from the incidence relations.
\begin{align}
\begin{aligned}
    2 \m^{\dot\a I} (\s) &= \left( z_0^{\dot\a \a} + v^{\dot\a \a} \s \right) \l_\a{}^I (\s) + \delta z^{\dot\a \a} (\s) \ell_\a{}^I + \delta z^{\dot\a \a} (\s) \l_\a{}^I (\s) \,,
    \\ 2 \bar{\m}_I{}^{\a} (\s) &= \bar{\l}_{I \dot\a} (\s) \left( \bar{z}_0^{\dot\a \a} + v^{\dot\a \a} \s \right) + \bar{\ell}_{I \dot\a} \delta \bar{z}^{\dot\a \a} (\s) + \bar{\l}_{I \dot\a} (\s) \delta \bar{z}^{\dot\a \a} (\s) \,.
\end{aligned}
\label{eq:incidence-again}
\end{align}
We use the positive frequency expansion,
\begin{align}
    \phi (x) = \int_k \phi(k) e^{ikx} \,,\quad  
    f(\s) = \int_\w f(\w) e^{-i \w \s} \,,
\end{align}
which relates the frequency space coefficients to annihilation modes and incoming momenta. The frequency space expression for the incidence relation turns out to be more useful
\begin{align}
\begin{aligned}
    \delta z^{\dot\a \a} (\w) &= \left[ 2 \m^{\dot\a I} - z_0^{\dot\a \b} \l_\b{}^I + i v^{\dot\a \b} \frac{\partial \l_\b{}^I}{\partial \w} - \int_{\w'} \delta z^{\dot\a \b} (\w') \l_\b{}^I (\w - \w') \right]
    (\ell^{-1})_I{}^\a \,,
    \\ 
    \delta \bar{z}^{\dot\a \a} (\w) &= (\bar{\ell}^{-1})^{\dot\a I} \left[ 2 \bar{\m}_I{}^{\a} - \bar{\l}_{I \dot\b} \bar{z}_0^{\dot\b \a} + i \frac{\partial \bar{\l}_{I \dot\b}}{\partial \w} v^{\dot\b \a} - \int_{\w'} \bar{\l}_{I \dot\b} (\w - \w') \delta \bar{z}^{\dot\b \a} (\w') \right] \,,
\end{aligned}
\end{align}
where we have suppressed the $\w$ dependence whenever it is obvious.

The free action in terms of the fluctuation fields becomes
\begin{align}
\begin{aligned}
    S_{\text{free}} &= \int \left[ \l_\a{}^I d \bar{\m}_I{}^\a + \bar{\l}_{I\dot \a} d\m^{\dot\a I} + \frac{\det(\l) + \det(\bar{\l}) + m (\ell^{-1})^\a{}_I \l_\a{}^I (\bar{\ell}^{-1})^{\dot\a J} \bar{\l}_{J \dot\a}}{2} d\s \right.
    \\ &\phantom{=} \left. \phantom{\frac{asdf}{2}} + \frac{\det(\l) (\bar{\ell}^{-1})^{\dot\a I} \bar{\l}_{I \dot\a} + (\ell^{-1})_I{}^\a \l_\a{}^I \det(\bar{\l})}{2} d \s + \frac{\det(\l) \det(\bar{\l})}{2m} d \s + \cdots \right] \,.
\end{aligned} \label{eq:free_action_exp_full}
\end{align}
where the ellipsis denotes constant and total derivative terms irrelevant for Feynman rules. The first line determines the 2pt functions, while the second line generates cubic and quartic vertices.

\subsubsection{Twistor propagators}

The quadratic part of the free action in frequency space can be written as
\begin{align}
\begin{aligned}
    i S_{2} &= \frac{i}{2} \int_{\w,\w'} \deltabar(\w' + \w) \begin{pmatrix}
        \l_\a {}^I (\w') & \bar{\m}_I{}^\a (\w') & \bar{\l}_{I \dot\a} (\w') & {\m}^{\dot\a I} (\w')
    \end{pmatrix}
    \\ &\phantom{=a} \times
    \begin{pmatrix}
        - \frac{1}{2} \e^{\a\b} \e_{IJ} & - i \w \delta_I^J \delta^\a_\b & \frac{m}{2} (\ell^{-1})_I{}^\a (\bar{\ell}^{-1})^{\dot\b J} & 0 \\ 
        + i \w \delta^I_J \delta_\a^\b & 0 & 0 & 0 \\
        \frac{m}{2} (\bar{\ell}^{-1})^{\dot\a I} (\ell^{-1})_J{}^\b & 0 & + \frac{1}{2} \e^{IJ} \e^{\dot\a \dot\b} & - i \w \delta^I_J \delta^{\dot\a}_{\dot\b} \\ 
        0 & 0 & + i \w \delta_I^J \delta_{\dot\a}^{\dot\b} & 0
    \end{pmatrix}
    \begin{pmatrix}
        \l_\b{}^J (\w) \\
        \bar{\m}_J{}^\b (\w) \\
        \bar{\l}_{J \dot\b} (\w) \\
        {\m}^{\dot\b J} (\w)
    \end{pmatrix} \,,
\end{aligned}
\label{quadratic-action}
\end{align}
where we used the delta support to convert $\w' \to - \w$. Inverting the quadratic action leads to the twistor propagators 
in the straight line background:
\begin{align}
\begin{aligned}
    \langle \l_\a^{~I} (\w') \bar{\m}_J^{~\b} (\w) \rangle &= - \frac{\delta_\a^\b \delta_J^I}{\w} \deltabar(\w' + \w) \,, \quad
    \langle \bar{\m}_I^{~\a} (\w') \bar{\m}_J^{~\b} (\w) \rangle = + \frac{ i }{2} \frac{\e_{IJ} \e^{\a\b}}{\w^2} \deltabar (\w' + \w) \,,
    \\ \langle \bar{\l}_{I \dot\a} (\w') {\m}^{\dot\b J} (\w) \rangle &= - \frac{\delta_{\dot\a}^{\dot\b} \delta_I^J}{\w} \deltabar (\w' + \w) \,, \quad 
    \langle {\m}^{\dot\a I} (\w') {\m}^{\dot\b J} (\w) \rangle = - \frac{i }{2} \frac{\e^{IJ} \e^{\dot\a \dot\b}}{\w^2} \deltabar (\w' + \w) \,,
    \\ \langle {\m}^{\dot\a I} (\w') \bar{\m}_J^{~\b} (\w) \rangle &= - \frac{i m}{2} \frac{(\bar{\ell}^{-1})^{\dot\a I} (\ell^{-1})^\b{}_J}{\w^2} \deltabar (\w' + \w) \,.
\end{aligned}
\label{propagator-all}
\end{align}
An $i 0^+$ prescription is needed to determine the causality flow of the 2pt functions; e.g. for $\w \to \w + i 0^+$ causality flows from $\w'$ to $\w$. Otherwise stated, we use time-symmetric $i0^+$ prescription in the calculations, which is the prescription relevant for computing the eikonal~\cite{Jakobsen:2021zvh}. We also remark that all position type 2pt functions (i.e. $\langle \m \m \rangle$, $\langle \bar{\m} \bar{\m} \rangle$, and $\langle \m \bar{\m} \rangle$) should be understood as
\begin{align}
\begin{aligned}
    \langle \bar{\m}_I^{~\a} (\w') \bar{\m}_J^{~\b} (\w) \rangle &= + \frac{ i \k^0 \det(\bar\ell)}{2} \frac{\e_{IJ} \e^{\a\b}}{\w^2} \deltabar (\w' + \w) \,, \\ \langle {\m}^{\dot\a I} (\w') {\m}^{\dot\b J} (\w) \rangle &= - \frac{i \k^0 \det(\ell)}{2} \frac{\e^{IJ} \e^{\dot\a \dot\b}}{\w^2} \deltabar (\w' + \w) \,,
    \\ \langle {\m}^{\dot\a I} (\w') \bar{\m}_J^{~\b} (\w) \rangle &= - \frac{i \k^0 \det(\ell) \det(\bar{\ell})}{2} \frac{(\bar{\ell}^{-1})^{\dot\a I} (\ell^{-1})^\b{}_J}{\w^2} \deltabar (\w' + \w) \,,
\end{aligned} \label{eq:pos_2pt_hidden_factor}
\end{align}
when we remove the gauge-fixing condition $\k^0 = \frac{1}{m}$ and background value determinant conditions $\det(\ell) = \det(\bar{\ell}) = m$. 
This will become relevant in discussions of causality cuts applied to the twistor model~\cite{Kim:causality-cut}.

Let us introduce a graphical notation for the propagators \eqref{propagator-all}. 
Without the mass-shell constraint, the worldline propagator would be simply
\begin{equation}
\begin{fmffile}{worldline-propagator-unconstrained}
    \parbox{80pt}{
    \begin{fmfgraph*}(50,50)
        \fmfstraight
        \fmfbottom{i1,o1}
        \fmftop{i2,o2}
        \fmfv{decor.shape=circle,decor.filled=empty,decor.size=10,label=$Z_A{}^I\; $,label.angle=180}{v2}
        \fmfv{decor.shape=circle,decor.filled=shaded,decor.size=10,label=$\bar{Z}_J{}^B\; $,label.angle=180}{v1} 
        \fmf{phantom}{i1,v1,o1}
        \fmf{phantom}{i2,v2,o2}
        \fmf{plain,tension=0.4}{v1,v2} 
    \end{fmfgraph*}}
    \hskip -1.5cm = \;\; 
    - \frac{\delta_A{}^B \delta_J{}^I}{\omega}
    \,,
\end{fmffile}
\label{twistor-propagator-1}
\end{equation}
which includes $\langle \l \bar{\m} \rangle$ and $\langle \bar{\l} \m \rangle$ propagators. 
The mass-shell constraint introduces additional propagators. We denote the  $\langle \m\m\rangle$, $\langle \bar{\m}\bar{\m} \rangle$ 
propagators as 
\begin{equation}
\begin{fmffile}{worldline-propagator-constrained}
    \parbox{80pt}{
     \begin{fmfgraph*}(50,50)
        \fmfstraight
        \fmfbottom{i1,o1}
        \fmftop{i2,o2}
        \fmfv{decor.shape=circle,decor.filled=empty,decor.size=10,label=$\mu^{\da I}\; $,label.angle=180}{a2}
        \fmfv{decor.shape=circle,decor.filled=empty,decor.size=10,label=$\mu^{\db J}\; $,label.angle=180}{a1}
        \fmfv{decor.shape=circle,decor.filled=full,decor.size=5}{t0}
         \fmf{phantom}{i1,a1,o1}
        \fmf{phantom}{i2,a2,o2}
        \fmf{plain,tension=0.4}{a1,t0,a2}
    \end{fmfgraph*}}
    \hskip -1.5cm = \;\; - \frac{i}{2} \frac{\e^{IJ} \e^{\da\db}}{\w^2} 
    \,, 
    \hskip 1cm  
     \parbox{80pt}{
     \begin{fmfgraph*}(50,50)
        \fmfstraight
        \fmfbottom{i1,o1}
        \fmftop{i2,o2}
        \fmfv{decor.shape=circle,decor.filled=shaded,decor.size=10,label=$\bar{\mu}_I{}^\a\; $,label.angle=180}{a2}
        \fmfv{decor.shape=circle,decor.filled=shaded,decor.size=10,label=$\bar{\mu}_J{}^\b\; $,label.angle=180}{a1}
        \fmfv{decor.shape=circle,decor.filled=full,decor.size=5}{t0}
         \fmf{phantom}{i1,a1,o1}
        \fmf{phantom}{i2,a2,o2}
        \fmf{plain,tension=0.4}{a1,t0,a2}
    \end{fmfgraph*}}
    \hskip -1.5cm = \;\;  +\frac{i}{2}  \frac{\e_{IJ} \e^{\a\b}}{\w^2} 
    \,.
\end{fmffile}
\label{twistor-propagator-2}
\end{equation}
The black dots in the middle remind us of the fact that these propagators originate from the $(\bar{\l}\bar{\l})$
and $(\l\l)$ vertices in the Lagrangian.\footnote{This is a valid interpretation of the 2pt functions; we may only regard \eqref{twistor-propagator-1} as the fundamental 2pt functions and understand the position type 2pt functions, \eqref{twistor-propagator-2} and \eqref{twistor-propagator-3}, as insertions of 2pt vertices between products of fundamental 2pt functions. See appendix \ref{app:GFprod} for regularisation of the divergences related to the symmetric $i0^+$ prescription of the 2pt functions.} 
Finally, we denote the $\langle \m \bar{\m} \rangle$ propagator by 
\begin{equation}
\begin{fmffile}{worldline-propagator-last}
    \parbox{80pt}{
     \begin{fmfgraph*}(50,50)
        \fmfstraight
        \fmfbottom{i1,o1}
        \fmftop{i2,o2}
        \fmfv{decor.shape=circle,decor.filled=empty,decor.size=10,label=$\mu^{\da I}\; $,label.angle=180}{a2}
        \fmfv{decor.shape=circle,decor.filled=shaded,decor.size=10,label=$\bar{\mu}_J{}^\b\; $,label.angle=180}{a1}
        \fmfv{decor.shape=square,decor.filled=full,decor.size=5}{t0}
         \fmf{phantom}{i1,a1,o1}
        \fmf{phantom}{i2,a2,o2}
        \fmf{plain,tension=0.4}{a1,t0,a2}
    \end{fmfgraph*}}
    \hskip -1.5cm = \;\; - \frac{im}{2} \frac{(\bar{\ell}^{-1})^{\dot\a I} (\ell^{-1})^\b{}_J}{\w^2} \,.
\end{fmffile}
\label{twistor-propagator-3}
\end{equation}
The black square in the middle is to show that this propagator comes from the 
$m(\ell^{-1})(\bar{\ell}^{-1})$ vertex in the Lagrangian.

\subsubsection{Vector 2-point functions}

Turning to the ``spacetime picture" where we organize diagrams 
in terms of $\d z$, $\d \bz$, 
the following 2pt functions will play a crucial role. 
\begin{align}
\begin{aligned}
    \langle \delta z^{\dot\a \a} (\w') \delta z^{\dot\b \b} (\w) \rangle &= - \frac{2 i \e^{\dot\a \dot\b} \e^{\a\b}}{m \w^2} \deltabar (\w' + \w) \,,
   \\ 
   \langle \delta \bar{z}^{\dot\a \a} (\w') \delta \bar{z}^{\dot\b \b} (\w) \rangle &= - \frac{2 i \e^{\dot\a \dot\b} \e^{\a\b}}{m \w^2} \deltabar (\w' + \w) \,,
    \\ 
    \langle \delta \bar{z}^{\dot\a \a} (\w') \delta z^{\dot\b \b} (\w) \rangle &= - \frac{2}{m} \left[ \left( \frac{i v^{\dot\a \a} v^{\dot\b \b}}{\w^2} + \frac{ v^{\dot\a \b} z_0^{\dot\b \a}}{\w'} + \frac{v^{\dot\a \b} \bar{z}_0^{\dot\b \a}}{\w} \right) \deltabar (\w' + \w) \right.
    \\ &\phantom{=} \left. \phantom{\frac{ v^{\dot\a \b} z_0^{\dot\b \a}}{\w'}} - i v^{\dot\a \b} v^{\dot\b \a} \left( \frac{1}{\w'} + \frac{1}{\w} \right) \deltabar' (\w' + \w) \right] \,, 
\end{aligned} \label{eq:zz_2pt_bare}
\end{align}
where $\deltabar'(x) = \frac{d}{dx} \deltabar (x)$ is the delta derivative. Note that terms proportional to $\w^{-2}$ in \eqref{eq:zz_2pt_bare} are contributions from the 2pt functions given in \eqref{eq:pos_2pt_hidden_factor}.
We have neglected the loop contribution to the $\langle \delta \bar{z} \delta z \rangle$ 2pt function,
\begin{align}
    \langle \delta \bar{z}^{\dot\a \a} (\w') \delta z^{\dot\b \b} (\w) \rangle \supset - \frac{4 v^{\dot\a \a} v^{\dot\b \b}}{m^2} \deltabar (\w' + \w) \int_{\w_1} \frac{1}{\w_1 (\w - \w_1)} \to 0 \,,
\end{align}
based on two reasons. First, $\hbar$ counting from dimensional analysis requires an extra $\hbar$ factor for this loop contribution compared to tree contributions given in \eqref{eq:zz_2pt_bare}. Second, the loop integral evaluates to zero if we assume invariance under shifts of the integration variable:\footnote{Although widely used in dimensional regularisation, this is not a trivial assumption; for example, ABJ anomalies evaluate to zero under this assumption for (divergent) loop integrals.}
\begin{align}
    \int_{\w'} \frac{1}{\w'(\w-\w')} = \frac{1}{\w} \int_{\w'} \left[ \frac{1}{\w'} - \frac{1}{\w' - \w} \right] \,.
\end{align}
The delta derivative contribution can be simplified by 
\begin{align}
\begin{aligned}
    \left( \frac{1}{\w'} + \frac{1}{\w} \right) \deltabar' (\w' + \w) &= \frac{\partial}{\partial \w} \left[ \frac{\deltabar(\w' + \w)}{\w'} \right] + \frac{\partial}{\partial \w'} \left[ \frac{\deltabar(\w' + \w)}{\w} \right]
    \\ &= \frac{\partial}{\partial \w} \left[ - \frac{\deltabar(\w' + \w)}{\w} \right] + \frac{\partial}{\partial \w'} \left[ \frac{\deltabar(\w' + \w)}{\w} \right]
    \\ &= \frac{1}{\w^2} \deltabar (\w' + \w) \,,
\end{aligned}
\end{align}
leading to
\begin{align}
\begin{aligned}
    \langle \delta \bar{z}^{\dot\a \a} (\w') \delta z^{\dot\b \b} (\w) \rangle &= - \frac{2}{m} \left[ \frac{i ( v^{\dot\a \a} v^{\dot\b \b} - v^{\dot\a \b} v^{\dot\b \a} )}{\w^2} + \frac{ v^{\dot\a \b} ( \bar{z}_0^{\dot\b \a} - z_0^{\dot\b \a} )}{\w} \right] \deltabar (\w' + \w) \\
    &= - \frac{2}{m} \left[ \frac{i \e^{\dot\a \dot\b} \e^{\a\b}}{\w^2} + \frac{ v^{\dot\a \b} ( \bar{z}_0^{\dot\b \a} - z_0^{\dot\b \a} )}{\w} \right] \deltabar (\w' + \w) \,,
\end{aligned}
\end{align}
so that the delta derivative contribution vanishes. 
We have used $v^2 = -1$ to simplify the second line; $v^{\dot\a \a} v^{\dot\b \b} - v^{\dot\a \b} v^{\dot\b \a} = - v^2 \e^{\dot\a \dot\b} \e^{\a\b} =  \e^{\dot\a \dot\b} \e^{\a\b}$.
Note that the term proportional to $\w^{-1}$ implies propagation of spin degrees of freedom $y^\m \propto z^\m - \bar{z}^\m$.

In the vector notation, the $\delta z$ and $\delta \bar{z}$ 2pt functions take the following form, 
\begin{align}
\begin{aligned}
    \langle \delta z^\m (\w') \delta z^\n (\w) \rangle &= \frac{i \eta^{\m\n}}{m \w^2} \deltabar (\w' + \w) \,,
    \\ 
    \langle \delta \bar{z}^\m (\w') \delta \bar{z}^\n (\w) \rangle &= \frac{i \eta^{\m\n}}{m \w^2} \deltabar (\w' + \w) \,,
    \\ \langle \delta \bar{z}^\m (\w') \delta z^\n (\w) \rangle &= \frac{i}{m} \left[ \frac{\eta^{\m\n}}{\w^2} + \frac{2( v^\m y_0^\n + y_0^\m v^\n  
    + i \ve^{\m\n\l\s} v_\l y_{0\s})}{\w } \right] \deltabar (\w' + \w) \,. 
\end{aligned} \label{eq:zz2pts}
\end{align}
It is useful to present the 2-point functions pictorially. 
To distinguish them from the twistor propagators \eqref{twistor-propagator-1}-\eqref{twistor-propagator-3}, we denote $\delta z^\mu $, $\delta \bar{z}^\mu$ by squares:
\begin{equation}
\begin{fmffile}{zz-or-zbarzbar-2pt}
    \parbox{80pt}{
     \begin{fmfgraph*}(50,50)
        \fmfstraight
        \fmfbottom{i1,o1}
        \fmftop{i2,o2}
        \fmfv{decor.shape=square,decor.filled=empty,decor.size=10,
            label=$\delta z^\mu \; $,label.angle=180}{a2}
        \fmfv{decor.shape=square,decor.filled=empty,decor.size=10,
            label=$\delta z^\nu\; $,label.angle=180}{a1}
        \fmfv{decor.shape=circle,decor.filled=full,decor.size=0}{t0}
         \fmf{phantom}{i1,a1,o1}
        \fmf{phantom}{i2,a2,o2}
        \fmf{plain,tension=0.4}{a1,t0,a2}
    \end{fmfgraph*}}
    \hskip -1.6cm = \; \langle \delta z^\mu  \delta z^\nu \rangle 
    \,, 
    \hskip 1cm  
     \parbox{80pt}{
     \begin{fmfgraph*}(50,50)
        \fmfstraight
        \fmfbottom{i1,o1}
        \fmftop{i2,o2}
        \fmfv{decor.shape=square,decor.filled=shaded,decor.size=10,
                label=$\delta \bar{z}^\mu\; $,label.angle=180}{a2}
        \fmfv{decor.shape=square,decor.filled=shaded,decor.size=10,
                label=$\delta \bar{z}^\nu\; $,label.angle=180}{a1}
        \fmfv{decor.shape=circle,decor.filled=full,decor.size=0}{t0}
         \fmf{phantom}{i1,a1,o1}
        \fmf{phantom}{i2,a2,o2}
        \fmf{plain,tension=0.4}{a1,t0,a2}
    \end{fmfgraph*}}
    \hskip -1.6cm = \;  \langle \delta \bar{z}^\mu  \delta \bar{z}^\nu \rangle 
    \,,
     \hskip 1cm  
     \parbox{80pt}{
     \begin{fmfgraph*}(50,50)
        \fmfstraight
        \fmfbottom{i1,o1}
        \fmftop{i2,o2}
        \fmfv{decor.shape=square,decor.filled=shaded,decor.size=10,
                label=$\delta \bar{z}^\mu\; $,label.angle=180}{a2}
        \fmfv{decor.shape=square,decor.filled=empty,decor.size=10,
                label=$\delta \bar{z}^\nu\; $,label.angle=180}{a1}
        \fmfv{decor.shape=circle,decor.filled=full,decor.size=0}{t0}
         \fmf{phantom}{i1,a1,o1}
        \fmf{phantom}{i2,a2,o2}
        \fmf{plain,tension=0.4}{a1,t0,a2}
    \end{fmfgraph*}}
    \hskip -1.6cm = \;  \langle \delta \bar{z}^\mu  \delta z^\nu \rangle 
    \,.
\end{fmffile}
\end{equation}
Finally, separating the position $x$ and the spin-length $y$, 
we get 
\begin{align}
    \begin{aligned}
    \langle \delta x^\m (\w') \delta x^\n (\w) \rangle &= \left[ \frac{i \eta^{\m\n}}{m \w^2} - \frac{1}{m\omega} \ve^{\m\n\l\s} v_\l y_{0\s}\right]\deltabar (\w' + \w) \,,
    \\ 
    \langle \delta y^\m (\w') \delta y^\n (\w) \rangle &= - \frac{1}{m\omega} \ve^{\m\n\l\s} v_\l y_{0\s} \deltabar (\w' + \w) \,,
    \\ 
    \langle \delta x^\m (\w') \delta y^\n (\w) \rangle 
    &= \frac{1}{m\w } \left( v^\m y_0^\n + y_0^\m v^\n  \right) \deltabar (\w' + \w) = - \langle \delta y^\m (\w') \delta x^\n (\w) \rangle \,.
\end{aligned} \label{eq:xxyy2pts}
\end{align}

\subsubsection{Higher order correlators}

For computations at 2PL or higher orders, we will need to evaluate the higher order correlators. We write the 2pt correlators as
\begin{align}
\begin{aligned}
    \langle \delta z^{\dot\a \a} (\w') \delta z^{\dot\b \b} (\w) \rangle &= - \frac{2 i \e^{\dot\a \dot\b} \e^{\a\b}}{m \w^2} \deltabar (\w' + \w) \,,
    \\ 
    \langle \delta \bar{z}^{\dot\a \a} (\w') \delta \bar{z}^{\dot\b \b} (\w) \rangle &= - \frac{2 i \e^{\dot\a \dot\b} \e^{\a\b}}{m \w^2} \deltabar (\w' + \w) \,,
    \\ 
    \langle \delta \bar{z}^{\dot\a \a} (\w') \delta z^{\dot\b \b} (\w) \rangle &= - \frac{2 i}{m} \left( \frac{\e^{\dot\a \dot\b} \e^{\a\b}}{\w^2} - \frac{2 v^{\dot\a \b} y_0^{\dot\b \a}}{\w} \right) \deltabar (\w' + \w) \,.
\end{aligned} \label{eq:zz_2pt_ref}
\end{align}
The connected part of the higher point correlators can be computed using the recursive substitutions
\begin{align}
\begin{aligned}
    \delta z^{\dot\a \a} (\w) &\to \int_{\w'} \delta z^{\dot\a \b} (\w') \times \left[ - \l_\b{}^I (\w - \w') (\ell^{-1})_I{}^\a \right] \,,
    \\ \delta \bar{z}^{\dot\a \a} (\w) &\to \int_{\w'} \left[ - (\bar{\ell}^{-1})^{\dot\a I} \bar{\l}_{I \dot\b} (\w - \w') \right] \times \delta \bar{z}^{\dot\b \a} (\w') \,,
\end{aligned} \label{eq:dz_exp}
\end{align}
the 2pt correlators
\begin{align}
\begin{aligned}
    - (\ell^{-1})_I{}^\a \langle \l_\b{}^I (\w') \delta \bar{z}^{\dot\g \g} (\w) \rangle &= - \frac{2 v^{\dot\g \a} \delta_\b^\g}{m \w} \deltabar (\w' + \w) \,,
    \\ - (\bar{\ell}^{-1})^{\dot\a I} \langle \bar{\l}_{I \dot\b} (\w') \delta {z}^{\dot\g \g} (\w) \rangle &= - \frac{2 v^{\dot\a \g} \delta_{\dot\b}^{\dot\g}}{m \w} \deltabar (\w' + \w) \,,
\end{aligned}
\end{align}
additional vertices from the free action
\begin{align}
\begin{aligned}
    i S_{\text{free},3} &= \frac{i}{2} \int d\s \left[ \det (\l) (\bar{\ell}^{-1})^{\dot\a I} \bar{\l}_{I \dot\a} + (\ell^{-1})_I{}^\a \l_\a{}^I \det (\bar{\l}) \right]
    \\ &= \frac{i}{2} \int_{\w', \w} \left( (\bar{\ell}^{-1})^{\dot\a I} \bar{\l}_{I \dot\a} (\w') \det (\l) [\w] + (\ell^{-1})_I{}^\a \l_\a{}^I (\w') \det (\bar{\l}) [\w] \right) \deltabar(\w' + \w) \,,
    \\ i S_{\text{free},4} &= \frac{i}{2m} \int d\s \left[ \det (\l) \det (\bar{\l}) \right] = \frac{i}{2m} \int_{\w' , \w} \det (\l) [\w'] \det (\bar{\l}) [\w] \deltabar(\w' + \w) \,,
\end{aligned}
\end{align}
where
\begin{align}
\begin{aligned}
    \det (\l) [\w] &= - \frac{\e^{\a\b} \e_{IJ}}{2} \int_{\w'} \l_\a{}^I (\w - \w') \l_\b{}^J (\w') \,,
    \\ \det (\bar{\l}) [\w] &= \frac{\e^{\dot\a \dot\b} \e^{IJ}}{2} \int_{\w'} \bar{\l}_{I \dot\a} (\w - \w') \bar{\l}_{J \dot\b} (\w') \,,
\end{aligned}
\end{align}
and determinant insertions to the 2pt correlators
\begin{align}
\begin{aligned}
    \langle \! \langle \delta {z}^{\dot\a \a} (\w_0) \delta {z}^{\dot\b \b} (\w_1) \det (\bar{\l}) [\w_2] \rangle \! \rangle &= \frac{4 \e^{\dot\a \dot\b} \e^{\a\b}}{m \w_0 \w_1} \deltabar (\w_0 + \w_1 + \w_2) \,,
    \\ \langle \! \langle \delta \bar{z}^{\dot\a \a} (\w_0) \delta \bar{z}^{\dot\b \b} (\w_1) \det ({\l}) [\w_2] \rangle \! \rangle &= \frac{4 \e^{\dot\a \dot\b} \e^{\a\b}}{m \w_0 \w_1} \deltabar (\w_0 + \w_1 + \w_2) \,.
\end{aligned} \label{eq:2pt_det_ins}
\end{align}
For example, the following 3pt correlator can be computed as
\begin{align*}
\begin{aligned}
    \langle \! \langle \delta \bar{z}^{\dot\a \a} (\w_0) \delta {z}^{\dot\b \b} (\w_1) \delta {z}^{\dot\g \g} (\w_2) \rangle \! \rangle &=  - { (\bar{\ell}^{-1})^{\dot\a I} } \int_{\w'} \langle \bar{\l}_{I \dot\delta} (\w_0 - \w') \delta {z}^{\dot\b \b} (\w_1) \rangle \langle \delta \bar{z}^{\dot\delta \a} (\w') \delta {z}^{\dot\g \g} (\w_2) \rangle
    \\ &\phantom{=} - {(\ell^{-1})_I{}^\b} \int_{\w'} \langle \l_\delta{}^I (\w_1 - \w') \delta \bar{z}^{\dot\a \a} (\w_0) \rangle \langle \delta {z}^{\dot\b \delta} (\w') \delta {z}^{\dot\g \g} (\w_2) \rangle
    \\ &\phantom{=asdf} + \left( \w_1 \leftrightarrow \w_2 \,,\, \b \leftrightarrow \g \,,\, \dot\b \leftrightarrow \dot\g \right)
    \\ &\phantom{=} + \frac{i}{2} \int_{\w'} {(\ell^{-1})_I{}^\delta} \langle \l_\delta{}^I (-\w') \delta \bar{z}^{\dot\a \a} (\w_0) \rangle
    \\ &\phantom{=asdfasdfasdf} \times \langle \! \langle \delta {z}^{\dot\b \b} (\w_1) \delta {z}^{\dot\g \g} (\w_2) \det (\bar{\l}) [\w'] \rangle \! \rangle \,,
\end{aligned}
\end{align*}
where the first three lines come from the expansion \eqref{eq:dz_exp} and the last line comes from the insertion \eqref{eq:2pt_det_ins}. The result partially simplifies to %
\begin{align}
\begin{aligned}
    \langle \! \langle \delta \bar{z}^{\dot\a \a} (\w_0) \delta {z}^{\dot\b \b} (\w_1) \delta {z}^{\dot\g \g} (\w_2) \rangle \! \rangle &= - \frac{4 i v^{\dot\a \b}}{m^2} \left( \frac{\e^{\dot\b \dot\g} \e^{\a \g}}{\w_0 \w_1 \w_2} + \frac{2 v^{\dot\b \g} y_0^{\dot\g \a}}{\w_1 \w_2} \right) \deltabar (\w_0 + \w_1 + \w_2)
    \\ &\phantom{=asdf} + \left( \w_1 \leftrightarrow \w_2 \,,\, \b \leftrightarrow \g \,,\, \dot\b \leftrightarrow \dot\g \right)
    \\ &\phantom{=} + \frac{4 i v^{\dot\a \a} \e^{\dot\b \dot\g} \e^{\b\g}}{m^2 \w_0 \w_1 \w_2} \deltabar (\w_0 + \w_1 + \w_2) \,,
\end{aligned} \label{eq:zb-z-z_3pt}
\end{align}
where the frequency exchange is only present for manifest symmetry. In vectorial notation the last term coming from determinant insertion cancels and simplifies to
\begin{align}
\begin{aligned}
    \langle \! \langle \delta \bar{z}^{\m} (\w_0) \delta {z}^{\n} (\w_1) \delta {z}^{\l} (\w_2) \rangle \! \rangle &= - \frac{4 i}{m^2 \w_1 \w_2} \left[ y_0^\m (\eta^{\n\l} + 2 v^\n v^\l ) + v^\m (v^\n y_0^\l + y_0^\n v^\l) \right. 
    \\ &\phantom{=} \left. \phantom{as} + i \e^{\m\n} [v, y_0] v^\l + i \e^{\m\l} [v, y_0] v^\n \right] \deltabar (\w_0 + \w_1 + \w_2) \,.
\end{aligned} \label{eq:zb-z-z_3pt_vec}
\end{align}
A similar calculation for the conjugate 3pt correlator yields
\begin{align}
\begin{aligned}
    \langle \! \langle \delta \bar{z}^{\dot\a \a} (\w_0) \delta \bar{z}^{\dot\b \b} (\w_1) \delta {z}^{\dot\g \g} (\w_2) \rangle \! \rangle &= \frac{4 i v^{\dot\a \g}}{m^2} \left( \frac{\e^{\dot\b \dot\g} \e^{\a \b}}{\w_0 \w_1 \w_2} + \frac{2 v^{\dot\b \a} y_0^{\dot\g \b}}{\w_0 \w_1} \right) \deltabar (\w_0 + \w_1 + \w_2)
    \\ &\phantom{=asdf} + \left( \w_0 \leftrightarrow \w_1 \,,\, \a \leftrightarrow \b \,,\, \dot\a \leftrightarrow \dot\b \right)
    \\ &\phantom{=} + \frac{4 i v^{\dot\g \g} \e^{\dot\a \dot\b} \e^{\a\b}}{m^2 \w_0 \w_1 \w_2} \deltabar (\w_0 + \w_1 + \w_2) \,,
\end{aligned} \label{eq:zb-zb-z_3pt}
\end{align}
which, in the vectorial notation, simplifies to
\begin{align}
\begin{aligned}
    \langle \! \langle \delta \bar{z}^{\m} (\w_0) \delta \bar{z}^{\n} (\w_1) \delta {z}^{\l} (\w_2) \rangle \! \rangle &= \frac{4 i}{m^2 \w_0 \w_1} \left[ y_0^\l (\eta^{\m\n} + 2 v^\m v^\n ) + v^\l (v^\m y_0^\n + y_0^\m v^\n) \right. 
    \\ &\phantom{=} \left. \phantom{as} - i \e^{\l\m} [v, y_0] v^\n - i \e^{\l\n} [v, y_0] v^\m \right] \deltabar (\w_0 + \w_1 + \w_2) \,.
\end{aligned} \label{eq:zb-zb-z_3pt_vec}
\end{align}
The following correlator may be of interest,
\begin{align}
    \langle \! \langle \delta y^{\m} (\w_0) \delta y^{\n} (\w_1) \delta y^{\l} (\w_2) \rangle \! \rangle &= - \frac{1}{m^2} \left[ \frac{y^\m (\eta^{\n\l} + v^\n v^\l)}{\w_1 \w_2} + (\text{cyc.}) \right] \deltabar (\w_0 + \w_1 + \w_2) \,,
\end{align}
where $(\text{cyc.})$ denotes cyclic permutation.

Note that $\delta z$ and $\delta \bar{z}$ variables do not obey Wick factorisation, e.g. $\langle (\delta z) (\delta \bar{z}) (\delta \bar{z}) \rangle \neq 0$. However, purely holomorphic/anti-holomorphic correlators such as $\langle (\delta z) (\delta z) \cdots (\delta z) \rangle$ do obey Wick factorisation, since $\delta z$($\delta \bar{z}$) is at most linear in $\m$($\bar{\m}$) and the correlators reduce to the correlators of the form $\langle \m \m \cdots \m \rangle$ or $\langle \bar{\m} \bar{\m} \cdots \bar{\m} \rangle$.

\subsubsection{Photon coupling vertex rules}

We begin with the spacetime picture where the vertex rules take a simple form that are easy to compare with other worldline models. 
Inserting the mode expansion of fluctuations into the interaction \eqref{eq:int_NJ_shift}, we get\footnote{Empty sum is zero and empty product is unity, i.e. $\sum_{i=1}^0 \# = 0$ and $\prod_{i=1}^0 \# = 1$.}
\begin{align} \label{eq:vertex_all}
    & q \int_{k, \{\w\} } A_\m^+ (k) ( v^\m) e^{i k \cdot z_0} \sum_{n=0}^\infty \frac{1}{n!} \deltabar \left( (k \cdot v) - \sum_{i=1}^n \w_i \right) \prod_{i=1}^n (i k) \cdot \delta z (\w_i)
    \\ & + q \int_{k, \{\w\} } A_\m^+ (k) (- i \w_{0} ) \delta z^\m (\w_{0}) e^{i k \cdot z_0} \sum_{n=0}^\infty \frac{1}{n!} \deltabar \left( (k \cdot v) - \sum_{i=0}^n \w_i \right) \prod_{i=1}^n (i k) \cdot \delta z (\w_i) \nn
    \\ &\phantom{asdf} + \left( A_\m^+(k) \to A_\m^-(k) \,,\, z_0^\m \to \bar{z}_0^\m \,,\, \delta z (\w_i) \to \delta \bar{z} (\w_i)  \right) \,. \nn
\end{align}
This expression is exact in $z_0$, $\bar{z}_0$. To obtain a result at a fixed order in the background spin-length $y_0$, we may simply set $z_0^\m = b_0^\m + i y_0^\m$ and $\bar{z}_0^\m = b_0^\m - i y_0^\m$ and expand in $y_0$. 

To compute Compton amplitudes or the 2PL eikonal, we only need terms up to linear order in fluctuations, 
\begin{align}
\begin{aligned}
    S_{\text{int}} &= q \int_k A_\m^+ (k) v^\m e^{i k \cdot z_0} \deltabar (k \cdot v)
    \\ &\phantom{=} + i q \int_{k,\w} A_\m^+ (k) \left[ v^\m k_\n - \w \delta^\m_\n \right] \delta z^\n (\w) e^{i k \cdot z_0} \deltabar [(k \cdot v) - \w]
    \\ &\phantom{asdf} + \left[ A_\m^+(k) \to A_\m^-(k) \,,\, z_0^\m \to \bar{z}_0^\m \,,\, \delta z (\w_i) \to \delta \bar{z} (\w_i) \right] \,.
\end{aligned}
\label{eq:vertex-linear-z}
\end{align}
Aside from the leading term proportional to $A^\pm \cdot v$, we can write the Feynman rules in terms of gauge-invariant field-strengths $F^\pm$, since the vertex rules with at least one $\delta z^\m (\s)$ fluctuation field\footnote{We argue using the time domain Feynman rules because the proof is simpler.} can be read out from the variational derivative
\begin{align}
    \frac{\delta S_{\text{int}}}{\delta[\delta z^\m (\s)]} &=
    q F_{\m\n}^+ [z_0 + v\s + \delta z(\s)] \left( v^\n + \frac{d(\delta z^\n (\s))}{d\s} \right) \,,
\end{align}
which only depends on the field-strength $F^+$. The same argument trivially generalises to the anti-holomorphic sector. As a demonstration, we write the interaction terms up to quadratic order in worldline perturbations as
\begin{align}
\begin{aligned}
    S_{\text{int}} &= q \int_k (A_k^+ \cdot v) e^{i k \cdot z_0} \deltabar(k \cdot v) + q \int_{k,\w} (\delta z_\w \cdot F_k^+ \cdot v) e^{i k \cdot z_0} \deltabar[(k \cdot v) - \w]
    \\ &\phantom{=} + \frac{q}{4} \int_{k, \w_1, \w_2} \hskip -15pt \left\{ \left[ (\delta z_1 \cdot F^+_k \cdot v)(i k \cdot \delta z_2) + (1 \leftrightarrow 2) \right] + i(\w_1 - \w_2) (\delta z_1 \cdot F^+_k \cdot \delta z_2) \right\}
    \\ &\phantom{=asdfasdfasdfasdf} \times e^{i k \cdot z_0} \deltabar[(k \cdot v) - \w_1 - \w_2 ]
    \\ &\phantom{asdf} + \left( A_\m^+(k) \to A_\m^-(k) \,,\, F_{\m\n}^+(k) \to F_{\m\n}^-(k) \,,\, z_0^\m \to \bar{z}_0^\m \,,\, \delta z (\w_i) \to \delta \bar{z} (\w_i) \right) \,.
\end{aligned} \label{eq:vertex_gi}
\end{align}
The expansion \eqref{eq:vertex_gi} is more useful than the expansion \eqref{eq:vertex_all} since photon propagators can be chosen to be free of Dirac string singularities.
See section~\ref{sec:photon-propagator} for more discussions on the photon propagator.

 
\subsection{Photon propagator} \label{sec:photon-propagator}

A worldline model of a charged particle 
provides a localised source for the electromagnetic field. 
Away from the sources, the photon propagates freely and the photon propagator is independent of the worldline model. 
However, since the NJ shift forces us to separate the self-dual and anti-self-dual parts of the photon field, 
we find it useful to recall some facts regarding how to split the propagator according to self-duality, which translates to the helicity of the photon at the quantum level. 

In our twistor model, the photon field couples to the particle worldline via the NJ shift \eqref{eq:int_NJ_shift} which we copy here:
\begin{align}
    S_{\text{int}} = q \int A^+_\m (z) dz^\m + q \int A^-_\m (\bar{z}) d\bar{z}^\m \,. 
    \label{eq:int_NJ_shift-copy}
\end{align}
This coupling may look unfamiliar to the readers. To gain some intuition, 
let us expand it to the quadratic order in $y$. The zeroth order term reproduces the standard minimal coupling for a non-spinning particle. 
The linear order term is 
\begin{align}
    S_\mathrm{int}^{(1)} = q \int \left[ \tilde{A}_\mu  \dot{y}^\mu + y^\m (\partial_\m \tilde{A}_\n)\dot{x}^\n \right] d\sigma \,. 
\end{align}
The appearance of $\dot{y}$ is a notable feature of the root-Kerr coupling. To linear order, we can remove it by integration by parts. 
Up to a total derivative, we find 
\begin{align}
  S_\mathrm{int}^{(1)} = q \int \left[ \tilde{F}_{\mu\nu} y^\mu \dot{x}^\nu \right] d\sigma  = 
  \frac{q}{2} \int \left[ \ve_{\mu\nu\r\s} y^\mu \dot{x}^\nu  F^{\r\s} \right] d\sigma\,. 
  \label{expanded-action-1}
  \end{align}
Starting from the quadratic order, it is impossible to remove all $\dot{y}$ factors. 
Up to a total derivative, we find 
\begin{align}
  S_\mathrm{int}^{(2)} = \frac{q}{2} \int \left[ \dot{y}^\m y^\n F_{\m\n} - (\partial_\m F_{\n\r}) y^\m y^\n \dot{x}^\r \right] d\sigma \,.
  \label{expanded-action-2}
\end{align}
We can continue this expansion and express all $S^{(n)}_\mathrm{int}$ $(n\ge 1)$ as Lorentz invariant products of $y^\m$, $\dot{y}^\m$, $\dot{x}^\m$, $F_{\m\n}$ and $\ve_{\m\n\r\s}$, with no reference to $A^\pm_\m$ at all. 
With this form of the action, the usual propagator for the photon field will suffice for all perturbative computations.

The beauty of the NJ shift \eqref{eq:int_NJ_shift-copy} is that we can perform computations exactly in $y$ without ever expanding in powers of $y$. 
A small price to pay is that we should use less familiar propagators written in terms of $A^\pm_\m$.

Our discussion is inspired by Zwanziger's (electromagnetic-duality covariant) two-potential formalism \cite{Zwanziger:1970hk} (see also refs.~\cite{Gubarev:1998ss,Shnir:2011zz,Terning:2020dzg,Moynihan:2020gxj}). 
But, we will not directly follow Zwanziger's formalism in that we never use two potentials or consider sources with net magnetic charges. 
We are interested in the long-distance interaction between two spatially localised sources. 
The interaction is captured by the integral, 
\begin{align}
   I_{12} =  \int  J_1^\m(x) \langle A_\m(x) A_\n(y) \rangle J_2^\n(y) \,. 
   \label{charge-charge-interaction}
\end{align}
We are doing classical physics, but we can use the propagator (Green's function) in a QFT notation, where $\langle A_\m A_\n \rangle$ is the 2-point function, which we take to be time-ordered for concreteness. 

Let us temporarily ignore the net (electric or magnetic) charges and focus on the dipole or higher multipole moments. 
For a magnetic dipole, it is well known that a long-distance observer cannot distinguish an 
Amp\`erian dipole (electric current loop) from a Gilbertian dipole (two opposing magnetic monopole charges). 
A similar story holds for an electric dipole and all higher electric/magnetic multipole moments. 
So, as far as the long-distance interaction is concerned, we can describe the \emph{same} source using \emph{either} an electric current \emph{or} a magnetic current. 

To switch between the two pictures, we recall that 
Maxwell's equations with both electric and magnetic sources read
\begin{align}
    \label{eq:maxwell-copy}
    d^\dagger F = J
    \,,\quad
    d^\dagger (\siF) = \J
    \,,
\quad 
    d^\dagger F
    := (\partial^\n F_{\m\n})\mem dx^\m
    \,.
\end{align}
Electric-magnetic duality states that
this set of equations is invariant under
\begin{align}
    \label{eq:emd-z4-copy}
    \text{EMD}\quad:\quad\,
    F
        \,\,\mapsto\,\,
    \siF
    \,,\quad
    (J,\J)
        \,\,\mapsto\,\,
    (\J,-J)
    \,.
\end{align}
It is natural to use the complex combinations of $F$ and $\siF$ that are eigenstates of $*$, 
\begin{align}
    \label{eq:FJcomplexes-copy}
    F^\pm
    := \minie\Big({
        F \pm i\mem \siF
    }\Big)
    \,,\quad
    J^\pm
    := 
    \minie\Big({
        J \pm i \J
    }\Big)
    \quad 
    \Longrightarrow 
    \quad 
    d^\dagger F^\pm  = J^\pm \,. 
\end{align}

For a given multipole, in the electric picture, we solve 
\begin{align}
        d^\dagger F = J
    \,,\quad
    d^\dagger (\siF) = 0
    \,, 
\end{align}
while in the magnetic picture, we solve 
\begin{align}
        d^\dagger F = 0
    \,,\quad
    d^\dagger (\siF) = \J
    \,.  
\end{align}
The two pictures are related such that $F$ away from the source is exactly the same. 
In other words, for a ``point-like" source, the difference between the two pictures 
is ultra-local (delta function supported). 

Depending on which picture we choose for each of the two sources, 
the integral \eqref{charge-charge-interaction} can take different forms, 
\begin{align}
\begin{split}
       I_{12} &=  \int  J_1(x) \langle A(x) A(y) \rangle J_2(y) = \int  \J_1(x) \langle A^\star(x) A^\star(y) \rangle \J_2(y) 
       \\
       &=  \int  J_1(x) \langle A(x) A^\star(y) \rangle \J_2(y) = \int  \J_1(x) \langle A^\star(x) A(y) \rangle J_2(y)  \,,    
\end{split}
\end{align}
where we suppressed the vector indices to avoid clutter. If we call $\langle A A \rangle$ ``electric-electric" propagator, 
we may call $\langle A^\star A^\star \rangle$ ``magnetic-magnetic", 
$\langle A A^\star \rangle$ ``electric-magnetic", etc. 
In the QFT approach to the propagators, which we will review shortly, 
we split the mode expansion according to the photon's helicity such that\footnote{The notation of this section is related to those in appendix~\ref{sec:conventions} as $\tilde{A} = - A^\star$. The minus sign originates from $\si F = - * F$.} 
\begin{align}
  A = A^+ + A^- \,, \quad A^\star = -i (A^+ - A^-) \,. 
\end{align}
Since $\langle A^+ A^+ \rangle = 0 = \langle A^- A^- \rangle $, it follows that 
\begin{align}
\begin{split}
        &\langle A A \rangle = \langle A^+ A^- \rangle +  \langle A^- A^+ \rangle = \langle A^\star A^\star \rangle \,,
    \\
    &\langle A A^\star \rangle = + i \left[ \langle A^+ A^- \rangle -  \langle A^- A^+ \rangle \right] = - \langle A^\star A \rangle \,. 
\end{split}
\end{align}

So far, our discussion has been general. Now let us focus on the multipole moments of a root-Kerr particle. 
As we saw in \eqref{expanded-action-1} and \eqref{expanded-action-2}, we find electric moments at $\CO(y^{2k})$ 
and magnetic moments at $\CO(y^{2k+1})$. 
This splitting is expected to be a generic feature of any parity-preserving spinning charged particle. 
The NJ shift \eqref{eq:int_NJ_shift-copy} suggests a hybrid approach
which uses the electric picture for the electric multipoles and the magnetic picture of the magnetic multipoles. 
We denote the currents by $J_e$ and $\J_o$ where $e$ and $o$ stand for even and odd, respectively. 

Contributions from different multipole moments simply add up to give 
\begin{align}
\begin{split}
       I_{12} &=  \int  J_{1,e}(x) \langle A(x) A(y) \rangle J_{2,e}(y) + \int  \J_{1,o}(x) \langle A^\star(x) A^\star(y) \rangle \J_{2,o}(y) 
       \\
       &\quad +  \int  J_{1,e}(x) \langle A(x) A^\star(y) \rangle \J_{2,o}(y) + \int  \J_{1,o}(x) \langle A^\star(x) A(y) \rangle J_{2,e}(y)  \,. 
\end{split}
\end{align}
Rewriting it in terms of $\langle A^\pm A^\mp \rangle$, we find 
\begin{align}
       I_{12} =  \int  J_{1,+}(x) \langle A^+(x) A^-(y) \rangle J_{2,-}(y) + \int  J_{1,-}(x) \langle A^-(x) A^+(y) \rangle J_{2,+} \,,
\end{align}
where, for each source, 
\begin{align}
    J_{\pm} = J_{e} \mp i \J_{o}  \,.
\end{align}
Comparing this with \eqref{eq:FJcomplexes-copy}, we note a slightly non-trivial ``metric" in the complex basis, 
\begin{align}
    J_\pm = 2 J^\mp \,.
\end{align}
Applying it to the root-Kerr particle, we have
\begin{align}
\begin{split}
    \label{eq:rkerr-Jpms-copy-2}
    \frac{1}{2} (J_-)^\m(x) = 
    (J^+)^\mu(x) &= 
        \frac{1}{2}
        q \int \dot{\bar{z}}^\mu(\sigma) \delta^4(x-\bar{z}(\sigma)) d \sigma
    \,,\\
        \frac{1}{2} (J_+)^\m(x) = 
    (J^-)^\mu(x) &= 
       \frac{1}{2} 
        q \int \dot{z}^\mu(\sigma) \delta^4(x-z(\sigma)) d \sigma
    \,.
\end{split}
\end{align}

To summarize, we took a long route to explain how the NJ shift \eqref{eq:int_NJ_shift-copy} can be understood in conventional descriptions of multipole moments, only to motivate a less familiar method;  
the most efficient way to compute the interaction between two root-Kerr particles 
is to use the ``helicity propagators" $\langle A^\pm A^\mp \rangle$.

\subsubsection{Helicity propagator} 

Let us present the result first and review the derivation. 
In terms of self-dual and anti-self-dual fields, 
the propagators are~\cite{Weinberg:1965rz} 
\begin{align}
\begin{split}
    \Delta^{+-}_{\m\n}(k) &:=
    \langle A_\mu^+(k) A_\nu^-(-k) \rangle = 
    \frac{i}{k^2 - i0^+} \left[ \frac{2 k_{(\m} n_{\n)} - (k \cdot n) \eta_{\m\n} + i \e_{\m\n\a\b}k^\a n^\b}{2 (k \cdot n)} \right]
    \,,\\
    \Delta^{-+}_{\m\n}(k) 
    &= \Delta^{+-}_{\n\m}(k)
    = [\Delta^{+-}_{\m\n}(k)]^*
    \,, 
    \quad 
    \Delta^{++}_{\m\n}(k) 
    = \Delta^{--}_{\m\n}(k) = 0
    \,,
\end{split}
\label{off-shell-propagator}
\end{align}
where $n_\m$ is an auxiliary reference vector, 
which we call a ``Dirac string".
The spinor notation offers a more compact expression. 
Pictorially, we denote the propagator as 
\begin{equation}
 \label{eq:Delta1}
\begin{fmffile}{photon-propagator}
    \parbox{80pt}{
    \begin{fmfgraph*}(50,30)
        \fmfstraight
        \fmfleft{i1,o1}
        \fmfright{i2,o2}
        \fmf{phantom}{i1,v1,o1}
        \fmf{phantom}{i2,v2,o2}
         \fmffreeze
        \fmf{wiggly,label=$k$,tension=0}{v1,v2} 
        \fmfv{decor.shape=square,decor.filled=empty,decor.size=10,
            label=$A^+_{\wrap{\a\da}} \;$,label.angle=180}{v1}
        \fmfv{decor.shape=square,decor.filled=shaded,decor.size=10,
            label=$\; A^-_\wrap{\b\db}$,label.angle=0}{v2}
    \end{fmfgraph*}}
    \; = \;  \Delta^{+-}_{\wrap{\a\da\b\db}}\hnem(k) \; = \; 
    \frac{
        n_{\wrap{\a\db}}\hhem k_{\wrap{\b\da}}
    }{n\mdot k}
    \frac{i}{k^2- i0^+} 
    \,.
\end{fmffile}
\end{equation}
For later purposes, we also note that
\begin{align}
\begin{split}
        \Delta^{+-}_{\m\n}(k) +   \Delta^{-+}_{\m\n}(k) &= \frac{-i}{k^2- i0^+} \left( \eta_{\m\n} - \frac{2k_{(\m} n_{\n)}}{k\cdot n} \right)\,,
        \\
         i \Delta^{+-}_{\m\n}(k) - i\Delta^{-+}_{\m\n}(k) &= \frac{-i \e_{\m\n\a\b}k^\a n^\b }{(k^2- i0^+)(k \cdot n)}  \,.
\end{split}
\label{photon-propagator-even-odd}
\end{align}

The chiral photon field $A_\m^\pm$ frequently appears in the form of the field strength tensor $F_{\m\n}^\pm = \partial_\m A_\n^\pm - \partial_\n A_\m^\pm$. We consider the combinations
\begin{align}
    \langle F_{\m\n}^\pm (k) A_\l^{\mp} (-k) \rangle v^\l \deltabar (k \cdot v) \,,\quad  
    \langle F_{\m\n}^\pm (k) F_{\l\s}^{\mp} (-k) \rangle \,. \nn
\end{align}
These 2pt functions can be expressed without the auxiliary reference vector $n_\m$, since they can be constructed from the non-chiral photon propagator \eqref{photon-propagator-even-odd} using the (anti-)self-dual projectors
\begin{align}
    (P^{\pm})_{\m\n}^{\a\b} := \frac{\delta_\m^\a \delta_\n^\b - \delta_\m^\b \delta_\n^\a \mp i \e_{\m\n}^{\phantom{\m\n}\a\b}}{4} \,,
    \quad 
    F_{\m\n}^\pm = (P^\pm)_{\m\n}^{\a\b} F_{\a\b} =: (P^\pm \cdot F)_{\m\n} \,.
\end{align}
resulting in 
\begin{align}
\begin{aligned}
    \langle F^\pm_{\m\n} (k) A^\mp_{\a} (-k) \rangle v^\a \deltabar(k \cdot v) &= (P^\pm)_{\m\n}^{\l\s} \langle F_{\l\s} (k) A_{\a} (-k) \rangle \rangle v^\a \deltabar(k \cdot v)
    \\ 
    &= \frac{ k_\m v_\n - v_\m k_\n \mp i \e_{\m\n\a\b} k^\a v^\b }{2 (k^2 - i0^+)} \deltabar(k \cdot v) \,,
\end{aligned} \label{eq:FAprop}
\end{align}
and
\begin{align}
\begin{aligned}
    \langle F^+_{\m\n} (k) F^-_{\a\b} (-k) \rangle &= (P^+)_{\m\n}^{\l\s} (P^-)_{\a\b}^{\g\delta} \langle F_{\l\s} (k) F_{\g\delta} (-k) \rangle
    \\ &= \frac{- i [\eta_{\m\a} k_\n k_\b - \eta_{\n\a} k_\m k_\b - \eta_{\m\b} k_\n k_\a + \eta_{\n\b} k_\m k_\a]}{2 (k^2 - i0^+)}
    \\ &\phantom{=as} + \frac{ - \left( k_\m \e_{\n\a\b\l} - k_\n \e_{\m\a\b\l} \right) k^\l + \left( k_\a \e_{\b\m\n\l} - k_\b \e_{\a\m\n\l} \right) k^\l}{4 (k^2 - i0^+)} \,,
\end{aligned} \label{eq:FFprop}
\end{align}
where we have dropped the ultra-local (non-pole-possessing) terms. These 2pt functions can also be derived from mode expansions of chiral photon fields.

We remark that all scattering observables and the classical eikonal (except for the 1PL eikonal) can be computed from the 2pt functions \eqref{eq:FAprop} and \eqref{eq:FFprop}, therefore the dependence on the ``Dirac string'' of \eqref{off-shell-propagator} is only superficial.

\subsubsection{Mode expansion for the propagator}

We can obtain the helicity propagator \eqref{off-shell-propagator} through an off-shell extension of the on-shell mode expansion in QFT. 
We define the polarisation vectors as ($k^0>0, n^0>0$) 
\begin{align}
\begin{aligned}
    \ve_\m^\pm (+k^0,\vec{k}) = \ve_\m^\pm (\vec{k}) \,,& \quad \ve_\m^\pm (-k^0,\vec{k}) = - \ve_\m^{\pm} (-\vec{k}) \,, 
    \\
    \ve_\m^+ (k) = \frac{[k| \s_\m |n \rangle}{\sqrt{2} \langle k n \rangle} \,, &\quad \ve_\m^- (k) = \frac{\langle k| \s_\m |n ]}{\sqrt{2} [ k n ]} \,, 
\end{aligned}
    \label{eq:pol_vec_cont}
\end{align}
such that $[\ve_\m^\pm (k)]^\ast = \ve_\m^\mp (-k)$. The mode expansion of the chiral photon fields $A_\m^{\pm} (x)$ are %
\begin{align}
\begin{aligned}
    A_\m^{\pm} (x) &= \int_{\vec{k}} \frac{1}{2 k^0} \left[ \ve_\m^\pm (\vec{k}) a_{\vec{k},\pm} e^{- i k^0 t + i \vec{k} \cdot \vec{x}} + \left[ \ve_\m^\mp (\vec{k}) \right]^\ast (a_{\vec{k},\mp})^\dagger e^{+ i k^0 t - i \vec{k} \cdot \vec{x}} \right]
    \\  &= \int_{\vec{k}} \frac{\ve_\m^\pm (k)}{2 k^0} \left[ a_{\vec{k},\pm} e^{- i k^0 t + i \vec{k} \cdot \vec{x}} - (a_{\vec{k},\mp})^\dagger e^{+ i k^0 t - i \vec{k} \cdot \vec{x}} \right] \,,
\end{aligned}
    \label{eq:chiral_photon_expansion}
\end{align}
where $k^0 = |\vec{k}|$ and $[A_\m^\pm (x)]^\dagger = A_\m^\mp (x)$. The creation-annihilation operators satisfy 
\begin{align}
   \left[ a_{\vec{k},h} , (a_{\vec{k}', h'})^\dagger \right] = 2k^0 \deltabar^{(3)}(\vec{k} - \vec{k}') \delta_{h, h'}\,, 
   \quad 
   h , h' = \pm \,.
\end{align}
The usual photon field and the dual photon field are given as (see appendix~\ref{sec:conventions})
\begin{align}
\begin{aligned}
      A_\m (x) &= A_\m^+ (x) + A_\m^- (x) \,,\, 
    \\
    \tilde{A}_\m(x) &= i A_\m^+ (x) - i A_\m^- (x) \,. 
\end{aligned}
\end{align}
The time-ordered 2pt functions are
\begin{align}
    \langle  A_\m^\pm (x) A_\n^\pm (0) \rangle &= 0 \,,
    \\ 
    \langle A_\m^+ (x) A_\n^- (0) \rangle &= \int_{\vec{k}} \frac{ \ve_\m^+(\vec{k}) [\ve_\n^+(\vec{k}) ]^\ast \Th(t) e^{- i |\vec{k}| t + i \vec{k} \cdot \vec{x}} + [ \ve_\m^-(\vec{k}) ]^\ast \ve_\n^-(\vec{k}) \Th(-t) e^{+ i |\vec{k}| t - i \vec{k} \cdot \vec{x}} }{2|\vec{k}|} \nn
    \\ &= \int_{\vec{k}} \frac{\Th(t) e^{- i |\vec{k}| t + i \vec{k} \cdot \vec{x}} + \Th(-t) e^{+ i |\vec{k}| t - i \vec{k} \cdot \vec{x}} }{2|\vec{k}|} \times \left[ - \ve_\m^+(k) \ve_\n^-(k) \right] \nn
    \\ &= \int_{k^\m} \frac{i [\ve_\m^+(k) \ve_\n^-(k)]}{k^2 - i 0^+} e^{i k \cdot x} \nn
    \\ &= \int_{k^\m} \frac{i}{k^2 - i0^+} \left[ \frac{2 k_{(\m} n_{\n)} - (k \cdot n) \eta_{\m\n} + i \e_{\m\n\a\b}k^\a n^\b}{2 (n \cdot k)} \right] e^{ik \cdot x} \,, \label{eq:chiralphoton2pt}
\end{align}
where we used the identity
\begin{align}
\begin{aligned}
    \int \frac{d\w}{2\pi} \frac{e^{- i \w t}}{- \w^2 + \vec{k}^2 - i 0^+} &= \int \frac{d\w}{2\pi} \frac{e^{- i \w t}}{(|\vec{k}| - i 0^+ - \w)(|\vec{k}| - i 0^+ + \w)}
    \\ &= \frac{+i}{2 |\vec{k}|} \left[ \Th(t) e^{- i |\vec{k}| t} + \Th(-t) e^{+ i |\vec{k}| t} \right] \,.
\end{aligned} \label{eq:2pt_onshell2offshell}
\end{align}

The computation can be repeated for 2pt functions of field strength tensors, which can be used to justify the propagators \eqref{eq:FAprop} and \eqref{eq:FFprop}. For example, \eqref{eq:FAprop} can be computed from the substitution
\begin{align*}
    \ve_\m^+(k) \ve_\n^-(k) \to 2 i [k_{[\m} \ve_{\n]}^+(k) \ve_{\l}^-(k)] v^\l \deltabar(k \cdot v) = \frac{i k_{[\m}[k|\s_{\n]}|n\rangle[n|v|k\rangle}{\langle k n \rangle [nk]} \deltabar(k \cdot v) \,,
\end{align*}
in the second line of \eqref{eq:chiralphoton2pt} before using the identity \eqref{eq:2pt_onshell2offshell}, where $k^\m = (|\vec{k}|, \vec{k})$ satisfies the on-shell condition $k^2 = 0$. Using the delta constraint, we can recast the numerator as
\begin{align*}
    \frac{i k_{[\m}[k|\s_{\n]}|n\rangle[n|v|k\rangle}{\langle k n \rangle [nk]} \deltabar(k \cdot v) &= - \frac{i}{4} [k|\s_\m \s_\n v|k \rangle \deltabar(k \cdot v)
    \\ &= - \frac{i}{2} \left( k_\m v_\n - k_\n v_\m - i \e_{\m\n\a\b} k^\a v^\b \right) \deltabar(k \cdot v) \,,
\end{align*}
which leads to the 2pt function \eqref{eq:FAprop} after off-shell continuation $k^2 \neq 0$ using \eqref{eq:2pt_onshell2offshell}.

With the modified definitions for the mode operators ($k^0 > 0$)
\begin{align}
    a_{k^\m,\pm} = a_{\vec{k},\pm} \,,\, a_{-k^\m, \pm} = - a_{- \vec{k}, \mp}^\dagger \,,\, k^\m = (k^0, \vec{k}) \,,
\end{align}
we can rewrite the mode expansion \eqref{eq:chiral_photon_expansion} as
\begin{align}
    A_\m^{\pm} (x) &= \int_k \deltabar(k^2) \, \ve_\m^{\pm} (k) \, a_{k,\pm} \, e^{ik \cdot x} \quad \stackrel{\text{off-shell}}{\longrightarrow} \quad \int_k A_\m^\pm (k) \, e^{i k \cdot x} \,,
\end{align}
which is the off-shell continued form used to obtain Feynman rules.

 
\section{Compton amplitude} \label{sec:Compton}

In this section, we compute the classical Compton amplitudes for a root-Kerr particle, 
and compare them with similar results in the literature.
We find perfect agreement to the linear order in spin (at $g=2$), whereas we find model-dependent discrepancy starting from 
the quadratic order in spin. 

\paragraph{3-point coupling}

The shift \eqref{eq:int_NJ_shift-copy} induces the 3-point coupling of an incoming positive helicity photon,
\begin{align}
    i A_3 \sim i q (\ve^+ \cdot v) e^{- k \cdot y} e^{i k \cdot b} \deltabar (k \cdot v) \,,
\end{align}
where $k^\m$ is photon momentum.  
This agrees with the black hole 3-point coupling in the literature. For example, appendix B. of \cite{Chen:2021kxt} gives the minimal coupling as
\begin{align}
    M_3^{\eta, s} &= M_3^{\eta, s=0} \times \exp[- \eta \k_0 (k \cdot a)] \,,
\end{align}
where $\eta = \pm 1$ is the helicity sign of the incoming massless quanta of momentum $k^\m$, $a^\m$ is the spin-length vector $(y^\mu = - a^\m)$, and $\k_0 = \eta_{00}$ is the metric convention parameter.

\subsection{Computation}

The Compton amplitudes will first be computed using the Feynman rules derived from the interaction term expansion \eqref{eq:vertex_all} to parallel the WQFT computations in the literature~\cite{Wang:2022ntx,Kim:2023drc}, which will be reorganised into a form that connects more naturally to the Feynman rules of the alternative expansion \eqref{eq:vertex_gi}.

\begin{figure}[htbp]
    \centering
\begin{subfigure}[b]{0.3\textwidth}
    \centering
    \begin{fmffile}{compton-a}
    \begin{fmfgraph*}(40,70)
        \fmfstraight
        \fmfleft{i1,a1,b1,o1}
        \fmfright{i2,a2,b2,o2}
        \fmfv{decor.shape=square,decor.filled=empty,decor.size=10}{a1}
        \fmfv{decor.shape=square,decor.filled=empty,decor.size=10}{b1}
        \fmfv{decor.shape=square,decor.filled=shaded,decor.size=10}{a2}
        \fmfv{decor.shape=square,decor.filled=shaded,decor.size=10}{b2}
        \fmf{dots}{i1,a1}
        \fmf{dots}{b1,o1}
        \fmf{plain,width=1}{a1,b1}
        \fmf{phantom}{i2,a2,b2,o2}
         \fmffreeze
        \fmf{wiggly,tension=0}{a1,a2} 
        \fmf{wiggly,tension=0}{b1,b2} 
    \end{fmfgraph*}
\end{fmffile}
    \caption{same helicity}
\end{subfigure}
\begin{subfigure}[b]{0.3\textwidth}
    \centering
    \begin{fmffile}{compton-c}
    \begin{fmfgraph*}(40,70)
        \fmfstraight
        \fmfleft{i1,a1,b1,o1}
        \fmfright{i2,a2,b2,o2}
        \fmfv{decor.shape=square,decor.filled=empty,decor.size=10}{a1}
        \fmfv{decor.shape=square,decor.filled=shaded,decor.size=10}{b1}
        \fmfv{decor.shape=square,decor.filled=shaded,decor.size=10}{a2}
        \fmfv{decor.shape=square,decor.filled=empty,decor.size=10}{b2}
        \fmf{dots}{i1,a1}
        \fmf{dots}{b1,o1}
        \fmf{phantom}{i2,a2,b2,o2}
        \fmf{plain,width=1}{a1,b1}
         \fmffreeze
        \fmf{wiggly,tension=0}{a1,a2} 
        \fmf{wiggly,tension=0}{b1,b2} 
    \end{fmfgraph*}
    \end{fmffile}
    \caption{opposite helicity}
\end{subfigure}
\caption{Diagrams for Compton amplitudes.}
\label{fig:compton-all}
\end{figure}
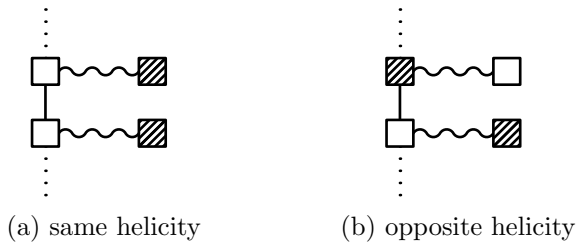

\subsubsection{Same helicity}

The diagram in Figure~\ref{fig:compton-all}(a) 
gives the same helicity amplitude. 
We need to compute the following 2-point function of linear $A^+$-fluctuation coupling terms \eqref{eq:vertex-linear-z}, where we substitute the $A^+_\m$ fields by the polarisation vectors of the external photons. 
\begin{align}
\begin{split}
     i A_4^{++} &= \left\langle V^+(\ve_3, k_3) \times  V^+( \ve_4, k_4) \right\rangle  
     \,,
     \\
     V^+(\ve, k) &=  - q \int_{k,\w} \ve_\m^+ (k) \left[ \k v^\m k_\n - \w \delta^\m_\n \right] \delta z^\n (\w) e^{i k \cdot z_0} \deltabar [\k (k \cdot v) - \w] \,.
\end{split}
\label{A4++schematics}
\end{align}
Using the $\langle \delta z \delta z \rangle$ 2-point function \eqref{eq:zz2pts} and contracting the tensor indices, we get
\begin{align}
\label{eq:ppCompton1}
    i A_4^{++} &= \frac{ i q^2}{m} e^{i (k_3 + k_4) \cdot z_0} \deltabar [(k_3 + k_4) \cdot v]
    \\ &\phantom{=} \times \left\{ \frac{(k_3 \cdot k_4)(\ve_3^+ \cdot v) (\ve_4^+ \cdot v)}{ (k_4 \cdot v)^2} + \frac{(\ve_3^+ \cdot k_4) (\ve_4^+ \cdot v) - (\ve_3^+ \cdot v) (\ve_4^+ \cdot k_3)}{ (k_4 \cdot v)} - (\ve_3^+ \cdot \ve_4^+) \right\} \,.\nn 
\end{align}
This is a simple shift of the non-spinning sector results~\cite{Kim:2023drc} by an exponential spin factor. 
The sign difference compared to (3.26) of the same reference comes from metric conventions. We may also write the amplitude in a gauge-invariant form as
\begin{align}
\label{eq:ppCompton2}
    i A_4^{++} &= \frac{ i q^2}{m} \frac{(v \cdot F_{3}^+ \cdot F_{4}^+ \cdot v)}{ (k_4 \cdot v)^2} e^{i (k_3 + k_4) \cdot z_0} \deltabar [(k_3 + k_4) \cdot v] \,,
\end{align}
where $F_{\m\n}^{\pm} = i (k_\m \ve_\n^\pm - k_\n \ve_\m^\pm)$ is the on-shell field strengths of the external photons and $(v \cdot F_1 \cdot F_2 \cdots ) = v_{\m_1} F_{1~~\m_2}^{~\m_1} F_{2~~\m_3}^{~\m_2} \cdots$ is a shorthand notation for a concatenation of tensor contractions. The expression \eqref{eq:ppCompton2} can be obtained directly from the Feynman rules corresponding to the alternative interaction term expansion \eqref{eq:vertex_gi}.
%

\subsubsection{Opposite helicity}

The opposite helicity Compton amplitude comes from 
the diagram in Figure~\ref{fig:compton-all}(b).
We may simplify the relevant expression in \eqref{eq:zz2pts} as
\begin{align}
    \langle \delta \bar{z}^\m (\w') \delta z^\n (\w) \rangle &\approx \frac{2i}{m} \left[ \frac{\eta^{\m\n}}{2 \w^2} + \frac{ i \e^{\m\n\l\s} v_\l y_{0\s} }{ \w}\right] \deltabar (\w' + \w) \,.
\end{align}
We discarded terms proportional to $v^\mu$ using the fact that the vertex rules at linear order in perturbations satisfy the ``Ward identity'' and vanish under the substitution $\delta z^\m \to v^\m$.
\begin{align}
\begin{aligned}
    & i q \int_{k,\w} A_\m^+ (k) \left[ \k v^\m k_\n - \w \delta^\m_\n \right] \delta z^\n (\w) e^{i k \cdot z_0} \deltabar [\k (k \cdot v) - \w]
    \\ & = - i \k q \int_{k,\w} A_\m^+ (k) (k \cdot v) \left[ \delta^\m_\n - \frac{v^\m k_\n}{k \cdot v} \right] \delta z^\n (\w) e^{i k \cdot z_0} \deltabar [\k (k \cdot v) - \w] \,.
\end{aligned}
\end{align}
Introducing the notation $a^{\m\n} = \e^{\m\n\a\b} v_\a y_{0\b}$, the resulting Compton amplitude becomes
\begin{align}
\label{eq:pmCompton1}
\begin{split}
       i A_4^{+-} &= \frac{ i q^2}{m} e^{i (k_3 \cdot z_0 + k_4 \cdot \bar{z}_0)} \deltabar [(k_3 + k_4) \cdot v] 
    \\ 
    &\phantom{=} \times \left\{ \frac{(k_3 \cdot k_4)(\ve_3^+ \cdot v) (\ve_4^- \cdot v)}{ (k_4 \cdot v)^2} + \frac{(\ve_3^+ \cdot k_4) (\ve_4^- \cdot v) - (\ve_3^+ \cdot v) (\ve_4^- \cdot k_3)}{ (k_4 \cdot v)} \right.  
    \\ 
    &\phantom{=} \left. \phantom{asdf} - (\ve_3^+ \cdot \ve_4^-)  + \frac{2 i (k_3 \cdot a \cdot k_4) (\ve_3^+ \cdot v) (\ve_4^- \cdot v)}{ (k_4 \cdot v)} - 2 i (\ve_3^+ \cdot a \cdot \ve_4^-) (k_4 \cdot v) \right. 
    \\ 
    &\phantom{=} \left. \phantom{\frac{(k_3 \cdot a \cdot k_4)}{ (k_4 \cdot v)}} + 2 i (\ve_3^+ \cdot a \cdot k_4) (\ve_4^- \cdot v) - 2 i (k_3 \cdot a \cdot \ve_4^-) (\ve_3^+ \cdot v) \right\} \,.  
\end{split}
\end{align}
Similar to \eqref{eq:ppCompton2}, the amplitude can be written in a gauge-invariant form as
\begin{align}
\label{eq:pmCompton2}
    i A_4^{+-} &= \frac{ i q^2}{m} \left\{ \frac{(v \cdot F_3^+ \cdot F_4^- \cdot v)}{ (k_4 \cdot v)^2} + 2i \frac{(v \cdot F_3^+ \cdot a \cdot F_4^- \cdot v)}{ (k_4 \cdot v)} \right\} e^{i (k_3 \cdot z_0 + k_4 \cdot \bar{z}_0)} \deltabar [(k_3 + k_4) \cdot v] \,,
\end{align}
which is more natural when the alternative interaction term expansion \eqref{eq:vertex_gi} is used for the Feynman rules.

\subsection{Comparison}

We compare our classical Compton amplitudes with existing results in the literature, setting $g=2$. 
The linear-in-spin amplitude should agree, since it is the universal part captured by the Thomas-Bargmann-Michel-Telegdi (TBMT) equation.
We may find model-dependent discrepancies starting from 
the quadratic order.  

The deviation at quadratic order in spin is an analogue of possible $R^2$ type couplings at $\CO(S^4)$ in the gravitational case~\cite{Bern:2022kto}. 
Such curvature-squared type couplings ($R^2$ or $F^2$) have an interpretation as contributions from induced multipole moments, requiring dimensionful coefficients for their correct normalisation; $[M L^4]$ for gravity and $[M^{-1} L^2]$ for electromagnetism at the leading order. We can introduce such operators without introducing any additional length scale in the case of spinning objects, since the spin-length vector $a^\m$ provides the necessary length scale, the spin order being $\CO(S^4)$ for gravity and $\CO(S^2)$ for electromagnetism for the leading order curvature-squared operators.

\subsubsection{Comparison with SUSY WQFT calculations}
To compare the two results \eqref{eq:ppCompton1} and \eqref{eq:pmCompton1} with those of ref.~\cite{Kim:2023drc}, we use the explicit polarisation vectors
\begin{align}
    \ve_{3\m}^+ = \frac{[3| \bar{\s}_\m |4\rangle}{\sqrt{2} \langle 34 \rangle} \,,\quad 
    \ve_{4\m}^- = \frac{[3| \bar{\s}_\m |4\rangle}{\sqrt{2} [43]} \,,\quad 
    \ve_{4\m}^+ = \frac{[4| \bar{\s}_\m |3\rangle}{\sqrt{2} \langle 43 \rangle} \,, \label{eq:polvec_def}
\end{align}
and the complex conjugation conditions
\begin{align}
    \left( \l_\a \right)^\ast = \text{sgn} (p^0) \bar{\l}_{\dot\a} \,.
\end{align}
We use $\text{sgn}(k_3^0)\text{sgn}(k_4^0) = -1$ because one of the massless photons has to be ingoing and the other has to be outgoing. 
In the rest frame of $v^\m = (1 , \vec{0})$ where $\w = k_3^0$ is the energy of the photon, $k^\m = k_3^\m + k_4^\m$ is the transfer momentum, and 
\begin{align}
\begin{gathered}
    (k_4 \cdot v) = - (k_3 \cdot v) = \w \,,\, |\langle 34 \rangle|^2 = \langle 34 \rangle [34] = {k^2} = 4 \w^2 \sin^2 (\th/2) = |[34]|^2 \,,\,
    \\ |[3|v|4\rangle|^2 = {-[3|v|4|v|3\rangle} = {4 (v \cdot k_4)^2 - k^2} = 4 \w^2 (1 - \sin^2 (\th/2)) = |[4|v|3\rangle|^2 \,,
\end{gathered}
\end{align}
where $\th$ is the scattering angle and we localised onto $(k_3 + k_4) \cdot v = 0$ for the second expression. 

The same helicity amplitude becomes
\begin{align}
    i A_4^{++} &= \frac{iq^2}{m} e^{- k \cdot y_0} \sin^2 (\th/2)
\end{align}
which is simple to evaluate because the $\langle 34 \rangle^{-2}$ factors out in the calculations, which we substitute by $|\langle 34 \rangle|^{-2}$. For the opposite helicity amplitude, we get %
\begin{align}
\begin{aligned}
    i A_4^{+-} 
    &= \frac{iq^2}{m} e^{(k_4-k_3) \cdot y_0} \frac{1}{-2k^2} \left\{ \frac{k^2 (n \cdot v)^2}{2 \w^2} + \frac{2 i \e[k_3, k_4, v, y_0] (n \cdot v)^2}{\w} - 2 i (k \cdot a \cdot n) (n \cdot v) \right\} \,,
\end{aligned}
\end{align}
where $n^\m = [3 | \bar{\s}^\m | 4 \rangle$ vector carries the helicity weights. Note that this form is \emph{manifestly shift-symmetric}; the expression is invariant under the shift of the spin vector by
\begin{align}
    S^\mu \to S^\mu + \xi k^\m / k^2 \,, \nn
\end{align}
where $\xi$ is an arbitrary parameter. The shift symmetry is one of the conjectures for tensor structures of spinning black holes~\cite{Aoude:2022trd,Bern:2022kto}.

Now we multiply the factor $\frac{[4|v|3\rangle^2}{|[4|v|3\rangle|^2}$ to compensate the helicity weights, and use the identity for $f^\m = - a^{\m\n} k_\n$, 
\begin{align}
\begin{aligned}
    [3|f|4|v|3 \rangle &= (k \cdot a \cdot n) [4|v|3\rangle
    \\ &= 2 \left( (k_3 \cdot f) (k_4 \cdot v) - (k_3 \cdot k_4) (f \cdot v) + (k_4 \cdot f) (k_3 \cdot v) - i \e[k_3, f, k_4, v] \right)
    \\ &= - 4 \w (k_3 \cdot a \cdot k_4) + 2 i \e[f , k_3, k_4, v]
    \\ &= - 4 \w \e[k_3, k_4, v, y_0] - i k^2 (k_3 - k_4) \cdot y_0
\end{aligned}
\end{align}
to obtain 
\begin{align}
\begin{aligned}
    i A_4^{+-} &= \frac{iq^2}{m} e^{(k_4-k_3) \cdot y_0} \left\{ - \cos^2 (\th/2) - (k_3 - k_4) \cdot y_0 + i \frac{\e[k_3, k_4, v, y_0]}{\w} \right\}
    \\ &= \frac{iq^2}{m} \left\{ - \cos^2 (\th/2) + ( \cos^2 (\th/2) - 1) (k_3 - k_4) \cdot y_0 + i \frac{\e[k_3, k_4, v, y_0]}{\w} + \CO(y_0^2) \right\} \,.
\end{aligned}
\end{align}
Setting $g=2$, the results in (3.27) of ref.~\cite{Kim:2023drc} for $h=1$ become\footnote{A factor of 2 has been removed from the results of ref.~\cite{Kim:2023drc}; the factor is due to overcounting $s$-channel and $u$-channel diagrams.}
\begin{subequations}
    \begin{align}
    i A_4^{+-} &= \frac{iq^2}{m} \left[ - \cos^2 (\th/2) + \sin^2 (\th/2) \k_0 [(k_1 - k_4) \cdot a] - \frac{i}{\w} (-\k_1) \e[k_1, k_4, a, v] \right] \,,\,
    \\ i A_4^{++} &= \frac{iq^2}{m}\sin^2 (\th/2) \left[ 1 - \k_0 [(k_1+k_4) \cdot a] \right] \,,
\end{align}
\end{subequations}
where $a^\m = S^\m / m$, and we have restored the convention parameters $\k_0$ and $\k_1$; ref.~\cite{Kim:2023drc} uses $\k_0 = \eta_{00} = +1$ and $\k_1 = \e_{0123} = -1$. The expressions match perfectly when we set $k_1^\m = k_3^\m$ and $y_0^\m = - a^\m$. 

\subsubsection{Comparison with amplitude calculations}
For the spin squared coupling, we compare our result with the minimal coupling amplitude constructed from BCFW recursion~\cite{Chen:2021kxt}
\begin{align}
\begin{aligned}
    A_4^{+-} &\propto \frac{[3|p_1|4\rangle^2}{(s-m^2)(u-m^2)} \exp \left[ - i \frac{k^\m [3|\s^\n|4\rangle S_{\m\n}}{\k_0 [3|p_1|4\rangle} \right] \,,
\end{aligned}
\end{align}
where we have restored the metric convention parameter $\k_0$. 
Matching the overall normalisation\footnote{As remarked in ref.~\cite{Kim:2023drc}, there is a mass factor difference between QFT amplitude results and WQFT amplitude results, which can be interpreted as the ratio of $\deltabar (k \cdot v)$ to $\deltabar(2 k \cdot p)$.} we write the amplitude as
\begin{align}
    iA_4^{+-} &= - \frac{iq^2}{m} \frac{(n \cdot v)^2}{4(k_4 \cdot v)^2} \exp \left[ i \frac{\e[k_3 + k_4, n , v, a]}{(n \cdot v)}\right] \,,
\end{align}
where we set $p_1^\m = m v^\m$ and take the classical limit for the Mandelstam invariants as 
\begin{align}
   s - m^2 = - (u - m^2) + \CO(\hbar^2) = 2 m (k_4 \cdot v) + \CO(\hbar^2) \,. 
\end{align} 
The exponent can be simplified to
\begin{align}
    \frac{i \e[k_3 + k_4, n , v, a]}{(n \cdot v)} = (k_3 - k_4) \cdot a - \frac{[(k_3 - k_4) \cdot v](n \cdot a)}{(n \cdot v)}
\end{align}
using the identity
\begin{align}
    i \ve^{\m\n\l\s} = \frac{( \s^\m \bar{\s}^{\n} \s^\l \bar{\s}^{\s} )_{\a}{}^{\a} - ( \s^\n \bar{\s}^{\l} \s^\s \bar{\s}^{\m} )_{\a}{}^{\a}}{-4}
\end{align}
and Schouten identities. 
It leads to the form of the amplitude used in HPET/HEFT~\cite{Aoude:2022trd,Aoude:2022thd,Aoude:2023vdk} and BHPT~\cite{Bautista:2021wfy,Bautista:2022wjf} approaches, 
\begin{align}
    i(A_4^{+-})_\mathrm{HEFT} &= - \frac{iq^2}{m} \frac{(n \cdot v)^2}{4(k_4 \cdot v)^2} e^{ (k_3 - k_4) \cdot a } \exp \left[ \frac{2(k_4 \cdot v)(n \cdot a)}{(n \cdot v)}\right] \label{eq:HEFTCompton1}
\end{align}
after localising onto $\deltabar[(k_3 + k_4)\cdot v]$, where the amplitude is regular up to order $\CO(a^2)$. We compare it to the classical Compton amplitude of our twistor model, 
\begin{align}
\begin{aligned}
    i A_4^{+-} &= \frac{iq^2}{m} e^{(k_3-k_4) \cdot a} \left\{ - \frac{(n \cdot v)^2}{4 (k_4 \cdot v)^2} + \frac{i \e[k_3, k_4, v, a] (n \cdot v)^2}{2(k_3 \cdot k_4)(k_4 \cdot v)} - \frac{i \e[k_3 + k_4, n, v, a] (n \cdot v)}{2(k_3 \cdot k_4)} \right\}
    \\ &= - \frac{iq^2}{m} \frac{(n \cdot v)^2}{4(k_4 \cdot v)^2} e^{ (k_3 - k_4) \cdot a } \left\{ 1 - \frac{2 (k_4 \cdot v)^2}{(k_3 \cdot k_4)} \left[ \frac{i \e[k_3, k_4, v, a]}{(k_4 \cdot v)} - \frac{i \e[k_3 + k_4, n, v, a]}{(n \cdot v)} \right] \right\}
\end{aligned} \nn
\end{align}
where we used $y_0^\m = - a^\m$. We can use the identities ($k = k_3 + k_4$)
\begin{align}
\begin{aligned}
    (v \cdot n) \e[k,k_4,v,a] &= (v \cdot k) \e[n,k_4,v,a] + (v \cdot k_4) \e[k,n,v,a]
    \\ &\phantom{=asdf} + (v \cdot v) \e[k,k_4,n,a] + (v \cdot a) \e[k,k_4,v,n]
    \\ &= (k_4 \cdot v) \e[k,n,v,a] - \e[k_3,k_4,n,a] \,,
\end{aligned}
\end{align}
and
\begin{align}
    i \e[k_3,k_4,n,a] &= (k_3 \cdot k_4) (n \cdot a) \,,
\end{align}
to write the twistor Compton amplitude as
\begin{align}
    i (A_4^{+-})_\mathrm{twistor} &= - \frac{iq^2}{m} \frac{(n \cdot v)^2}{4(k_4 \cdot v)^2} e^{ (k_3 - k_4) \cdot a } \left\{ 1 + \frac{2 (k_4 \cdot v) (n \cdot a)}{(n \cdot v)} \right\} \,.
    \label{eq:pmCompton3}
\end{align}
The difference between \eqref{eq:HEFTCompton1} and \eqref{eq:pmCompton3} is a non-pole-possessing term,\footnote{Adding \eqref{eq:pmCompton_diff} (without $\CO(a^3)$ corrections) to \eqref{eq:pmCompton3} results in the HPET amplitude (3.9) of ref.~\cite{Aoude:2022trd} with all free parameters $c_j^{(n)}$ set to zero. JWK would like to thank Kays Haddad for pointing this out.}
\begin{align}
    i \D A_4^{+-} &= - \frac{iq^2}{m} e^{(k_3 - k_4)\cdot a} (n \cdot a)^2 + \CO(a^3) \,. \label{eq:pmCompton_diff}
\end{align}
This term can be reverse-engineered to find a $\CO(a^2)$ worldline contact term that generates this contribution. From the definition of the polarisation vector \eqref{eq:polvec_def} we relate the $n^\m$ vectors to the polarisation vectors as
\begin{align}
    n^\m n^\n = - 4 (k_3 \cdot k_4) \ve_3^{+\m} \ve_4^{-\n} \,.
\end{align}
To ensure the Ward identity, we use the substitution rule
\begin{align}
\begin{aligned}
    n^\m n^\n &\to - 4 \left[ (k_3 \cdot k_4) \ve_3^{+\m} \ve_4^{-\n} - (\ve_3^+ \cdot k_4) k_3^\m \ve_4^{-\n} - (k_3 \cdot \ve_4^-) \ve_3^{+\m} k_4^{\n} + (\ve_3^+ \cdot \ve_4^-) k_3^\m k_4^\n \right]
\end{aligned} \label{eq:n2ve4pmCompton}
\end{align}
with symmetrisation if necessary. We may also write this substitution as
\begin{align}
    n_\m n_\n &\to 4 \eta^{\a\b} F_{\a\m}^{+} (k_3) F_{\b\n}^{-} (k_4) \,,\, F_{\m\n}^{\pm} (k) := i k_\m \ve^{\pm}_\n (k) - i k_\n \ve^{\pm}_\m (k) \,,
\end{align}
where $F^{\pm}_{\m\n} (k)$ is the momentum space mode coefficient of the field strength 2-form $F^{\pm} = dA^{\pm}$. 
We can now attribute the difference \eqref{eq:pmCompton_diff} to the worldline contact term
\begin{align}
    i S_{\text{cont}} &= i \int \frac{4q^2}{m} \left[ y \cdot F^+ (z) \cdot F^- (\bar{z}) \cdot y \right] d\s + \CO(y^3) \,.
    \label{yFFy-contact}
\end{align}
As we observed in section~\ref{sec:NJ-eom}, the combination $(y\cdot F^+\cdot F^-\cdot y)$ is an inevitable consequence of the zig-zag symplectic perturbation theory. Our twistor model differs from other models which do not carry $(y\cdot F^+\cdot F^-\cdot y)$ terms.

The Compton amplitude \eqref{eq:pmCompton3} can also be compared to predictions of higher-spin gauge symmetry~\cite{Cangemi:2023ysz}. In the notations of ref.~\cite{Cangemi:2023ysz}, the non-scalar part of \eqref{eq:pmCompton3} can be written as $e^x ( 1 - w )$, which differs from the result (6.61) of ref.~\cite{Cangemi:2023ysz} reproduced below
\begin{align}
    e^x \cosh z - w e^x \frac{\sinh z}{z} + \frac{w^2 - z^2}{2} E(x,y,z) \,,
\end{align}
where
\begin{align*}
\begin{gathered}
    x = - (k_4 - k_3) \cdot a \,,\, y = - (k_3 + k_4) \cdot a \,,\, z = - |a| v_1 \cdot (k_4 - k_3) \,,\\ w = - \frac{(n \cdot a)[v \cdot (k_4 - k_3)]}{(n \cdot v)} \,,
\end{gathered}
\end{align*}
and
\begin{align*}
    E(x,y,z) &= \frac{e^y - e^x \cosh z + (x-y) e^x \frac{\sinh z}{z}}{(x-y)^2 - z^2} + (y \to -y) \,.
\end{align*}
While this amplitude is quite different from the twistor worldline prediction \eqref{eq:pmCompton3}, the amplitude shares the same $e^x$ factor conjectured to be responsible for the singularity structure of the aligned-spin one-loop eikonal \eqref{eq:one-loop_eik_as_full}. %

 
\section{Conservative dynamics from WQFT} \label{sec:Eikonal}

In this section, we revisit the scattering observables of the binary system from the WQFT perspective. 
We computed the 1PL and 2PL observables in section~\ref{sec: scattering observables} by solving the equations of motion
and extracted the classical eikonal along the way. 
One advantage of the WQFT approach is that it allows us to compute the eikonal before computing observables. 
Using two approaches to compute the same eikonal serves as a consistency check. 
Besides, we specialise to the aligned spin configurations and evaluate the Fourier integrals explicitly 
to obtain the position space expressions for the classical eikonal up to the 2PL order. 

\subsection{1PL observables and eikonal}

Since the complex coordinate $z^\mu = x^\mu + i y^\mu$ includes the position and the spin  (recall the sign $a^\mu = -y^\mu$), 
we can compute the velocity kick $\Delta v^\mu$ and the spin kick $\Delta y^\mu$ both from the expectation values $\langle\!\langle \delta z_1^\m (\w) \rangle \! \rangle$ and $\langle \! \langle \delta \bar{z}_1^\m (\w) \rangle \! \rangle$.
For particle 1, the expectation value  $\langle\!\langle \delta z_1^\m (\w) \rangle \! \rangle$ is given by 
\begin{align}
\begin{aligned}
    \langle \! \langle \delta z_1^\m (\w) \rangle \! \rangle &= - q_1 q_2 \int_{k,k',\w'} \left[ \langle \delta z_1^\m (\w) \delta z_1^\n (\w') \rangle \langle F_{\n\l}^+ (k) A_\a^- (k') \rangle e^{i k \cdot z_{1} + i k' \cdot \bar{z}_{2}} \right.
    \\ &\phantom{=asdfasdfasdfasdf} \left.{} + \langle \delta z_1^\m (\w) \delta \bar{z}_1^\n (\w') \rangle \langle F_{\n\l}^- (k) A_\a^+ (k') \rangle e^{i k \cdot \bar{z}_{1} + i k' \cdot {z}_{2}} \right]
    \\ &\phantom{=asdfasdfasdfasdfasdfasdf} \times v_1^\l v_2^\a \deltabar [(k \cdot v_1) - \w'] \deltabar(k' \cdot v_2) \,, 
\end{aligned}
\end{align}
where $z_{1,2}$, $\bar{z}_{1,2}$ in the exponents are understood as background values.
This expression is free of Dirac string singularity when we use the propagator \eqref{eq:FAprop}. 

The velocity kick and the spin kick can be computed from the expectation values as
\begin{align}
    \D_{(1)} v_1^\m &= \Re \lim_{\t \to \infty} \frac{d \delta z_1^\m (\t)}{d\t} = \Re \int_{\t} \frac{d^2 \delta z_1^\m (\t)}{d\t^2} = \Re \int_{\t,\w} (-\w^2) \delta z_1^\m (\w) e^{- i \w \t} \nn
    \\ &= \Re \lim_{\w \to 0} (-\w^2) \delta z_1^\m (\w) \,, \label{eq:vkick}
    \\ 
    \D_{(1)} y_1^\m &= \Im \lim_{\t \to \infty} \delta z_1^\m (\t) = \Im \int_{\t} \frac{d \delta z_1^\m (\t)}{d\t} = \Im \int_{\t,\w} (-i\w) \delta z_1^\m (\w) e^{-i\w\t} \nn
    \\ &=  \Im \lim_{\w \to 0} (-i\w) \delta z_1^\m (\w) \,. \label{eq:skick}
\end{align}
This may be viewed as the worldline version of the LSZ reduction formula~\cite{Lehmann:1954rq}; we can expect that the equivalent of the $S$-matrix equivalence theorem~\cite{Chisholm:1961tha,Kamefuchi:1961sb} will also hold for worldline observables in WQFT. 
After some algebra, we can write the velocity kick as
\begin{align}
\begin{aligned}
    \D_{(1)} v_1^\m &= - \frac{q_1 q_2}{2m_1} \int_{k} \frac{i \gamma k^\m - \e^{\m}[k,v_1,v_2]}{k^2} e^{k \cdot y} e^{i k \cdot b} \deltabar(k \cdot v_1) \deltabar(k \cdot v_2)
    \\ &\phantom{=} - \frac{q_1 q_2}{2m_1} \int_{k} \frac{i \gamma k^\m + \e^{\m}[k,v_1,v_2]}{k^2} e^{- k \cdot y} e^{i k \cdot b} \deltabar(k \cdot v_1) \deltabar(k \cdot v_2) \,.
\end{aligned} \label{eq:1PM_impulse}
\end{align}
Similarly, we can write the spin kick as
\begin{align}
\begin{aligned}
     \D_{(1)} y_1^\m &= + \frac{q_1 q_2}{2m_1} \int_{k} \frac{i (v_2 \cdot y_1) k^\m - i (k \cdot y_1) v_2^\m - v_1^\m \e[k, v_{1}, v_{2}, y_1] - \gamma \e^{\m}[ k, v_1, y_1]}{k^2}
    \\ &\hskip 2cm \times e^{i k \cdot (b - iy)} \deltabar(k \cdot v_1) \deltabar(k \cdot v_2)
    \\ 
    &\phantom{=} + \frac{q_1 q_2}{2m_1} \int_{k} \frac{i (v_2 \cdot y_1) k^\m - i (k \cdot y_1) v_2^\m + v_1^\m \e[ k, v_{1}, v_{2}, y_1] + \gamma \e^{\m}[ k, v_1, y_1]}{k^2}
    \\ &\hskip 2cm \times e^{i k \cdot (b + iy)} \deltabar(k \cdot v_1) \deltabar(k \cdot v_2) \,.
\end{aligned}
\end{align}
Both $\D_{(1)} v_1^\m$ and $\D_{(1)} y_1^\m$ 
agree with the results of section~\ref{sec: scattering observables} as expected.

\subsubsection{1PL eikonal}

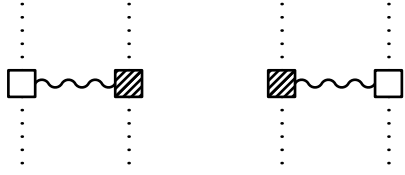
\begin{figure}[htbp]
    \centering
\begin{fmffile}{1PM-a}
    \begin{fmfgraph*}(40,60)
        \fmfstraight
        \fmfleft{i1,a1,o1}
        \fmfright{i2,a2,o2}
        \fmfv{decor.shape=square,decor.filled=empty,decor.size=10}{a1}
        \fmfv{decor.shape=square,decor.filled=shaded,decor.size=10}{a2}
        \fmf{dots}{i1,a1,o1}
        \fmf{dots}{i2,a2,o2}
         \fmffreeze
        \fmf{wiggly,tension=0}{a1,a2} 
    \end{fmfgraph*}
\end{fmffile}
\hskip 1.5cm 
\begin{fmffile}{1PM-b}
    \begin{fmfgraph*}(40,60)
        \fmfstraight
        \fmfleft{i1,a1,o1}
        \fmfright{i2,a2,o2}
        \fmfv{decor.shape=square,decor.filled=shaded,decor.size=10}{a1}
        \fmfv{decor.shape=square,decor.filled=empty,decor.size=10}{a2}
        \fmf{dots}{i1,a1,o1}
        \fmf{dots}{i2,a2,o2}
         \fmffreeze
        \fmf{wiggly,tension=0}{a1,a2} 
    \end{fmfgraph*}
\end{fmffile}
    \caption{Diagrams contributing to the 1PL eikonal.}
    \label{fig:1PM-diagrams}
\end{figure}

\noindent
The eikonal is evaluated as the sum over the diagrams in Figure~\ref{fig:1PM-diagrams}:
\begin{align}
\begin{aligned}
    i \chi_{(1)} &= - q_1 q_2 \int_{k_\perp } v_1^\m v_2^\n \left[ e^{i k \cdot (z_1 - \bar{z}_2)} \Delta^{+-}_{\m\n}(k) + e^{i k \cdot (\bar{z}_1 - {z}_2)} \Delta^{-+}_{\m\n}(k) \right]  \,, 
\end{aligned}
\end{align}
where $\Delta^{+-}_{\m\n}(k)$ and $\Delta^{-+}_{\m\n}(k)$ are the helicity propagators \eqref{off-shell-propagator}. 
We write the integrand as
\begin{align}
\begin{aligned}
    & v_1^\m v_2^\n \left[ e^{- k \cdot y} \Delta^{+-}_{\m\n}(k) + e^{+ k \cdot y} \Delta^{-+}_{\m\n}(k) \right] 
    \\ 
    &\qquad = \cosh(k \cdot y) v_1^\m v_2^\n \left[ \Delta^{+-}_{\m\n}(k) + \Delta^{-+}_{\m\n}(k) \right] 
    \\ 
    &\qquad \quad \phantom{=} 
    + \frac{\sinh(k \cdot y)}{(k \cdot y)} y^\l v_1^\m v_2^\n (i k_\l) \left[ i \Delta^{+-}_{\m\n}(k) - i \Delta^{-+}_{\m\n}(k)\right] \,.
\end{aligned}
\label{1PM-eikonal-identity}
\end{align}
Using \eqref{photon-propagator-even-odd}, we can simplify the $(\cosh)$ term slightly and write 
\begin{align}
   \cosh(k \cdot y) v_1^\m v_2^\n \left[ \Delta^{+-}_{\m\n}(k) + \Delta^{-+}_{\m\n}(k) \right] =  
   \cosh(k \cdot y) (v_1 \cdot v_2)  \frac{-i}{k^2}  \,.
\end{align}
We remind the readers that we are treating terms proportional to $(v_1\cdot k)$, $(v_2\cdot k)$ 
or $k^2$ as zero.
The $(\sinh)$ term is more interesting.  
\begin{align}
\begin{aligned}
    & y^\l v_1^\m v_2^\n (i k_\l) \left[ i \Delta^{+-}_{\m\n}(k) - i \Delta^{-+}_{\m\n}(k)\right] 
    \\ 
     &= y^\l v_1^\m v_2^\n (i k_\l) \left[ i \Delta^{+-}_{\m\n}(k) - i \Delta^{-+}_{\m\n}(k)\right] - 
     y^\l v_1^\m v_2^\n (i k_\m) \left[ i \Delta^{+-}_{\l\n}(k) - i \Delta^{-+}_{\l\n}(k)\right] 
    \\
    &= y^\l v_1^\m v_2^\n  \e_{\l\m}{}^{\a\b} (i k_\a) \left[ \Delta^{+-}_{\b\n}(k) + \Delta^{-+}_{\b\n}(k) \right] 
    = i \e[k,v_1,v_2,y] \frac{-i}{k^2}   \,.
\end{aligned}
\label{1PM-eikonal-identity-2}
\end{align}
The term added to the second line vanishes due to $\deltabar(v_1\cdot k)$; its purpose is to anti-symmetrise in $\l$, $\m$ indices. 
The equality between the second line and the third line follows from \eqref{photon-propagator-even-odd} and the 4d Schouten identity \eqref{4d-schouten}.
In the end, we obtain
\begin{align}
    \chi_{(1)} &= - q_1 q_2 \int_{k_\perp} \left[ \cosh(k\cdot y) \gamma - i \frac{\sinh(k\cdot y)}{k\cdot y}  \epsilon[k,v_1,v_2,y]\right] \frac{ e^{ik\cdot b}}{k^2}   \,.
    \label{1PL-eikonal-WQFT}
\end{align}
in agreement with \eqref{1PL-eikonal}. 
It is free of the Dirac string ambiguity as expected.

Next, we perform the Fourier integral explicitly and obtain 
\begin{align}
\begin{split}
     \chi_{(1)} &= \frac{q_1 q_2 \g}{4\pi \sqrt{\g^2 - 1}} \left[ \frac{1}{\e} + \Re \left( \log \frac{(b^\m + i y_\perp^\m)^2}{b_0^2} \right) \phantom{+ \left(\frac{y_\perp^2 + \sqrt{y_\perp^2 - (b \cdot y_\perp)^2}}{y_\perp^2 - \sqrt{y_\perp^2 - (b \cdot y_\perp)^2}} \right)} \right. \nn
    \\ 
    &\phantom{=asdf} \left. \phantom{asdf} - \frac{\e[b,v_1,v_2,y_\perp]}{2 \g \sqrt{b^2 y_\perp^2 - (b \cdot y_\perp)^2}} \log \left(\frac{b^2 + y_\perp^2 + 2 \sqrt{b^2 y_\perp^2 - (b \cdot y_\perp)^2}}{b^2 + y_\perp^2 - 2 \sqrt{b^2 y_\perp^2 - (b \cdot y_\perp)^2}} \right) \right] \,,
\end{split}
\end{align}
where $D = 4 - 2 \e$, $b_0^2$ is the dimensional regularisation parameter absorbing all regularisation artefacts (factors of $\pi$ and $\g_E$, etc.), and 
\begin{align}
    y_\perp^\m = y^\m + \left(\frac{\g(y \cdot v_2) - (y \cdot v_1)}{\g^2 - 1} \right) v_1^\m + \left(\frac{\g(y \cdot v_1) - (y \cdot v_2)}{\g^2 - 1} \right) v_2^\m
\end{align} 
is the projection of $y^\m$ onto the impact parameter space defined by $b \cdot v_1 = b \cdot v_2 = 0$. 
In the aligned spin configuration, $y_\perp^\m = y^\m$ and $(b \cdot y) = 0$, the eikonal simplifies even further,
\begin{align}
    \chi_{(1,\text{aligned})} &= \frac{q_1 q_2 \g}{4\pi \sqrt{\g^2 - 1}} \left[ \frac{1}{\e} + \log \frac{b^2 - y^2}{b_0^2} - \frac{\e[b,v_1,v_2,y]}{\g |b| |y|} \log \left(\frac{|b| + |y|}{|b| - |y|} \right) \right] \,, \label{eq:tree_eikonal_as}
\end{align}
where $|b| = \sqrt{b^2}$ and $|y| = \sqrt{y^2}$. Note that the aligned spin eikonal develops a logarithmic singularity at $b^2 = y^2$; the eikonal ``knows'' that classical spin is a finite-size effect and the point particle approximation breaks down when the two bodies are too close to each other.

\subsection{2PL eikonal}

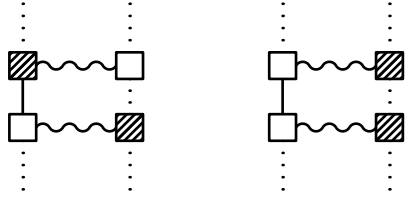
\begin{figure}[htbp]
    \centering
\begin{fmffile}{2PM-a}
    \begin{fmfgraph*}(40,70)
        \fmfstraight
        \fmfleft{i1,a1,b1,o1}
        \fmfright{i2,a2,b2,o2}
        \fmfv{decor.shape=square,decor.filled=empty,decor.size=10}{a1}
        \fmfv{decor.shape=square,decor.filled=shaded,decor.size=10}{b1}
        \fmfv{decor.shape=square,decor.filled=shaded,decor.size=10}{a2}
        \fmfv{decor.shape=square,decor.filled=empty,decor.size=10}{b2}
        \fmf{dots}{i1,a1}
        \fmf{dots}{b1,o1}
        \fmf{dots}{i2,a2,b2,o2}
        \fmf{plain,width=1}{a1,b1}
         \fmffreeze
        \fmf{wiggly,tension=0}{a1,a2} 
        \fmf{wiggly,tension=0}{b1,b2} 
    \end{fmfgraph*}
\end{fmffile}
\hskip 1.5cm 
\begin{fmffile}{2PM-b}
    \begin{fmfgraph*}(40,70)
        \fmfstraight
        \fmfleft{i1,a1,b1,o1}
        \fmfright{i2,a2,b2,o2}
        \fmfv{decor.shape=square,decor.filled=empty,decor.size=10}{a1}
        \fmfv{decor.shape=square,decor.filled=empty,decor.size=10}{b1}
        \fmfv{decor.shape=square,decor.filled=shaded,decor.size=10}{a2}
        \fmfv{decor.shape=square,decor.filled=shaded,decor.size=10}{b2}
        \fmf{dots}{i1,a1}
        \fmf{dots}{b1,o1}
        \fmf{plain,width=1}{a1,b1}
        \fmf{dots}{i2,a2,b2,o2}
         \fmffreeze
        \fmf{wiggly,tension=0}{a1,a2} 
        \fmf{wiggly,tension=0}{b1,b2} 
    \end{fmfgraph*}
\end{fmffile}
    \caption{Diagrams contributing to the 2PL eikonal, up to the exchange of the two particles and the overall flip of holomorphy/helicity.}
    \label{fig:2PM-diagrams}
\end{figure}

The diagrams relevant for the 2PL eikonal are shown in Figure~\ref{fig:2PM-diagrams}.

\paragraph{Building up the eikonal integrand}
The integrand can be constructed by replacing the field strength tensors $F^\pm_{i\m\n}$ of the Compton amplitudes \eqref{eq:ppCompton2} and \eqref{eq:pmCompton2} by the linearised source contribution from the other particle using the propagator \eqref{eq:FAprop},
\begin{align}
\begin{aligned}
    F_{i \m\n}^{+} &\to \frac{(k\wedge v_2)_{\m\n} - i \e_{\m\n}[ k_i , v_2] }{2 k_i^2} \times i q_2 e^{- i k_i \cdot \bar{z}_2} \deltabar (k_i \cdot v_2) \,,
    \\ F_{i \m\n}^{-} &\to \frac{(k\wedge v_2)_{\m\n} + i \e_{\m\n}[ k_i , v_2] }{2 k_i^2} \times i q_2 e^{- i k_i \cdot {z}_2} \deltabar (k_i \cdot v_2) \,,
\end{aligned}
\end{align}
attaching symmetry factors, integrating over photon momenta $\int_{k_3,k_4}$, summing over helicity configurations, and summing over worldline permutation $1 \leftrightarrow 2$. 
\begin{align}
    i \chi_{(2)} &= i \int_{k_3,k_4} \left[ I^{++} + 2 I^{+-} + I^{--} \right] + (1 \leftrightarrow 2) \,,
\end{align}
where we used the fact that $I^{+-} = I^{-+}$. 
%

\paragraph{Same helicity integrand}

The integrand turns out to be quite simple
\begin{align}
I^{++} = - \frac{ (q_1 q_2)^2 }{8 m_1} e^{i(k_3+k_4)\cdot (z_{1} - \bar{z}_{2})} \frac{\deltabar(v_1\cdot k_3 + v_1 \cdot k_4) \deltabar(v_2 \cdot k_3)\deltabar(v_2 \cdot k_4)}{(v_1\cdot k_3)(v_1 \cdot k_4)} \frac{k_3\cdot k_4}{k_3^2 k_4^2} \,. 
\end{align}
At the level of the integrand, it is clear that $I^{++} + I^{--}$ agrees with the $\cosh[(k+\ell)\cdot y]$ term in \eqref{2PL-eikonal-final} upon the identification $(k_3 , k_4)_\mathrm{here} \leftrightarrow (k,\ell)_\mathrm{there}$.  

We can use Passarino-Veltman reduction to rewrite the integrand as
\begin{align}
\begin{aligned}
    \frac{k_3\cdot k_4}{(v_1\cdot k_3)(v_1 \cdot k_4)k_3^2 k_4^2} &= \frac{1}{2} \left[ \frac{(k_3 + k_4)^2}{(v_1\cdot k_3)(v_1 \cdot k_4)k_3^2 k_4^2} - \frac{k_3^2 + k_4^2}{(v_1\cdot k_3)(v_1 \cdot k_4)k_3^2 k_4^2} \right]
    \\ &= \frac{-1}{2} \left[ \frac{k^2}{(v_1\cdot k_3)^2 k_3^2 (k-k_3)^2} - \frac{2}{(v_1\cdot k_3)^2 k_3^2} \right]
\end{aligned}
\end{align}
where we used the symmetry between $k_3$ and $k_4$, and then used the condition $\deltabar[(k_3 + k_4) \cdot v_1]$. 
We also set $k^\m = k_3^\m + k_4^\m$. We are left with evaluation of the integral (in $D = 4 - 2\e$ dimensions for regularisation)
\begin{align}
\begin{aligned}
    \CI^{++} (k,v) &:= \int_{k_3,k_4} \frac{k_3\cdot k_4  \, e^{i (k_3 + k_4) \cdot z}}{(v_1\cdot k_3)(v_1 \cdot k_4) k_3^2 k_4^2} \deltabar[(k_3 + k_4) \cdot v_1] \deltabar[(k_3 + k_4) \cdot v_2] \deltabar(k_3 \cdot v_2)
    \\ 
    &= - \frac{1}{2} \int_{k_\perp} \, e^{i k \cdot z}\int_{k_3} \left[ \frac{k^2}{(v_1\cdot k_3)^2 k_3^2 (k-k_3)^2} - \frac{2}{(v_1\cdot k_3)^2 k_3^2} \right] \deltabar (k_3 \cdot v_2) \,.
\end{aligned}
\end{align}
Then we separate the integral argument into $k_{3\parallel}$ and $k_{3\perp}$, such that $k_{3\perp} \cdot v_2 = 0$. The $d k_{3\parallel}$ integral is trivial due to $\deltabar(k_3 \cdot v_2)$, and we get
\begin{align}
    \int d^{D-1} k_{3 \perp} \left[ \frac{k^2}{(v_{1 \perp} \cdot k_{3\perp})^2 k_{3\perp}^2 (k-k_{3\perp})^2} - \frac{2}{(v_{1\perp} \cdot k_{3\perp})^2 k_{3\perp}^2} \right] \,,
\end{align}
where $v_{1\perp}^\m = v_1^\m + (v_1 \cdot v_2) v_2^\m = v_1^\m - \g v_2^\m$ is the projection of $v_1^\m$ onto the orthogonal space such that $v_{1\perp} \cdot v_2 = 0$. The projection for $k^\m$ is not needed due to $\deltabar(k \cdot v_2)$ constraint. The remaining integrals evaluate to zero when using the master integral \eqref{eq:MI}, which is consistent with vanishing same helicity sector contributions for the triangle coefficient in amplitude calculations~\cite{Guevara:2017csg,Guevara:2018wpp,Chung:2018kqs}.

\paragraph{Opposite helicity integrand}
Evaluating the relevant diagrams, and factoring out the common denominator as  
\begin{align}
\begin{aligned}
    I^{+-} &= - \frac{ (q_1 q_2)^2 }{8 m_1} e^{i(k_3+k_4) \cdot b} \frac{\deltabar(v_1\cdot k_3 + v_1 \cdot k_4) \deltabar(v_2 \cdot k_3)\deltabar(v_2 \cdot k_4)}{k_3^2 k_4^2 (v_1 \cdot k_3) (v_1 \cdot k_4)}
    \\ &\phantom{=asdfasdf} \times  e^{ (k_4 - k_3) \cdot (y_1 + y_2)} N[k_3,k_4,v_1,v_2,y_1] \,,
\end{aligned}
\end{align}
we find that the numerator, organized in powers of $\gamma$, is given as 
\begin{align}
    \label{2PL-eikonal-numerator}
    N &= 2 \g^2 \left( i (v_1 \cdot k_4) \e[k_3,k_4,v_1,y_1] + (k_3 \cdot k_4) \right)
    \\ &\phantom{=} + 2 i \g \left( \e[k_3,k_4,v_1,v_2] - 2i (k_3 \cdot k_4) (v_1 \cdot k_4) (v_2 \cdot y_1) - (v_1 \cdot k_4)^2 \e[(k_3 + k_4), v_1, v_2, y_1] \right) \nn
    \\ &\phantom{=} - (k_3 \cdot k_4) + 2 (v_1 \cdot k_4)^2 \left[ 1 + (y_1 \cdot k_3) - (y_1 \cdot k_4) \right] + 2 i (v_1 \cdot k_4) (v_2 \cdot y_1) \e[k_3,k_4,v_1,v_2] \,. \nn
\end{align}
It is straightforward to show that it agrees with the opposite helicity integrand in \eqref{2PL-eikonal-final}. 

To evaluate the Fourier integrals, it is convenient to reorganise the numerator as
\begin{align}
\begin{aligned}
    N 
    &= \g^2 \left[ k^2 - k_3^2 - k_4^2 \right] + 2 i \g^2 (v_1 \cdot k_4) \e[k_3,k,v_1,y_1]
    \\ &\phantom{=} + 2 i \g \e[k_3,k,v_1,v_2] + 2 \g (v_2 \cdot y_1) (v_1 \cdot k_4) \left[ k^2 - k_3^2 - k_4^2 \right]
    \\ &\phantom{=} + 2 i \g (v_1 \cdot k_3) (v_1 \cdot k_4) \e[k, v_1, v_2, y_1]
    \\ &\phantom{=} - \frac{1}{2} \left[ k^2 - k_3^2 - k_4^2 \right] - 2 (v_1 \cdot k_3) (v_1 \cdot k_4) \left[ 1 + (y_1 \cdot k_3) - (y_1 \cdot k_4) \right]
    \\ &\phantom{=} + 2 i (v_1 \cdot k_4) (v_2 \cdot y_1) \e[k_3,k,v_1,v_2] \,. 
\end{aligned}
\end{align}

Inspecting the master integral \eqref{eq:MI} we find that $k_3^2$ and $k_4^2$ of the numerator will evaluate to zero ($\l_1 = 0$ or $\l_2 = 0$ condition) and can be thrown away.  
We organise the integrand as
\begin{align}
\begin{aligned}
    I^{+-} &= \frac{ (q_1 q_2)^2 }{8 m_1} e^{i k \cdot (b - i y)} \deltabar(v_1 \cdot k) \deltabar(v_2 \cdot k) \frac{e^{ - 2 k_3 \cdot y} N[k_3,k - k_3,v_1,v_2,y_1] \deltabar(v_2 \cdot k_3)}{k_3^2 (k-k_3)^2 (v_1 \cdot k_3)^2} \,,
    \\ y^\m &= y_1^\m + y_2^\m \,,
\end{aligned} \label{eq:one-loop_eik_integrand_def}
\end{align}
where we use $k_3$ as the loop momentum. 
Performing the $d k_{3\parallel} = dk_3^0$ integral we get
\begin{align}
    \int \frac{d^D k_3}{(2\pi)^D} \frac{e^{ - 2 k_3 \cdot y} N \, \deltabar(v_2 \cdot k_3)}{k_3^2 (k-k_3)^2 (v_1 \cdot k_3)^2} &= \int \frac{d^{D-1} k_{3\perp}}{(2\pi)^{D-1}} \frac{e^{ - 2 k_3 \cdot y_\perp} N_\perp }{k_3^2 (k-k_3)^2 (v_{1\perp} \cdot k_3)^2} \label{eq:1-loop_pm_integral}
\end{align}
where $y_\perp^\m = y_1^\m + y_2^\m + v_2^\m (y_1 \cdot v_2)$ and the effective numerator is
\begingroup
\allowdisplaybreaks
\begin{align}
    N_\perp &= \left[\g^2 - 1/2 \right] k^2 - 2 \g (v_2 \cdot y_1) k^2 (v_{1\perp} \cdot k_3) \nn
    \\ &\phantom{=} + 2 \left[ \frac{i \g \e[k, v_1, v_2, y_1]}{\g^2 -1} + 1 \right] (v_{1\perp} \cdot k_3)^2 \nn
    \\ &\phantom{=} + 4 (v_{1\perp} \cdot k_3)^2 \left( \left[y_1 - \frac{k (k \cdot y_1)}{k^2} + v_2 (y_1 \cdot v_2) \right] \cdot k_3 \right) \nn
    \\ &\phantom{=} - 2 i \left\{ \g^2 \left( \left[ \e^\m [k,v_1,y_1] + \frac{v_{1\perp}^\m \g \e[k,v_1,v_2,y_1]}{\g^2 - 1} \right] + v_{2\m} \e [v_2,k,v_1,y_1] \right) \right. \nn
    \\ &\phantom{=asdfasdf} \left. \phantom{\left( \left[ \frac{v_{1\perp}^\m }{\g^2 } \right] \right)} + (v_2 \cdot y_1) \e_\m[k,v_1,v_2] \right\} (v_{1\perp} \cdot k_3) k_3^\m \nn
    \\ &\phantom{=} + 2 i \g \e_\m [k,v_1,v_2] k_3^\m \,. \label{eq:1-loop_pm_integrand}
\end{align}
\endgroup
The remaining integral can be evaluated using the list of integrals in appendix~\ref{app:MIs}. We present the results in the ancillary file \texttt{loopdata.dat.m}. 
Including the overall $e^{+k \cdot y}$ factor from \eqref{eq:one-loop_eik_integrand_def}, the integral is consistent with the QED amplitude coefficients provided by ref.~\cite{Bern:2023ity} to bilinear order in spins, under the conditions $C_i = 1$, $D_i = 0$, and covariant SSC. We also present the full eikonal as a formal power series in the ancillary file \texttt{eikonaldata.dat.m}. While the expressions by themselves do not provide any insight, they greatly simplify in the aligned spin configuration, which we present next.

\subsubsection{Aligned spin}

Let us simplify the expression by going to the aligned spin configuration. For aligned spin we have the conditions $y_1^\m \propto y_2^\m$ and $y \cdot v_{1,2} = 0$. We introduce the ratio parameter $\z$ defined by $y_1^\m = \z y^\m$; it follows that $y_2^\m = (1 - \z) y^\m$ and $y_\perp^\m = y^\m$. This reduces the expression to a single infinite sum and we get
\begin{align}
\begin{aligned}
    &\int \frac{d^4 k_3}{(2 \pi)^4} \frac{e^{ - 2 k_3 \cdot y} N \, \deltabar(v_2 \cdot k_3)}{k_3^2 (k-k_3)^2 (v_1 \cdot k_3)^2} = \int \frac{d^{3} k_{3\perp}}{(2 \pi)^3}\frac{ e^{ - 2 k_3 \cdot y_\perp} N_\perp }{k_3^2 (k-k_3)^2 (v_{1\perp} \cdot k_3)^2}
    \\ &= \frac{e^{- k \cdot y} }{4 (k^2)^{\frac{1}{2}}} \sum_{m=0}^\infty \frac{\left( -\frac{k^2 y^2}{2} \right)^m}{m!} \left[ \left( \frac{2 \g^2 -1}{\g^2 - 1} - \z \right) \frac{I_{m+1}(k \cdot y)}{(k \cdot y)^{m-1}} \right.
    \\ &\phantom{=} \left. \phantom{ \left( \frac{2 \g^2}{\g^2} \right) } + \left( \frac{\g^2 (4m+1) - 2m -1}{\g^2 - 1} - 2 m \z + \frac{i \g (\z - 2)\e[k,v_1,v_2,y] }{ (\g^2 - 1)} \right) \frac{I_{m}(k \cdot y)}{(k \cdot y)^{m}} \right] \,,
\end{aligned}
\end{align}
where $I_n (x)$ is the modified Bessel function of the first kind. The impact parameter space integral can be organised as
\begingroup
\allowdisplaybreaks
\begin{align}
    \chi^{+-} &= \int_{k_3, k_4} I^{+-} = \frac{(q_1 q_2)^2}{32 m_1} \int \frac{d^4 k}{(2 \pi)^4} \deltabar(v_1 \cdot k) \deltabar(v_2 \cdot k) \frac{\CI^{+-} e^{i k \cdot b}}{\sqrt{k^2}} \nn
    \\ &= \frac{(q_1 q_2)^2}{32 m_1 \sqrt{\g^2 - 1}} \int \frac{d^2 k_E}{(2 \pi)^2} \frac{\CI^{+-} e^{i (k_E \cdot b)}}{\sqrt{k_E^2}} \,, \label{eq:2PMeikPM}
    \\ \CI^{+-} &= \sum_{m=0}^\infty \frac{\left( -\frac{k^2 y^2}{2} \right)^m}{m!} \left[ \left( \frac{2 \g^2 -1}{\g^2 - 1} - \z \right) \frac{I_{m+1}(k \cdot y)}{(k \cdot y)^{m-1}} \right. \nn
    \\ &\phantom{=} \left. \phantom{ as } + \left( \frac{\g^2 (4m+1) - 2m -1}{\g^2 - 1} - 2 m \z + \frac{i \g (\z - 2)\e[k,v_1,v_2,y] }{ (\g^2 - 1)} \right) \frac{I_{m}(k \cdot y)}{(k \cdot y)^{m}} \right] \,, \label{eq:2PMeikPMint}
\end{align}
\endgroup
where the factor $i \e[k,v_1,v_2,y]$ can be traded for the derivative operator $\e^\m [v_1, v_2, y] \frac{\partial}{\partial b^\m}$. Imposing the additional constraint\footnote{This additional constraint conforms to the usage of ``aligned spin'' in the literature, where the orbital angular momentum is also aligned with the spin direction.} $y \cdot b = 0$ simplifies the expression further and yields,
\begin{align}
    \chi^{+-} &= \frac{(q_1 q_2)^2}{64 m_1 \sqrt{\g^2 - 1}} \left( \frac{1}{\pi (b^2 - y^2)^{3/2}} \left[ b^2 + \frac{\g^2 (1 - \z) + \z}{\g^2 - 1} \, y^2 \right] \right. \nn
    \\ &\left. \phantom{asdf \frac{1 - (1 + 2 \z) (\frac{y^2}{b^2})}{\pi \sqrt{b^2}}} + \frac{(\z - 2)\g}{\pi (\g^2 - 1)} \e^\m [v_1, v_2, y] \frac{\partial}{\partial b^\m} \frac{1}{\sqrt{b^2 -y^2}} \right) \,,
\end{align}
which has a singularity structure $(b^2 - y^2)^{- 3/2} = (b^2 - (a_1 + a_2)^2)^{- 3/2}$ that was not visible in the original perturbative spin expansion. The full aligned-spin 2PL eikonal is
\begin{align}
    \chi_{(2,\text{aligned})} &= \frac{(q_1 q_2)^2 \left( b^2 - \frac{(\z - 2)\g}{(\g^2 - 1)} \, \e[b, v_1, v_2, y] + \frac{\g^2 (1 - \z) + \z}{\g^2 - 1} \, y^2 \right)}{32 \pi m_1 \sqrt{\g^2 - 1} \, (b^2 - y^2)^{3/2}} + \left( 1 \leftrightarrow 2 \right) \,, \label{eq:one-loop_eik_as_full}
\end{align}
where symmetrisation is implemented by $\{ m_1 \to m_2 \,,\, v_1^\m \leftrightarrow v_2^\m \,,\, b^\m \to - b^\m \,,\, \z \to 1 - \z \}$.

We remark that the singularity is still present in the spinless probe limit $\z \to 0$, and since spin-dependence of the eikonal integrand \eqref{eq:one-loop_eik_integrand_def} enters only through the exponential factors $e^{+ k \cdot y} e^{- 2 k_3 \cdot y}$ in this limit, the singularity structure $(b^2 - y^2)^{-3/2}$ seems to be a consequence of the ``Newman-Janis shift'' of the integrand, which shifts the displacement between the two worldlines by an imaginary spin sum vector $\pm i y^\m = \mp i (a_1^\m + a_2^\m)$. Note that similar singularity structures at 2PM aligned-spin scattering were reported in the gravitational case; the spinning-spinless eikonal has the form $(b^2 - a^2)^{-3/2}$~\cite{Aoude:2022thd},\footnote{JWK would like to thank Kays Haddad for bringing this reference to attention.} and Kerr-(spinning) probe eikonal has the form $a_p^k (b^2 - a_b^2)^{-3/2-k}$ where $a_b$ is the spin parameter of the Kerr background and $a_p$ is the probe spin ($k \le 2$)~\cite{Damgaard:2022jem}\footnote{Ref.~\cite{Damgaard:2022jem} reports the scattering angle which scales as $a_p^k (b^2 - a_b^2)^{-5/2-k} \sim \partial_b [a_p^k (b^2 - a_b^2)^{-3/2-k}]$.} which can be viewed as an artifact of expanding the singularity $(b^2 - (a_b + a_p)^2)^{-3/2}$. If the ``Newman-Janis shift'' of the integrand persists at higher loop orders, we can conjecture that the singularity structure of the spinless probe scattering $(b^2 - a_b^2 )^{-3n/2}$ from $n$-loop contributions reported by ref.~\cite{Damgaard:2022jem} generalises to the singularity structure $(b^2 - (a_b + a_p)^2)^{-3n/2}$. As remarked when comparing twistor worldline Compton amplitudes with that of higher spin gauge symmetry~\cite{Cangemi:2023ysz}, it would also be interesting to check whether same singularity structures appear in the eikonal when the Compton amplitude has exponential dependence on spin, a feature that is also shared by the Compton amplitude construction in ref.~\cite{Bjerrum-Bohr:2023iey}.

We also consider axial scattering $y^\m \propto b^\m$, which is independent of the sign of $y \cdot b$ because the Fourier integrand \eqref{eq:2PMeikPMint} contains only even powers of $k \cdot y$. The $i\e[k,v_1,v_2,y]$ contribution drops out due to the condition $y^\m \propto b^\m$, and the Fourier transform \eqref{eq:2PMeikPM} evaluates to
\begin{align}
    \chi^{+-} &= \frac{(q_1 q_2)^2 \, \sqrt{b^2}}{32 \pi^2 m_1 (\g^2 - 1)^{3/2}} \left[ \frac{\g^2 ( \z - 1) - \z }{b^2}  K\left(- \frac{y^2}{b^2}\right) - \frac{ \g^2 (\z - 2) - ( \z - 1) }{b^2 + y^2}  E\left( - \frac{y^2}{b^2} \right) \right] \,,
\end{align}
where $K(x)$ and $E(x)$ are the complete elliptic integrals of the first and second kind. The full result is
\begin{align}
    \chi_{(2,\text{axial})} &= \frac{(q_1 q_2)^2 \, \sqrt{b^2}}{16 \pi^2 m_1 (\g^2 - 1)^{3/2}} \left[ \frac{\g^2 ( \z - 1) - \z }{b^2}  K\left(- \frac{y^2}{b^2}\right) - \frac{ \g^2 (\z - 2) - ( \z - 1) }{b^2 + y^2}  E\left( - \frac{y^2}{b^2} \right) \right] \nn
    \\ &\phantom{=asdf} + (1 \leftrightarrow 2) \,, \label{eq:one-loop_eik_ax_full}
\end{align}
which, unlike the aligned-spin case \eqref{eq:one-loop_eik_as_full}, develops a \emph{logarithmic} singularity at $b^2 = 0$ and another singularity at the unphysical impact parameter $b^2 = - y^2$.  The results \eqref{eq:one-loop_eik_as_full} and \eqref{eq:one-loop_eik_ax_full} can be reproduced from the full eikonal given in the ancillary file \texttt{eikonaldata.dat.m} by taking the corresponding configurations and resumming the series expansion in $y^2 / b^2$.

Before ending this section, we remark that the LSZ-like formulae \eqref{eq:vkick} and \eqref{eq:skick} can also be applied to 2PL scattering observables, where retarded worldline propagators are used instead~\cite{Jakobsen:2022psy}. The 2PL observables can be separated into the \emph{eikonal part} (the same diagrams with symmetric worldline $i0^+$ prescription) and the \emph{causality cut part} (the contributions from changing the worldline $i0^+$ prescription), where the eikonal part computes $\{ \chi_{(2)} , O \}$ and the causality cut part computes $\frac{1}{2} \{ \chi_{(1)} , \{ \chi_{(1)} , O \} \}$. This computation serves as a consistency check of the calculations in section~\ref{sec:2PL spin kick}. The separation of the observables into the eikonal part and the causality cut part can be shown to be a more general phenomenon that holds in Hamiltonian worldline models~\cite{Kim:causality-cut}.

\section{Discussion} \label{sec:discussion}

The (ambi-)twistor model for electromagnetically interacting spinning particles was studied in this manuscript, which has the advantage that it is one of the simplest descriptions of charged spinning particles where spin effects can be tracked to arbitrarily high orders. Using (dynamical) Newman-Janis shift as the only input for generating all-orders-in-spin interactions, it was found that the spin effects can be resummed to simple expressions in special kinematic configurations; in the aligned-spin case \eqref{eq:one-loop_eik_as_full} and in the axial scattering case \eqref{eq:one-loop_eik_ax_full}. Also, the model was used to confirm the interpretation, up to the 2PL order, of the classical eikonal as the generator of canonical transformations that map the incoming scattering states to outgoing scattering states.

Despite the disparities between electromagnetic and gravitational interactions, the similarities between the singularity structures of the spin-resummed electromagnetic eikonal \eqref{eq:one-loop_eik_as_full}, $\chi_{(2)} \propto (b^2 - (a_1 + a_2)^2)^{-3/2}$, and the probe limit Kerr scattering reported by ref.~\cite{Damgaard:2022jem}, $\th_{(2)} \sim \partial_b \chi_{(2)} \propto \partial_b [(b^2 - a_b^2)^{-3/2}]$, provides further evidence that using the total spin length vector $a_+^\m = a_1^\m + a_2^\m$ as the spin parameter of the effective Kerr metric\textemdash an ansatz motivated by leading order PN Hamiltonian results~\cite{Vines:2016qwa}\textemdash in the effective-one-body approach~\cite{Khalil:2023kep} is the preferable choice for resumming spin effects.\footnote{This is not the unique choice considered in the literature. A comparison of different choices for the spin parameter of the effective Kerr metric can be found in ref.~\cite{Khalil:2020mmr}.} On the other hand, one-loop results only correspond to leading order effects in the mass-ratio expansion~\cite{Vines:2018gqi}, therefore the singularity structures resembling that of the background-probe calculation~\cite{Damgaard:2022jem} could be a coincidence of the leading order mass-ratio expansion. Whether novel singularity structures arise at NLO in mass-ratio expansion will only be answered by pushing the computations to two-loops and higher orders, and may point us to new directions in resumming spin effects. Of course, studying the gravitationally interacting case is also necessary to confirm that such singularity structures are also present in gravitating binary black holes.

When viewing the classical eikonal as the generator of canonical transformations, it would be interesting to understand what it means to analytically continue the scattering generator to bound dynamics. The boundary-to-bound map for the radial action~\cite{Kalin:2019rwq,Kalin:2019inp} suggests that the continuation is a finite time-evolution generator that advances the system by one radial period, e.g. the periastron passing is sent to the next periastron passing. If this interpretation is correct, then we may argue that separability of the Hamilton-Jacobi equations is not necessary for the existence of the bound orbit counterpart of the classical eikonal, although its determination by analytic methods may only be possible when Hamilton-Jacobi equations are separable~\cite{Gonzo:2023goe,Gonzo:2024zxo}.

Apart from the obvious future direction\textemdash massive twistor worldline in gravitational fields\textemdash there are several other directions that would be interesting to expand upon. One future research direction would be to explore whether recent attempts to resum analytic results for gravitational scattering of spinning black holes~\cite{Rettegno:2023ghr, Buonanno:2024vkx} can be improved using the singularity structures of \eqref{eq:one-loop_eik_as_full} and their conjectured generalisation to higher loops $\chi_{(n)} \propto (b^2 - (a_1 + a_2)^2)^{-3n/2}$.

Another direction would be making the (WQFT approach to the) model live up to its name; \emph{quantisation}. Since the twistor model has a simple set of constraints, the standard BRST-BFV methods should be applicable. For small values of quantised spin, say $1/2$ or $1$, we expect the results to agree with the standard QFT of massive spinning fields. The attempt to quantise the model for higher spin may shed new light on the complication with massive higher spin fields. Comparison of the approach with chiral models for massive higher spin fields~\cite{Ochirov:2022nqz} would also be an interesting study.

While the fundamental variables of our model are twistors, the physical observables (and the classical eikonal) were given entirely in terms of the gauge invariant $(x,y,p)$ variables. Some intermediate steps of the computations, such as the ones in appendix~\ref{app:iteration}, tend to be quite lengthy and not particularly illuminating. 
The computations may become vastly simplified when full advantage of the twistor variables is taken. To do so, it would be crucial to use massless twistor variables 
for the photon fields as well. Bailey's twistor propagator \cite{Bailey:1985a}, and Guevara's holomorphic classical limit \cite{Guevara:2017csg} and twistor reconstruction \cite{Guevara:2021yud} could provide clues for further progress.  

We remark that iterated action of the classical eikonal can be understood as causality cuts, which computes contributions associated to changing the $i0^+$ prescription of the worldline propagators from time-symmetric to retarded; in the WQFT formalism scattering observables are computed using retarded propagators~\cite{Jakobsen:2022psy}, and changing the $i0^+$ prescription of the worldline propagators from retarded to symmetric generates (nested) Poisson brackets which reorganises the scattering observable $\D O$ as the action of the scattering generator $e^{\{ \chi, \bullet \}} O$~\cite{Kim:causality-cut}. 
A direct consequence is that the longitudinal impulse at 2PL order is related to the $i0^+$ prescription of the worldline propagators, which could be an interesting observation for understanding the $i0^+$ prescription affecting the definition of the impact parameter used in one-loop waveform results~\cite{Brandhuber:2023hhy,Herderschee:2023fxh,Elkhidir:2023dco,Georgoudis:2023lgf,Caron-Huot:2023vxl,Bini:2023fiz,Georgoudis:2023eke}.

Finally, it would be interesting to generalise the concept of the classical eikonal to massless fields. 
Such an extended eikonal would place massive particles and massless fields on an equal footing, 
and may help us clarify to what extent we can identify the eikonal as the classical shadow of the quantum $S$-matrix.

\acknowledgments
JWK would like to thank Alessandra Buonanno, Kays Haddad, Gustav Jakobsen, Gustav Mogull, Raj Patil, Harald Pfeiffer, Lorenzo Pompili, and Jan Steinhoff for valuable discussions.
SL is grateful to Hojin Lee, Kanghoon Lee, Sungjay Lee, Chia-Hsien Shen, Tianheng Wang and Piljin Yi for discussions. 
SL would also like to thank Korea Institute for Advanced Study for hospitality where a large part of this work was done.
The work of SL is supported by the National Research Foundation of Korea grant NRF-2019R1A2C2084608.

\newpage 
\appendix

\section{Conventions} \label{sec:conventions}

\subsection*{Vector}

Flat metric and Levi-Civita tensor, 
\begin{align}
    \eta_{\mu\nu} = \mathrm{diag}(-,+,+,+) \,, 
    \quad 
    \varepsilon_{0123} = +1 \,.
\end{align}
Electromagnetism without spin,  
\begin{align}
    \partial^\mu F_{\mu\nu} = - J_\nu \,,
    \quad 
    m \frac{du^\mu}{d\tau} = q F^{\mu\nu} u_\nu \,.
\end{align}
Hodge star acting on a two-form,
\begin{align}
    (*F)_{\mu\nu} = \frac{1}{2} \varepsilon_{\mu\nu\rho\sigma} F^{\rho\sigma} \,.
\end{align}
Self-dual and anti-self-dual parts of a two-form, 
\begin{align}
    F^\pm = \frac{1}{2} (F \mp i \! * \! F) 
    \quad \Longrightarrow \quad 
    * (F^\pm) = \pm i (F^\pm) \,.
\end{align}
We may also use $\tilde{F}_{\mu\nu} = (*F)_{\mu\nu}$. 
If we define $A^\pm$ and $\tilde{A}$ by $F^\pm = dA^\pm$ and $\tilde{F} = d \tilde{A}$, 
\begin{align}
    A_\m = A^+_\m + A^-_\m \,,
    \quad 
    \tilde{A}_\m = i ( A^+_\m - A^-_\m ) \,.
\end{align}

\subsection*{Spinor}


We follow the conventions of ref.~\cite{Kim:2021rda} to a large extent, where $|\l \rangle$ spinors are associated to \emph{incoming} negative helicity states. 
An important difference is that we define 
\begin{align}
    v_{\a\dot\a} := v_\m \s^\m_{\a \dot\a} \,,
    \quad 
    v^\mu = -\frac{1}{2} (\bar{\s}^\mu)^{\da\a} v_{\a\da} 
\end{align}
for \emph{all} vectorial quantities, nullifying the exception for $x^\mu$ made in ref.~\cite{Kim:2021rda}. 
To compare with references where the metric $\eta^{\mu\nu}$ and/or 
the Levi-Civita tensor $\varepsilon_{\mu\nu\rho\sigma}$ carry the opposite sign ({\it e.g.} ref.~\cite{Chung:2018kqs}), 
an invariant way to express conversion between spinor and Lorentz indices is to introduce the parameters $\k_0 = \eta_{00}$ and $\k_1 = \ve_{0123}$:
\begin{align}
    p_{\a \dot\a} &= p_\m \s^\m_{\a \dot\a} = \left\{
    \begin{aligned}
        & \k_0 |p\rangle_\a [p|_{\dot\a} && p^2 = 0
        \\ & \k_0 |p^I\rangle_\a [p_I|_{\dot\a} && p^2 = \k_0 m^2
    \end{aligned} \right.
    \\ v_{\a \dot\a} w^{\dot\a \a} &= 2 \k_0 (v \cdot w)
    \\ 
    ( \s^\m \bar{\s}^{\n} \s^\l \bar{\s}^{\s} )_{\a}{}^{\a} &= 2 ( \eta^{\m\n} \eta^{\l\s} - \eta^{\m\l} \eta^{\n\s} + \eta^{\m\s} \eta^{\n\l} - i \k_1 \ve^{\m\n\l\s})
    \label{eq:ssss-iden}
\end{align}
The invariant tensor satisfies the complex conjugation relation
\begin{align}
    \e^{\dot\a \dot\g} \e^{\b \delta} \left[ \s^\m_{\g \dot\delta} \right]^\ast = \bar{\s}^{\m \dot\a \b}
\end{align}
which is useful for evaluating complex conjugation of vectors.

\subsection*{Twistor}

The two major differences from ref.~\cite{Kim:2021rda} are 
\begin{align}
    (x^{\da\a})_\mathrm{here} = (-2) (x^{\da\a})_\mathrm{there} \,,
    \quad 
    (\mu^{\dot\a I}, \bar{\mu}_{I}{}^{\a})_\mathrm{here} = - (\mu^{\dot\a I}, \bar{\mu}_{I}{}^{\a})_\mathrm{there} \,.
\end{align}
These changes propagate to all other equations. For example, 
the incidence relations read 
\begin{align}
    \m^{\dot\a I} = \frac{1}{2} z^{\dot\a \b} \l_{\b}{}^{I} \,,
    \quad \bar{\m}_{I}{}^{\a} = \frac{1}{2} \bar{\l}_{I \dot\b} \bar{z}^{\dot\b\a} \,,
    \label{eq:incidence_new}
\end{align}
where we define the complex conjugate relations as 
\begin{align}
    \bar{z}^\m = \left[z^\m \right]^\ast \Rightarrow \bar{z}^{\dot\a \b} = \left[ z^{\dot\b \a}\right]^\ast \,,
    \quad 
    \bar{\m}_{I}{}^{\a} = \left[ \m^{\dot\a I} \right]^\ast \,,\, \bar{\l}_{I \dot\a} = \left[ \l_\a{}^{I} \right]^\ast \,.
\end{align}

The defining Poisson brackets are 
\begin{align}
\begin{gathered}
    \{ x^\m, p_\n \} = \delta^\m_\n \;\; \Rightarrow \;\; \{ x^{\dot\a \a} , p_{\b \dot\b} \} = - 2 \delta^{\dot\a}_{\dot\b} \delta^\a_\b \,,
    \\ 
    \{ \bar{\m}_{I}{}^{\a} , \l_\b{}^{J} \} = \delta^\a_\b \delta_I^J \,,\quad \{ \m^{\dot\a I} , \bar{\l}_{J \dot\b} \} = \delta_{\dot\b}^{\dot\a} \delta^I_J \,.
\end{gathered} \label{eq:def_PB_new}
\end{align}
The consistency of the defining brackets can be confirmed from the relations
\begin{align}
    x^{\dot\a \a} = \frac{z^{\dot\a \a} + \bar{z}^{\dot\a \a}}{2} \,,\quad  
    p_{\a \dot\a} = - \l_\a{}^{I} \bar{\l}_{I \dot\a} \,. \label{eq:twistor2xvar}
\end{align}

To determine the relation between $y^\m$ and the spin-length vector $a^\m = S^\m /m$, we note the Poisson brackets of the rotation generators
\begin{align}
    \{ J^{23} , J^{31} \} = J^{12} 
    \;\; \Leftrightarrow \;\;
    \{ J^1 , J^2 \} = J^3
\end{align}
and leverage the calculation to demand that
\begin{align}
    \{ S^\m , S^\n \}_\ast = (- \k_1) \ve^{\a \m\n\l} \frac{(- p_\a)}{m} S_\l 
    \;\; \Leftrightarrow \;\; 
    \{ a^\m , a^\n \}_\ast = \frac{\k_1}{m^2} \ve^{\a \m\n\l} p_\a a_\l
\end{align}
where $\{ \bullet, \bullet \}_\ast$ is the Dirac bracket and $\k_1 = \ve_{0123}$. 
The end result is the standard convention for the orientation of $a^\mu$: 
\begin{align}
    \{ a^1 , a^2 \}_* = + \frac{a^3}{m} \,. 
\end{align} 
On the twistor side, from the Poisson brackets we find
\begin{align}
    \{ z^\m , \bar{z}^\n \} &= \frac{-2i}{m} \left[ y^\m v^\n - \eta^{\m\n} (y \cdot v) + v^\m y^\m + i \k_1 \ve^{\m\n\a\b} y_\a v_\b \right]
\end{align}
which implies 
\begin{align}
    \{ y^\m , y^\n \} = \frac{\k_1}{m} \ve^{\m\n\a\b} y_\a v_\b \;\; \Rightarrow \;\; \{ y^1, y^2 \} = - \k_1 \ve^{1230} \frac{y^3}{m} = - \frac{y^3}{m} \,.
\end{align}
Thus we have to set $y^\mu = - a^\mu$. 
%


\section{List of integrals}
\subsection{Master one-loop integral} \label{app:MIs}
We compute the Euclidean loop integral $(k \cdot v = 0)$
\begin{align*}
    \int \frac{ d^D \ell_E \,\, e^{2 \ell_E \cdot a}}{(\ell_E^2)^{\l_1} [(k - \ell_E)^2]^{\l_2} (2 v \cdot \ell_E - i 0^+)^{\l_3} } \,,
\end{align*}
using the identities
\begin{align}
\begin{aligned}
    \frac{1}{\a^{\l}} &= \frac{1}{\G(\l)} \int_0^\infty dt \, t^{\l - 1} e^{- \a t} \,,
    \\ \frac{1}{(\a - i0^+)^\l} &= \frac{i^\l}{\G (\l)} \int_0^\infty dt \, t^{\l - 1} e^{- i (\a - i0^+) t} \,.
\end{aligned}
\end{align}
After substitution, we have the integral
\begin{align*}
    &\frac{i^{\l_3}}{\G(\l_1)\G(\l_2)\G(\l_3)} \int_0^\infty dt_1 dt_2 dt_3 \, t_1^{\l_1 -1} t_2^{\l_2 - 1} t_3^{\l_3 -1} \exp \left[ - \frac{t_1 t_2}{t_1 + t_2} k^2 - \frac{t_3^2}{t_1 + t_2} v^2 \right]
    \\ &\phantom{asdfas} \times \exp \left[ \frac{2 t_2}{t_1 + t_2} (k \cdot a) + \frac{- 2 i t_3}{t_1 + t_2} (v \cdot a) + \frac{a^2}{t_1 + t_2} \right] \int_{\ell_E'} e^{-(t_1 + t_2) (\ell_E')^2}
\end{align*}
where $\ell_E'$ is the shifted loop integration variable. We perform the Gaussian integral and expand the exponential of the second line. Evaluating the gamma function and beta function integrals, we get
\begin{align}
    & \int \frac{ \m^{2\e} d^D \ell_E \,\, e^{2 \ell_E \cdot a}}{(\ell_E^2)^{\l_1} [(k - \ell_E)^2]^{\l_2} (2 v \cdot \ell_E - i 0^+)^{\l_3} } \nn
    \\ &= \frac{i^{\l_3} \pi^{D/2} \m^{2\e}}{2 \G(\l_1) \G(\l_2) \G(\l_3) (k^2)^{\l_1 + \l_2 + \frac{\l_3}{2} - \frac{D}{2}} (v^2)^{\frac{\l_3}{2}}} \nn
    \\ &\phantom{=} \times \sum_{l,m,n = 0}^\infty \frac{(2 k \cdot a)^l \left( - 2 i (v \cdot a) \sqrt{\frac{k^2}{v^2}} \right)^m (k^2 a^2)^n}{l! m! n!} \frac{\G(\l_1 + \l_2 + \frac{\l_3}{2} - \frac{m}{2} - n - \frac{D}{2}) \G (\frac{\l_3 + m}{2})}{\G(D - \l_1 - \l_2 - \l_3 + l + m + 2n)} \nn
    \\ &\phantom{=asdfasdf} \times \G(\frac{D}{2} - \l_2 - \frac{\l_3}{2} + \frac{m}{2} + n) \, \G(\frac{D}{2} - \l_1 - \frac{\l_3}{2} + l + \frac{m}{2} + n) \,, \label{eq:MI}
\end{align}
where $\m^{2\e}$ is the mass scale required for dimensional regularisation $D = 3 - 2\e$. It is easy to verify that for $\l_1 = 0$ or $\l_2 = 0$ the integral vanishes, and for $\l_3 = 0$ that the integral localises onto $m = 0$.\footnote{The value for $\l_3 = 0$ should be understood as a limiting value $\l_3 \to 0$, where $\frac{\G(\l_3/2)}{\G(\l_3)} \to 2$.} Setting $a = 0$ we only keep $l = m = n = 0$ of the sum, and recover the Euclidean version of (10.25) of ref.~\cite{Smirnov:2012gma}. The divergence of the integral for non-positive integral values of $(\l_1 + \l_2 + \frac{\l_3}{2} - \frac{m}{2} - n - \frac{D}{2}) \in \mathbb{Z}^{\le 0}$ is harmless since the result formally becomes non-negative integral powers of $k^2$, which vanishes under the impact parameter space integral $\int_k e^{i k \cdot b}$ for $b^\m \neq 0$. The master integral \eqref{eq:MI} can be viewed as a \emph{tensor integral generating function}, e.g. the vector integral can be evaluated as
\begin{align}
    \int \frac{ d^D \ell_E \,\, \ell_E^\mu \, e^{2 \ell_E \cdot a}}{(\ell_E^2)^{\l_1} [(k - \ell_E)^2]^{\l_2} (2 v \cdot \ell_E - i 0^+)^{\l_3} } = \frac{1}{2} \frac{\partial}{\partial a_\m} \int \frac{ d^D \ell_E \,\, e^{2 \ell_E \cdot a}}{(\ell_E^2)^{\l_1} [(k - \ell_E)^2]^{\l_2} (2 v \cdot \ell_E - i 0^+)^{\l_3} } \,. \nn
\end{align}
This can be used to check consistency of \eqref{eq:MI}, e.g. $v^\m \frac{\partial}{\partial a^\m} \Leftrightarrow (\l_3 \to \l_3 - 1)$. Such consistency relations could be used to bootstrap tensor integral generating functions~\cite{Feng:2022hyg}.

In all the integrals listed in this appendix, we only keep the non-analytic terms in $k^2$ and drop dimensional regularisation artefacts ($\CO(\e^{-1})$ and $\CO(\e)$). All $\log(k^2)$-dependent terms of the integrals vanish for time-symmetric $i0^+$ prescription, which is equivalent to taking the real part of the integrals.

\paragraph{Special cases: Scalar integrals}
$\l_1 = \l_2 = 1$, $\l_3 = 2$, and $D=3-2\e$ with extra $k^2$.
\begin{align}
    & k^2 \int \frac{ d^D \ell_E \,\, \mu^{2\e} e^{2 \ell_E \cdot a}}{(\ell_E^2) (k - \ell_E)^2 (2 v \cdot \ell_E - i 0^+)^{2} } \nn
    \\ &= \frac{\pi^{3}}{2 (k^2)^{1/2} v^2} \sum_{l,m,n = 0}^\infty \frac{\G(l+m+n-\frac{1}{2}) (2 k \cdot a)^l \left( k^2 \frac{(v \cdot a)^2}{v^2} \right)^m (- k^2 a^2)^n}{\G(l+1) \G (m + \frac{1}{2}) \G (n+1) \G(l+2m+2n - 1)} \nn
    \\ &\phantom{=} + \frac{i \pi^2 (v \cdot a) \log (k^2)}{2 (v^2)^{3/2} } \sum_{l,m,n=0}^\infty \frac{\G(l+m+n) (2 k \cdot a)^l \left( k^2 \frac{(v \cdot a)^2}{v^2} \right)^m (- k^2 a^2)^n}{\G(l+1) \G (m + 1) \G (n+1) \G(l+2m+2n)} \nn
    \\ &= \frac{\pi^{5/2}}{2 (k^2)^{1/2} v^2} \sum_{l,m = 0}^\infty \frac{\G(l+m-\frac{1}{2}) (2 k \cdot a)^l (- k^2 a^2)^m \, {}_2F_1 (1,-m;\frac{1}{2};\frac{(v \cdot a)^2}{v^2 a^2})}{\G(l+1) \G (m+1) \G(l+2m - 1)} \nn
    \\ &\phantom{=} + \frac{i \pi^2 (v \cdot a) \log (k^2)}{2 (v^2)^{3/2} } \sum_{l,n=0}^\infty \frac{\G(l+n) (2 k \cdot a)^l (- k^2 a^2)^n  \left( 1 - \frac{(v \cdot a)^2}{a^2 v^2} \right)^n}{\G(l+1) \G (n+1) \G(l+2n)} \,. \label{eq:MI_c1}
\end{align}
$\l_1 = \l_2 = \l_3 = 1$ and $D=3-2\e$ with extra $k^2$.
\begin{align}
    & k^2 \int \frac{ d^D \ell_E \,\, \mu^{2\e} e^{2 \ell_E \cdot a}}{(\ell_E^2) (k - \ell_E)^2 (2 v \cdot \ell_E - i 0^+) } \nn
    \\ &= \frac{\pi^{3} (v \cdot a) (k^2)^{1/2}}{2 v^2} \sum_{l,m,n = 0}^\infty \frac{\G(l+m+n+\frac{1}{2}) (2 k \cdot a)^l \left( k^2 \frac{(v \cdot a)^2}{v^2} \right)^m (- k^2 a^2)^n}{\G(l+1) \G (m + \frac{3}{2}) \G (n+1) \G(l+2m+2n + 1)} \nn
    \\ &\phantom{=} + \frac{i \pi^2 \log (k^2)}{2 (v^2)^{1/2} } \sum_{l,m,n=0}^\infty \frac{\G(l+m+n) (2 k \cdot a)^l \left( k^2 \frac{(v \cdot a)^2}{v^2} \right)^m (- k^2 a^2)^n}{\G(l+1) \G (m + 1) \G (n+1) \G(l+2m+2n)} \nn
    \\ &= \frac{\pi^{5/2} (v \cdot a) (k^2)^{1/2}}{v^2} \sum_{l,m = 0}^\infty \frac{\G(l+m+\frac{1}{2}) (2 k \cdot a)^l (- k^2 a^2)^m \, {}_2F_1(1,-m;\frac{3}{2};\frac{(v \cdot a)^2}{v^2 a^2})}{\G(l+1) \G (m+1) \G(l+2m + 1)} \nn
    \\ &\phantom{=} + \frac{i \pi^2 \log (k^2)}{2 (v^2)^{1/2} } \sum_{l,n=0}^\infty \frac{\G(l+n) (2 k \cdot a)^l (- k^2 a^2)^n  \left( 1 - \frac{(v \cdot a)^2}{a^2 v^2} \right)^n}{\G(l+1) \G (n+1) \G(l+2n)} \,. \label{eq:MI_c2}
\end{align}
$\l_1 = \l_2 = 1$, $\l_3 = 0$, and $D=3$.
\begin{align}
    \int \frac{ d^D \ell_E \,\, e^{2 \ell_E \cdot a}}{(\ell_E^2) (k - \ell_E)^2 } &= \frac{\pi^{5/2}}{(k^2)^{1/2}} \sum_{l,n = 0}^\infty \frac{\G(l+n+\frac{1}{2}) (2 k \cdot a)^l (- k^2 a^2)^n}{\G(l+1) \G (n+1) \G(l+2n + 1)} \,. \label{eq:MI_c3}
\end{align}
$\l_1 = \l_2 = 1$, $\l_3 = -1$, and $D=3$.
\begin{align}
    \int \frac{ d^D \ell_E \,\, (2 v \cdot \ell_E) \, e^{2 \ell_E \cdot a}}{(\ell_E^2) (k - \ell_E)^2 } &= - 2 \pi^{5/2} (v \cdot a) (k^2)^{1/2} \sum_{l,m = 0}^\infty \frac{\G(l+m+\frac{3}{2}) (2 k \cdot a)^l (- k^2 a^2)^m}{\G(l+1) \G (m+1) \G(l+2m + 3)} \,. \label{eq:MI_c4}
\end{align}

\paragraph{Special cases: Vector integrals}
Assume $f \cdot k = f \cdot v = 0$. $\l_1 = \l_2 = \l_3 = 1$ and $D=3-2\e$.
\begin{align}
    & \int \frac{ d^D \ell_E \,\, \mu^{2\e} (f \cdot \ell_E) \, e^{2 \ell_E \cdot a}}{(\ell_E^2) (k - \ell_E)^2 (2 v \cdot \ell_E - i 0^+) } \nn
    \\ &= - \frac{\pi^{3} (v \cdot a) (f \cdot a) (k^2)^{1/2}}{2 v^2} \sum_{l,m,n = 0}^\infty \frac{\G(l+m+n+\frac{3}{2}) (2 k \cdot a)^l \left( k^2 \frac{(v \cdot a)^2}{v^2} \right)^m (- k^2 a^2)^n}{\G(l+1) \G (m + \frac{3}{2}) \G (n+1) \G(l+2m+2n + 3)} \nn
    \\ &\phantom{=} - \frac{i \pi^2 (f \cdot a) \log (k^2)}{2 (v^2)^{1/2} } \sum_{l,m,n=0}^\infty \frac{\G(l+m+n+1) (2 k \cdot a)^l \left( k^2 \frac{(v \cdot a)^2}{v^2} \right)^m (- k^2 a^2)^n}{\G(l+1) \G (m + 1) \G (n+1) \G(l+2m+2n+2)} \nn
    \\ &= - \frac{\pi^{5/2} (v \cdot a) (f \cdot a) (k^2)^{1/2}}{v^2} \sum_{l,m = 0}^\infty \frac{\G(l+m+\frac{3}{2}) (2 k \cdot a)^l (- k^2 a^2)^m \, {}_2F_1(1,-m;\frac{3}{2};\frac{(v \cdot a)^2}{v^2 a^2})}{\G(l+1) \G (m+1) \G(l+2m + 3)} \nn
    \\ &\phantom{=} - \frac{i \pi^2 (f \cdot a) \log (k^2)}{2 (v^2)^{1/2} } \sum_{l,n=0}^\infty \frac{\G(l+n+1) (2 k \cdot a)^l (- k^2 a^2)^n  \left( 1 - \frac{(v \cdot a)^2}{a^2 v^2} \right)^n}{\G(l+1) \G (n+1) \G(l+2n+2)} \,. \label{eq:MI_c5}
\end{align}
$\l_1 = \l_2 = 1$, $\l_3 = 2$, and $D=3-2\e$.
\begin{align}
    & \int \frac{ d^D \ell_E \,\, \mu^{2\e} (f \cdot \ell_E) \, e^{2 \ell_E \cdot a}}{(\ell_E^2) (k - \ell_E)^2 (2 v \cdot \ell_E - i 0^+)^{2} } \nn
    \\ &= - \frac{\pi^{3} (f \cdot a)}{2 (k^2)^{1/2} v^2} \sum_{l,m,n = 0}^\infty \frac{\G(l+m+n+\frac{3}{2}) (2 k \cdot a)^l \left( k^2 \frac{(v \cdot a)^2}{v^2} \right)^m (- k^2 a^2)^n}{\G(l+1) \G (m + \frac{1}{2}) \G (n+1) \G(l+2m+2n + 3)} \nn
    \\ &\phantom{=} - \frac{i \pi^2 (f \cdot a) (v \cdot a) \log (k^2)}{2 (v^2)^{3/2} } \sum_{l,m,n=0}^\infty \frac{\G(l+m+n+1) (2 k \cdot a)^l \left( k^2 \frac{(v \cdot a)^2}{v^2} \right)^m (- k^2 a^2)^n}{\G(l+1) \G (m + 1) \G (n+1) \G(l+2m+2n+2)} \nn
    \\ &= - \frac{\pi^{5/2} (f \cdot a)}{2(k^2)^{1/2} v^2} \sum_{l,m = 0}^\infty \frac{\G(l+m+\frac{1}{2}) (2 k \cdot a)^l (- k^2 a^2)^m \, {}_2F_1(1,-m;\frac{1}{2};\frac{(v \cdot a)^2}{v^2 a^2})}{\G(l+1) \G (m+1) \G(l+2m + 1)} \nn
    \\ &\phantom{=} - \frac{i \pi^2 (f \cdot a) (v \cdot a) \log (k^2)}{2 (v^2)^{3/2} } \sum_{l,n=0}^\infty \frac{\G(l+n+1) (2 k \cdot a)^l (- k^2 a^2)^n  \left( 1 - \frac{(v \cdot a)^2}{a^2 v^2} \right)^n}{\G(l+1) \G (n+1) \G(l+2n+2)} \,. \label{eq:MI_c6}
\end{align}

\subsection{Fourier transform integral} \label{app:FTs}
The Fourier transform to impact parameter space is given as
\begin{align}
    \int \frac{d^D k_E}{(2 \pi)^D} \frac{e^{i (k_E \cdot b)}}{[k_E^2 \mp i 0^+]^\l} &= \frac{\G (\frac{D}{2} - \l) }{2^{2\l} \pi^{\frac{D}{2}} \G (\l)} \frac{1}{ \left( b^2 \pm i0^+ \right)^{\frac{D}{2} - \l} } \,, \label{eq:FT_MI}
\end{align}
where we assumed $b^2 \in \IR$ and included $i0^+$ prescription for convergence. The $i0^+$ prescription can be dropped since there is no branch cut ambiguity for Euclidean signature.

Fourier transforms with numerators can be evaluated using differentiation; $k_E^\m \leftrightarrow - i \frac{\partial}{\partial b_\m}$. Repeated numerator factors can be computed as directional derivatives, i.e.
\begin{align}
\begin{aligned}
    \int \frac{d^D k_E}{(2 \pi)^D} \frac{(k_E \cdot a)^l e^{i (k_E \cdot b)}}{[k_E^2]^\l} &= l! \left. \int \frac{d^D k_E}{(2 \pi)^D} \frac{e^{i (k_E \cdot [b-ia])}}{[k_E^2]^\l} \right|_{\CO(a^l)}
    \\ &= \left. \frac{l! \G (\frac{D}{2} - \l) }{2^{2\l} \pi^{\frac{D}{2}} \G (\l)} \frac{1}{ \left[ (b-ia)^2 \right]^{\frac{D}{2} - \l} } \right|_{\CO(a^l)}
    \\ &= \frac{l! \G (\frac{D}{2} - \l) }{2^{2\l} \pi^{\frac{D}{2}} \G (\l)} \frac{i^l C_l^{(\frac{D}{2} - \l)}(\frac{(a \cdot b)}{(a^2 b^2)^{1/2}}) (a^2)^{l/2}}{ \left( b^2 \right)^{\frac{D}{2} - \l + \frac{l}{2}} } \,,
\end{aligned} \label{eq:FT_MI_c1}
\end{align}
where $C_n^{(\l)}(x)$ is the ultraspherical/Gegenbauer polynomial. Differentiation in $\l$ can be used to compute Fourier transform for logarithms,
\begin{align}
\begin{aligned}
    & \int \frac{d^D k_E}{(2 \pi)^D} [k_E^2]^\l (k_E \cdot a)^l \log(k_E^2) e^{i (k_E \cdot b)} = l! \left. \frac{\partial}{\partial \l} \int \frac{d^D k_E}{(2 \pi)^D} [k_E^2]^\l e^{i (k_E \cdot [b-ia])} \right|_{\CO(a^l)}
    \\ &\phantom{asdfasdf} = - \frac{(-4)^\l \l! \G (\frac{D}{2} + \l) \G (l + 1)}{\pi^{\frac{D}{2}}} \frac{i^l C_l^{(\frac{D}{2} + \l)}(\frac{(a \cdot b)}{(a^2 b^2)^{1/2}}) (a^2)^{l/2}}{ \left( b^2 \right)^{\frac{D}{2} + \l + \frac{l}{2}} } \,,
\end{aligned} \label{eq:FT_MI_c2}
\end{align}
where $\l \in \IZ^{\ge 0}$ is assumed.

\paragraph{Special cases: aligned spin}
Only even powers of $l$ are relevant for the aligned spin configuration, where $a \cdot b = 0$. The master integral \eqref{eq:FT_MI_c1} reduces to
\begin{align}
\begin{aligned}
    \int \frac{d^D k_E}{(2 \pi)^D} \frac{(k_E \cdot a)^{2l} e^{i (k_E \cdot b)}}{[k_E^2]^\l} &= \frac{\G (\frac{D}{2} - \l) }{2^{2\l} \pi^{\frac{D}{2}} \G (\l)} \frac{(\frac{D}{2} - \l)_l }{ \left( b^2 \right)^{\frac{D}{2} - \l} } \frac{(2l)!}{l!} \left( \frac{a^2}{b^2} \right)^l \,,
\end{aligned} \label{eq:FT_MI_c3}
\end{align}
where $(a)_n$ is the Pochhammer symbol. For axial scattering we set $a \cdot b = \pm \sqrt{a^2 b^2}$, which leads to
\begin{align}
\begin{aligned}
    \int \frac{d^D k_E}{(2 \pi)^D} \frac{(k_E \cdot a)^{2l} e^{i (k_E \cdot b)}}{[k_E^2]^\l} &= \frac{\G (D + 2l - 2\l) }{2^{D-1} \pi^{\frac{D - 1}{2}} \G (\l) \G(\frac{D+1}{2} - \l)} \frac{ 1 }{ \left( b^2 \right)^{\frac{D}{2} - \l} } \left( \frac{- a^2}{b^2} \right)^l \,.
\end{aligned} \label{eq:FT_MI_c4}
\end{align}

 
\section{2PL computations}
\label{app:iteration}

\subsection{Longitudinal part of the momentum kick} 

Our goal is to show that the longitudinal part of $\D_{(2)}p_1^\m$ 
agrees with 
\begin{align}
   \D_{(2)}p_1^\m|_\mathrm{iter} =  \frac{1}{2} \{ \chi_{(1)} , \{ \chi_{(1)} , p_1^\m \} \} = \frac{1}{2} \{ \chi_{(1)} , \D_{(1)} p_1^\m \} \,.
   \label{p1-iter-appendix}
\end{align}

\paragraph{Part 1} 

In the spin-less case, all iteration terms contain $\deltabar'$. That is no longer true when spin is turned on. 
Let us first focus on the new terms not containing $\deltabar'$. 

Using the following results as building blocks,  
\begin{align}
\begin{split}
     \{k\cdot b, \ell \cdot b\} = 
     \{k\cdot y, \ell \cdot y\} &=  \frac{\e[k,\ell ,y_1,v_1]}{m_1}  + \frac{\e[k,\ell ,y_2,v_2]}{m_2} \,,
     \\
     \left\{ k\cdot b , -\e^\m[\ell, v_1, v_2] \right\} &= \frac{\e^{\m}[k,\ell,v_2]}{m_1} + \frac{\e^{\m}[k,\ell,v_1]}{m_2} \,,
      \\
    \left\{\epsilon[k,v_1,v_2,y], \ell\cdot b \right\}  &= \frac{\e[k,\ell ,y,v_2]}{m_1}  + \frac{\e[k,\ell ,y,v_1]}{m_2}\,, 
    \\
    \left\{\epsilon[k,v_1,v_2,y], \ell\cdot y \right\} &= (k\cdot\ell) \left(\frac{v_2\cdot y_1}{m_1} - \frac{v_1\cdot y_2}{m_2}\right)  \,,
\end{split}
\end{align}
we collect four contributions to \eqref{p1-iter-appendix}:
\begin{align}
\begin{split}
    A_1^\m &= (i\ell^\m) \gamma^2 \left\{ \cosh(k\cdot y) e^{ik\cdot b} , \cosh(\ell\cdot y) e^{i\ell\cdot b} \right\} 
    \\
    &= (i\ell^\m) \gamma^2 \{k\cdot y, \ell \cdot y\} \sinh(k\cdot y) \sinh(\ell \cdot y) e^{iq\cdot b} 
     \\
    &\quad - (i\ell^\m) \gamma^2 \{k\cdot b, \ell \cdot b\}  \cosh(k\cdot y) \cosh(\ell \cdot y) e^{iq\cdot b} \,,
\end{split}
\end{align}
\begin{align}
\begin{split}
    A_4^\m &= \left\{ - i \epsilon[k,v_1,v_2,y] \frac{\sinh(k\cdot y)}{k\cdot y}   e^{ik\cdot b} , - \e^{\m}[\ell,v_1,v_2] \sinh(\ell\cdot y)  e^{i\ell\cdot b} \right\}
    \\
    &= i \epsilon[k,v_1,v_2,y] \e^{\m}[\ell,v_1,v_2] \{k\cdot y, \ell \cdot y\} \left[ \frac{\cosh(k\cdot y)}{k\cdot y} -\frac{\sinh(k\cdot y)}{(k\cdot y)^2}   \right] \cosh(\ell\cdot y) e^{iq\cdot b} 
     \\
    &\quad - i \epsilon[k,v_1,v_2,y] \e^{\m}[\ell,v_1,v_2] \{k\cdot b, \ell \cdot b\} \frac{\sinh(k\cdot y)}{k\cdot y}  \sinh(\ell\cdot y) e^{iq\cdot b} 
    \\
    &\quad + \epsilon[k,v_1,v_2,y]  \left( \frac{\e^{\m}[k,\ell,v_2]}{m_1} + \frac{\e^{\m}[k,\ell,v_1]}{m_2} \right) \frac{\sinh(k\cdot y)}{k\cdot y}  \sinh(\ell\cdot y)  e^{iq\cdot b}
    \\
    &\quad -  \left( \frac{\e[k,\ell ,y,v_2]}{m_1}  + \frac{\e[k,\ell ,y,v_1]}{m_2} \right) \e^{\m}[\ell,v_1,v_2] \frac{\sinh(k\cdot y)}{k\cdot y}  \sinh(\ell\cdot y)  e^{iq\cdot b}
    \\
    &\quad + i (k\cdot\ell) \left(\frac{v_2\cdot y_1}{m_1} - \frac{v_1\cdot y_2}{m_2}\right) \e^{\m}[\ell,v_1,v_2]  \frac{\sinh(k\cdot y)}{k\cdot y} \cosh(\ell\cdot y) e^{iq\cdot b}\,,
\end{split}
\end{align}
\begin{align}
\begin{split}
    A_2^\m &= \gamma \left\{ \cosh(k\cdot y) e^{ik\cdot b}, - \e^{\m}[\ell,v_1,v_2] \sinh(\ell\cdot y)  e^{i\ell\cdot b} \right\}
    \\
    &= - \gamma \,\e^{\m}[\ell,v_1,v_2] \{k\cdot y, \ell \cdot y\}  \sinh(k\cdot y) \cosh(\ell \cdot y) e^{iq\cdot b}
    \\
    &\quad + \gamma \,\e^{\m}[\ell,v_1,v_2] \{k\cdot b, \ell \cdot b\}  \cosh(k\cdot y) \sinh(\ell \cdot y) e^{iq\cdot b}
    \\
    &\quad + i \gamma \left( \frac{\e^{\m}[k,\ell,v_2]}{m_1} + \frac{\e^{\m}[k,\ell,v_1]}{m_2} \right) \cosh(k\cdot y) \sinh(\ell \cdot y) e^{iq\cdot b}  \,,
    \end{split}
\end{align}
\begin{align}
\begin{split}
    A_3^\m &= (i\ell^\m) \gamma \left\{ - i \epsilon[k,v_1,v_2,y] \frac{\sinh(k\cdot y)}{k\cdot y}   e^{ik\cdot b}, \cosh(\ell\cdot y) e^{i\ell\cdot b}\right\}
    \\
    &= \gamma \,\ell^\m \epsilon[k,v_1,v_2,y] \{k\cdot y, \ell \cdot y\}  \left[ \frac{\cosh(k\cdot y)}{k\cdot y} -\frac{\sinh(k\cdot y)}{(k\cdot y)^2}   \right] \sinh(\ell\cdot y) e^{iq\cdot b} 
    \\
    &\quad - \gamma \,\ell^\m \epsilon[k,v_1,v_2,y] \{k\cdot b, \ell \cdot b\}  \frac{\sinh(k\cdot y)}{k\cdot y}  \cosh(\ell\cdot y) e^{iq\cdot b} 
    \\
    &\quad + \gamma \,\ell^\m  (k\cdot\ell) \left(\frac{v_2\cdot y_1}{m_1} - \frac{v_1\cdot y_2}{m_2}\right) \frac{\sinh(k\cdot y)}{k\cdot y} \sinh(\ell\cdot y) e^{iq \cdot b}
    \\
    &\quad + i \gamma\, \ell^\m \left( \frac{\e[k,\ell ,y,v_2]}{m_1}  + \frac{\e[k,\ell ,y,v_1]}{m_2} \right) \frac{\sinh(k\cdot y)}{k\cdot y} \cosh(\ell\cdot y) e^{iq\cdot b}  \,,
\end{split}
\end{align}
As usual, we may work in the probe limit $(m_1/m_2 \rightarrow 0)$, where we get 
\begin{align}
\begin{split}
    m_1 A_1^\m 
    &= -(i\ell^\m) \gamma^2 \e[k,\ell ,y_1,v_1] \cosh[(k-\ell) \cdot y] e^{iq\cdot b} \,, 
    \\ 
    m_1 A_4^\m 
    &= -i \epsilon[k,v_1,v_2,y] \e^{\m}[\ell,v_1,v_2] \e[k,\ell ,y_1,v_1] \frac{\sinh(k\cdot y)}{(k\cdot y)^2}   \cosh(\ell\cdot y) e^{iq\cdot b}  
    \\
    &\quad + i \epsilon[k,v_1,v_2,y] \e^{\m}[\ell,v_1,v_2] \e[k,\ell ,y_1,v_1]  \frac{\cosh[(k-\ell) \cdot y]}{k\cdot y}    e^{iq\cdot b}  
    \\
    &\quad + \epsilon[k,v_1,v_2,y]  \e^{\m}[k,\ell,v_2] \frac{\sinh(k\cdot y)}{k\cdot y}  \sinh(\ell\cdot y)  e^{iq\cdot b}
    \\
    &\quad -   \e[k,\ell ,y,v_2]   \e^{\m}[\ell,v_1,v_2] \frac{\sinh(k\cdot y)}{k\cdot y}  \sinh(\ell\cdot y)  e^{iq\cdot b}
    \\
    &\quad + i (k\cdot\ell) (v_2\cdot y_1) \e^{\m}[\ell,v_1,v_2]  \frac{\sinh(k\cdot y)}{k\cdot y} \cosh(\ell\cdot y) e^{iq\cdot b}\,,
\end{split}
\end{align}
\begin{align}
\begin{split}
    m_1 A_2^\m 
    &= - \gamma \,\e^{\m}[\ell,v_1,v_2] \e[k,\ell ,y_1,v_1]  \sinh[(k-\ell) \cdot y] e^{iq\cdot b}
    \\
    &\quad + i \gamma  \e^{\m}[k,\ell,v_2]\cosh(k\cdot y) \sinh(\ell \cdot y) e^{iq\cdot b}  \,,
    \\
    m_1 A_3^\m 
    &= - \gamma \,\ell^\m \epsilon[k,v_1,v_2,y] \e[k,\ell ,y_1,v_1] \frac{\sinh(k\cdot y)}{(k\cdot y)^2} \sinh(\ell\cdot y) e^{iq\cdot b} 
    \\
    &\quad - \gamma \,\ell^\m \epsilon[k,v_1,v_2,y] \e[k,\ell ,y_1,v_1] \frac{\sinh[(k-\ell)\cdot y]}{k\cdot y}  e^{iq\cdot b} 
    \\
    &\quad + \gamma \,\ell^\m  (k\cdot\ell) (v_2\cdot y_1) \frac{\sinh(k\cdot y)}{k\cdot y} \sinh(\ell\cdot y) e^{iq \cdot b}
    \\
    &\quad + i \gamma\, \ell^\m  \e[k,\ell ,y,v_2]   \frac{\sinh(k\cdot y)}{k\cdot y} \cosh(\ell\cdot y) e^{iq\cdot b}  \,.
\end{split}
\end{align}
We can simplify $A_2^\m$ a bit and write 
\begin{align}
\begin{split}
    m_1 A_2^\m 
    &= - \gamma \, \ell^\m (k\cdot \ell)(v_2\cdot y_1)   \sinh[(k-\ell) \cdot y] e^{iq\cdot b}
    \\
    &\quad \rred{ + i \gamma  \e^{\m}[k,\ell,v_2]\cosh(k\cdot y) \sinh(\ell \cdot y) e^{iq\cdot b} }  \,,
\end{split}
\label{red-12}
\end{align}
Next, using \eqref{kdoty-identity}, we try to remove $(k\cdot y)$ factors in the denominators of $A_3^\m$ and $A_4^\m$:
\begin{align}
\begin{split}
    m_1 A_3^\m 
    &= - \gamma \,\ell^\m  (k\cdot\ell) (v_2\cdot y_1)  \sinh[(k-\ell) \cdot y] e^{iq\cdot b}
    \\
    &\quad \rred{ + i \gamma\, \ell^\m  \e[k,\ell ,y,v_2]   \frac{\sinh(k\cdot y)}{k\cdot y} \cosh(\ell\cdot y) e^{iq\cdot b}  }\,,
\\
       m_1 A_4^\m 
    &= (i\ell^\m)  (v_2\cdot y_1)  \e[k, \ell,v_1,v_2]  \cosh[(k-\ell) \cdot y]  e^{iq\cdot b}  
    \\
    &\quad \rred{+ \epsilon[k,v_1,v_2,y]  \e^{\m}[k,\ell,v_2] \frac{\sinh(k\cdot y)}{k\cdot y}  \sinh(\ell\cdot y)  e^{iq\cdot b}  }
    \\
    &\quad \rred{ -   \e[k,\ell ,y,v_2]   \e^{\m}[\ell,v_1,v_2] \frac{\sinh(k\cdot y)}{k\cdot y}  \sinh(\ell\cdot y)  e^{iq\cdot b} }\,.
\end{split}
\label{red-34}
\end{align}

We have enumerated all terms \emph{not} containing $\deltabar'$. 
Now we split them into two parts: the ``y-part" and the ``v-part". The former is linear in $y_1$ while the latter  (marked red in the equations above) is independent of $y_1$. The $k\cdot y$ factor in the denominator is to be cancelled against an $\epsilon[\cdot, \cdot, \cdot, y]$ factor in the numerator.

The y-part gives fairly simple expressions:
\begin{align}
\begin{split}
    m_1 A_{1^+4}^\m|_y &= (i\ell^\m) (- \gamma^2 \e[k,\ell ,y_1,v_1] + (v_2 \cdot y) \e[k,\ell,v_1,v_2] ) \cosh[(k-\ell)\cdot y] e^{iq\cdot b} \,, 
     \\
       m_1 A_{2+3}^\m|_y &= (i\ell^\m) (2 i \g) (v_2\cdot y_1) (k\cdot \ell) \sinh[(k-\ell)\cdot y] e^{iq\cdot b} \,. 
\end{split}
\end{align}
There agree perfectly with the longitudinal part of \eqref{2PL-impulse-y}, \eqref{2PL-impulse-y-N} in the main text.

\paragraph{Part 2} 

The v-part is more involved, as it gets combined with the $\deltabar'$ terms.
In the probe limit, the $\deltabar'$ factor comes from
\begin{align}
\begin{split}
     \left\{ e^{ik\cdot b} \deltabar(k\cdot v_1) ,  e^{i\ell\cdot b} \deltabar(\ell \cdot v_1)  \right\} 
    \; \rightarrow \;\; & i (k\cdot \ell) \frac{e^{iq\cdot b}}{m_1} \left[ \deltabar(k\cdot v_1) \deltabar'(\ell\cdot v_1) - \deltabar'(k\cdot v_1) \deltabar(\ell\cdot v_1) \right]   
\\
&= i (k\cdot \ell) \frac{e^{iq\cdot b}}{m_1} \deltabar(q\cdot v_1) \deltabar'(\ell\cdot v_1) \,.
\end{split}
\end{align}
This factor is to be multiplied by 
\begin{align}
\begin{split}
        B^\m &= \left[  \gamma \cosh(k\cdot y) - i \epsilon[k,v_1,v_2,y]\frac{\sinh(k\cdot y)}{k\cdot y}   \right] 
        \\
        &\quad \times \left[(i\ell^\m) \g \cosh(\ell\cdot y) -\e^\m[\ell,v_1,v_2] \sinh(\ell\cdot y)  \right] 
        \\
        &= B_1^\m + B_2^\m + B_3^\m +B_4^\m  \,,
\end{split}
\end{align}
where 
\begin{align}
\begin{split}
        B_1^\m 
        &=(i\ell^\m) \g^2 \cosh(k\cdot y) \cosh(\ell\cdot y) \,,
        \\
        B_4^\m &= i \epsilon[k,v_1,v_2,y] \e^\m[\ell,v_1,v_2] \frac{\sinh(k\cdot y)}{k\cdot y}  \sinh(\ell\cdot y) \,,
        \\
        B_2^\m &=  - \g \e^\m[\ell,v_1,v_2] \cosh(k\cdot y) \sinh(\ell\cdot y) \,,
        \\
        B_3^\m &= \g \ell^\m \epsilon[k,v_1,v_2,y]\frac{\sinh(k\cdot y)}{k\cdot y} \cosh(\ell\cdot y)  \,.
\end{split}
\end{align}
Multiplying them by $i(k\cdot \ell)$, and hiding $e^{iq\cdot b} \deltabar(q\cdot v_1) \deltabar'(\ell\cdot v_1)$ for now, we get 
\begin{align}
\begin{split}
   D_1^\m 
        &= - \ell^\m  \g^2 (k\cdot \ell) \cosh(k\cdot y) \cosh(\ell\cdot y) \,,
        \\
        D_4^\m &= -  (k\cdot \ell)  \epsilon[k,v_1,v_2,y] \e^\m[\ell,v_1,v_2] \frac{\sinh(k\cdot y)}{k\cdot y}  \sinh(\ell\cdot y) \,,
        \\
        D_2^\m &=  -i  \g (k\cdot \ell) \e^\m[\ell,v_1,v_2] \cosh(k\cdot y) \sinh(\ell\cdot y) \,,
        \\
        D_3^\m &= i \g \ell^\m (k\cdot \ell) \epsilon[k,v_1,v_2,y]\frac{\sinh(k\cdot y)}{k\cdot y} \cosh(\ell\cdot y)  \,.  
\end{split}
\end{align}
Now, we bring the red colored terms from \eqref{red-12} and \eqref{red-34} and apply the identity 
\begin{align}
    \delta(x) = - x \delta'(x) \,.
\end{align}
After pulling out some overall factor, we record the results as
\begin{align}
\begin{split}
    C_1^\m &= 0 \,,
    \\
    C_4^\m 
    &= -(\ell\cdot v_1) \epsilon[k,v_1,v_2,y]  \e^{\m}[k,\ell,v_2] \frac{\sinh(k\cdot y)}{k\cdot y} \sinh(\ell\cdot y)  
    \\
    &\quad  - (k\cdot v_1)    \e[k,\ell ,y,v_2]   \e^{\m}[\ell,v_1,v_2] \frac{\sinh(k\cdot y)}{k\cdot y} \sinh(\ell\cdot y)  \,, 
    \\
    C_2^\m &= -  i \gamma (\ell\cdot v_1)  \e^{\m}[k,\ell,v_2]\cosh(k\cdot y) \sinh(\ell \cdot y)    \,,
    \\
    C_3^\m 
    &= +  i \gamma (k\cdot v_1)  \ell^\m  \e[k,\ell ,y,v_2]   \frac{\sinh(k\cdot y)}{k\cdot y} \cosh(\ell\cdot y)  \,.
\end{split}
\end{align}
For $C_3^\m$ and the second line of $C_4^\m$, we used $\deltabar(k\cdot v_1 + \ell\cdot v_1)$ to replace $(\ell\cdot v_1)$ by $-(k\cdot v_1)$. 
Merging all the $C$-terms and the $D$-terms, 
we obtain the final result in perfect agreement with the longitudinal part of $\D_{(2v)} p^\m$ in \eqref{2PL-impulse-vo}, \eqref{2PL-impulse-vs} in the main text.

\subsection{Transverse part of the spin kick} \label{app:spin kick more}

In section~\ref{sec:2PL spin kick} of the main text, we showed how to compute the 2PL spin kick. 
In this appendix, we give some details of the computation and confirm that 
the transverse part of the spin kick agrees with the eikonal formula, 
\begin{align}
\begin{split}
    \{ \chi_{(n)} , y_1^\m \} &= \frac{1}{m_1} \left[ v_1^\mu y_1^\nu  \frac{\partial}{\partial x_1^\nu} 
    +  \e^{\m\n}[ v_1, y_1] \frac{\partial}{\partial y_1^\n } \right] \chi_{(n)} \,.
\end{split}
\label{eikonal-to-impulse-copy2}
\end{align}
We begin with the overall structure of the 2PL spin kick \eqref{2PL spin kick overview}:
\begin{align}
    \Delta_{(2)} y_1^\m = \frac{(q_1q_2)^2}{m_1^2} \int_{q_\perp} e^{iq\cdot b} \int_\ell \frac{\deltabar(v_2\cdot \ell)}{k^2\ell^2(ik\cdot v_1+0^+)^2} \CN^\m \,.
    \label{2PL spin kick overview copy}
\end{align}
The numerator $\CN^\m$ can be computed separately for each term in \eqref{many-terms-for-y}. 
For (a) and (b) terms, 
we also distinguish the same/opposite helicity contributions. 
\begin{align}
\begin{split}
      \CN^\m_{(2ao)} =  \left( \mathrm{ch}\!\boxminus C_{ao} +  \mathrm{sh}\!\boxminus S_{ao}  \right)^\m \,,
      &\quad 
       \CN^\m_{(2as)}  =  \left( \mathrm{ch}\!\boxplus C_{as} +  \mathrm{sh}\!\boxplus  S_{as}  \right)^\m \,,
    \\  
     \CN^\m_{(2bo)} = \left( \mathrm{ch}\!\boxminus C_{bo} +  \mathrm{sh}\!\boxminus S_{bo} \right)^\m \,,
     &\quad 
         \CN^\m_{(2bs)} = \left( \mathrm{ch}\!\boxplus  C_{bs} +  \mathrm{sh}\!\boxplus  S_{bs}  \right)^\m \,,
          \\
      \CN^\m_{(2c)} =  \left( \mathrm{ch}\!\boxminus C_c +  \mathrm{sh}\!\boxminus S_c  \right)^\m \,,
    &\quad 
     \CN^\m_{(2d)} = \left( \mathrm{ch}\!\boxminus C_d +  \mathrm{sh}\!\boxminus S_d  \right)^\m \,,
\end{split}
\label{C-S-12-copy}
\end{align}
The same helicity terms are  
\begin{align}
\begin{split}
C^\m_{as} &= \frac{1}{2} (ik\cdot v_1)\left[ k^\m (\ell \cdot y_1) - \ell^\m (k\cdot y_1) -y_1^\m (k\cdot \ell) \right] \,,
\quad
S^\m_{as} = \frac{i}{2}(ik\cdot v_1) \e^\m[k,\ell,y_1] \,, 
\\
C^\m_{bs} &= \frac{i}{2}\left[ \ell^\m (v_2\cdot y_1) - v_2^\m (\ell \cdot y_1) \right] \gamma (k\cdot \ell) + \frac{i}{2} \e^\m[\ell,v_2,y_1]\e[k,\ell,v_1,v_2] \,,
\\
S^\m_{bs} &= \frac{1}{2} \left[ \ell^\m (v_2\cdot y_1) - v_2^\m (\ell \cdot y_1) \right]\e[k,\ell,v_1,v_2] -\frac{1}{2} \e^\m[\ell,v_2,y_1]\gamma (k\cdot \ell) \,.
\end{split}
\end{align}
Among the opposite helicity terms, (a) and (b) terms are linear in $y_1$:
\begin{align}
\begin{split}
C^\m_{ao} &= \frac{1}{2} (ik\cdot v_1) \left[ - k^\m (\ell \cdot y_1) - \ell^\m (k\cdot y_1) + y_1^\m (k\cdot \ell) + 2 v_2^\m (v_2\cdot y_1) (k\cdot \ell) \right] \,,
\\
S^\m_{ao} &=  \frac{i}{2} (ik\cdot v_1)\left[  v_2^\m \e[k,\ell,y_1,v_2] +\e^\m[k,\ell,v_2] (v_2\cdot y_1)\right] \,,
\\
C^\m_{bo} &= \frac{i}{2}\left[ \ell^\m (v_2\cdot y_1) - v_2^\m (\ell \cdot y_1) \right] \gamma (k\cdot \ell) - \frac{i}{2} \e^\m[\ell,v_2,y_1]\e[k,\ell,v_1,v_2]\,,
\\ 
S^\m_{bo} &=  \frac{1}{2} \left[ \ell^\m (v_2\cdot y_1) - v_2^\m (\ell \cdot y_1) \right]\e[k,\ell,v_1,v_2] +\frac{1}{2} \e^\m[\ell,v_2,y_1]\gamma (k\cdot \ell) \,,
\end{split}
\end{align}
whereas (c), (d) terms are quadratic in $y_1$:
\begin{align}
\begin{split}
   C^\m_c &= (ik\cdot v_1) i \left[  \left[ \ell^\m (v_2\cdot y_1) - v_2^\m (\ell\cdot y_1) \right] \e[k,\ell,y_1,v_2] - \e^\m[\ell,v_2,y_1](k\cdot \ell)(v_2\cdot y_1) \right] \,,
      \\
      S^\m_c &= (ik\cdot v_1) \left[  \left[ \ell^\m (v_2\cdot y_1) - v_2^\m (\ell\cdot y_1) \right] (k\cdot \ell)(v_2\cdot y_1) - \e^\m[\ell,v_2,y_1] \e[k,\ell,v_2,y_1]  \right]\,, 
\\
      C^\m_d &= (ik\cdot v_1)^2 \left[ (\ell\cdot v_1) (v_2\cdot y_1) + \gamma (\ell\cdot y_1) \right] \e^\m[k,v_2,y_1] 
      \\
      &\qquad + (ik\cdot v_1)^2 \e[\ell,v_1,v_2,y_1] \left[ k^\m (v_2\cdot y_1) - v_2^\m (k\cdot y_1) \right]   \,,
      \\
      S^\m_d &= (ik\cdot v_1)^2 i \left[ (\ell\cdot v_1)(v_2\cdot y_1) + \gamma (\ell\cdot y_1) \right]\left[ k^\m (v_2\cdot y_1) - v_2^\m (k\cdot y_1) \right]  
      \\
      &\qquad 
      - (ik\cdot v_1)^2 i \e[\ell,v_1,v_2,y_1] \e^\m[k,v_2,y_1] \,.
\end{split}
\end{align}

\paragraph{Same helicity sector}

The 2PL eikonal \eqref{2PL-eikonal-final} contains a single term in the same helicity sector, so its contribution to the spin kick should be also quite simple. Indeed, after a lot of cancellations, we get 
\begin{align}
    \begin{split}
        C^\m_{as} + C^\m_{bs} &= - \frac{i}{2} (k\cdot \ell) \left[ v^\m_1 (\ell\cdot y_1)- \textcolor{red}{y^\m_1 (\ell\cdot v_1)}   \right] \,,
        \\
        S^\m_{as} + S^\m_{bs} &= - \frac{1}{2} (k\cdot \ell) \e^\m[v_1,y_1,\ell] \,.
    \end{split}
\end{align}
Not all terms contribute to the transverse part; $\Delta_{(2)}y^\m|_\mathrm{tr}$ should be orthogonal to $y_1^\m$. The non-orthogonal term, marked red in the equation above, is projected out upon symmetrisation under the exchange operation $k\leftrightarrow \ell$, 
\begin{align}
     (\ell\cdot v_1)  (k\cdot \ell) \cosh\boxplus 
    &\;\; \rightarrow \;\;  
    (q\cdot v_1)  (k\cdot \ell)\cosh\boxplus = 0 \,.
\end{align}
The equality hold in the $(q_\perp)$ integral. 
What is left after the symmetrisation is 
\begin{align}
        C^\m_{as+bs}|_\mathrm{tr} = - \frac{i}{2} (k\cdot \ell)  v^\m_1 (\ell\cdot y_1) \,,
        \quad
        S^\m_{as+bs}|_\mathrm{tr} = - \frac{1}{2} (k\cdot \ell) \e^\m[v_1,y_1,\ell] \,.
\end{align}
These match the expectation from the 2PL eikonal \eqref{eikonal-to-impulse-copy2} perfectly.

\paragraph{Opposite helicity sector} 

We treat the (a)-(b) group and the (c)-(d) group separately.
In the (a)-(b) group, partial cancellations leave us with 
\begin{align}
    \begin{split}
        C^\m_{ao} + C^\m_{bo} &= i (k\cdot \ell) \g \left[ \ell^\m (v_2\cdot y_1) - v_2^\m(\ell\cdot y_1)  \right] + \frac{i}{2} (k\cdot \ell) v_1^\m (\ell\cdot y_1)
        \\
        &\quad 
        +i \ell^\m (\ell\cdot v_1)  (k\cdot y_1)  - \frac{i}{2} (k\cdot \ell) (\ell\cdot v_1) \textcolor{red}{\left[ 2 v_2^\m (v_2\cdot y_1) +  y_1^\m  \right]}  \,,
        \\
        S^\m_{ao} + S^\m_{bo} &=  \left[ \ell^\m (v_2\cdot y_1) - v_2^\m(\ell\cdot y_1) \right] \e[k,\ell,v_1,v_2] 
        + \frac{1}{2} (k\cdot \ell) \e^\m[\ell,v_1,y_1]  
        \\
        &\quad + \textcolor{red}{v_2^\m (\ell\cdot v_1) \e[k,\ell,y_1,v_2]} \,.
    \end{split}
\end{align}
Again, the non-orthogonal terms, marked red in the equations above, are projected out upon symmetrisation under the exchange $k\leftrightarrow \ell$.
We are left with 
\begin{align}
     \begin{split}
        C^\m_{ao+bo}|_\mathrm{tr} &= i (k\cdot \ell) \g \left[ q^\m (v_2\cdot y_1) - v_2^\m(q\cdot y_1)  \right] + \frac{i}{2} (k\cdot \ell) v_1^\m (q\cdot y_1) 
        \\
        &\quad + \frac{i}{2}[(k-\ell)\cdot v_1] \left[ k^\m (\ell \cdot y_1) - \ell^\m (k \cdot y_1) \right] \,,
        \\
        S^\m_{ao+bo}|_\mathrm{tr} &=  \left[ q^\m (v_2\cdot y_1) - v_2^\m(q\cdot y_1) \right] \e[k,\ell,v_1,v_2] 
        - \frac{1}{2} (k\cdot \ell) \e^\m[k-\ell,v_1,y_1]   \,.
    \end{split}
\end{align}
On the other hand, the transverse spin kick derived from the eikonal by \eqref{eikonal-to-impulse-copy2} contains 
\begin{align}
     \begin{split}
        C^\m_\chi|_\mathrm{tr} &=  i (q\cdot y_1)  v_1^\m \left[ - (\g^2-1/2) (k\cdot \ell) +  (k\cdot v_1)(\ell \cdot v_1)  \right] 
        \\
        &\quad + i\g \e^\m[ k-\ell, v_1,y_1] \e[k,\ell,v_1,v_2] 
        \\
        &\quad - i[(k-\ell)\cdot v_1] \e^\m[v_1,v_2,y_1] \e[k,\ell,v_1,v_2] 
        \\
        &\quad - \frac{i}{2} [(k-\ell)\cdot v_1] \e^{\m\n}[v_1,y_1] \e_\n[k,\ell,v_1] \,,
        \\
        S^\m_\chi|_\mathrm{tr} 
        &=   - v_1^\m (q\cdot y_1) \g \e[k,\ell,v_1,v_2] 
        \\
        &\quad  - (\g^2-1/2) (k\cdot \ell) \e^\m[k-\ell,v_1,y_1]  
        + \g [(k-\ell)\cdot v_1]  (k\cdot \ell) \e^\m[v_1,v_2,y_1] \,.
    \end{split} 
\end{align}
Despite appearances, things do match as expected, 
\begin{align}
 C^\m_{ao+bo}|_\mathrm{tr} = C^\m_\chi|_\mathrm{tr} \,.
 \quad 
   S^\m_{ao+bo}|_\mathrm{tr} = S^\m_\chi|_\mathrm{tr} \,.
\end{align}

We can repeat the same exercise for the (c)-(d) group. The computations are even lengthier and not particularly illuminating, 
so we omit the details here.


\section{Regularisation for the product of time-symmetric Green's functions} \label{app:GFprod}
The usual time-symmetric $i0^+$ prescription for $\w^{-2}$ propagators is
\begin{align}
    \frac{1}{2} \left( \frac{1}{(\w + i0^+)^2} + \frac{1}{(\w - i0^+)^2} \right) \quad \Leftrightarrow \quad - \frac{1}{2} |\s| \,, \label{eq:GF1}
\end{align}
where we use the positive frequency expansion $f(\s) = \int_\w f(\w) e^{- i \w \s}$. The square of the time-symmetric $\w^{-1}$ propagator is given as
\begin{align}
    \left[ \frac{1}{2} \left( \frac{1}{\w + i0^+} + \frac{1}{\w - i0^+} \right) \right]^2 \quad \Leftrightarrow \quad - \frac{1}{4} |\s| + \frac{e^{- 0^+ \times |\s|}}{4 0^+} = \frac{1}{4 0^+} - \frac{1}{2} |\s| + \CO[(0^+)^1]\,, \label{eq:GF2}
\end{align}
where we expanded the expression as a Laurent series in the regulator $0^+$ and kept up to $\CO[(0^+)^0]$ terms. Employing the ``minimal subtraction'' scheme for the regulator $0^+$, we throw out the divergent term in $0^+$ and conclude that the propagators \eqref{eq:GF1} and \eqref{eq:GF2} are equivalent as distributions.

The reason \eqref{eq:GF2} has a divergent contribution compared to \eqref{eq:GF1} is because it should be understood as the convolution of the time-symmetric $\w^{-1}$ propagator in the time domain,
\begin{align}
    \frac{1}{2} \left( \frac{1}{\w + i0^+} + \frac{1}{\w - i0^+} \right) \quad \Leftrightarrow \quad - \frac{i}{2} \text{sgn} (\s) \,, \label{eq:GF3}
\end{align}
where $\text{sgn} (\s)$ is the sign function. Unlike the retarded/advanced Green's functions given by the Heaviside step function, the convolution of \eqref{eq:GF3} with itself diverges due to the ``infinite volume'' of the real line, which manifests itself as the $(0^+)^{-1}$ divergence in \eqref{eq:GF2}.

\newpage
\bibliographystyle{JHEP}
\bibliography{reference}

\end{document}